\documentclass{article}
\usepackage[utf8]{inputenc}
\usepackage{graphicx}
\usepackage{tikz-timing}
\usepackage{hyperref}
\usepackage{pdfpages}
\usepackage{lineno}
\usepackage{array}
\usepackage{amsmath}
\usepackage{adjustbox}
\usepackage{isotope}
\usepackage{tikz}
\usepackage{caption}
\usepackage{booktabs}
\usepackage{verbatim}
\usetikzlibrary{circuits.logic.US,positioning,shapes,fit,calc,arrows}
\usepackage{circuitikz}
\usepackage[capitalise,nameinlink,noabbrev]{cleveref}
\usepackage[capitalise]{cleveref}
\usepackage{paralist}
\usepackage{lineno}

\usepackage{float}
\usepackage{subfig}

\RequirePackage[left=1.in,right=1.in,top=1.in,bottom=1.in]{geometry}

\graphicspath{ {pictures/} }

\newcommand{\degree}    {^{\circ}}

\title{Accelerator Neutrino Neutron Interaction Experiment (ANNIE): Physics Phase Proposal}

\begin{document}
\maketitle


{ }
\vspace{1.0in}
{ }

\begin{center}
{\Large \bf
Phase~II Physics Proposal for T-1063:\\
The Accelerator Neutrino Neutron Interaction Experiment (ANNIE)\\
\medskip
\medskip
\
\bigskip
}
\vspace{0.6in}
 \centerline{\today}
\end{center}

\begin{center}
\includegraphics[width=1.0 \linewidth]{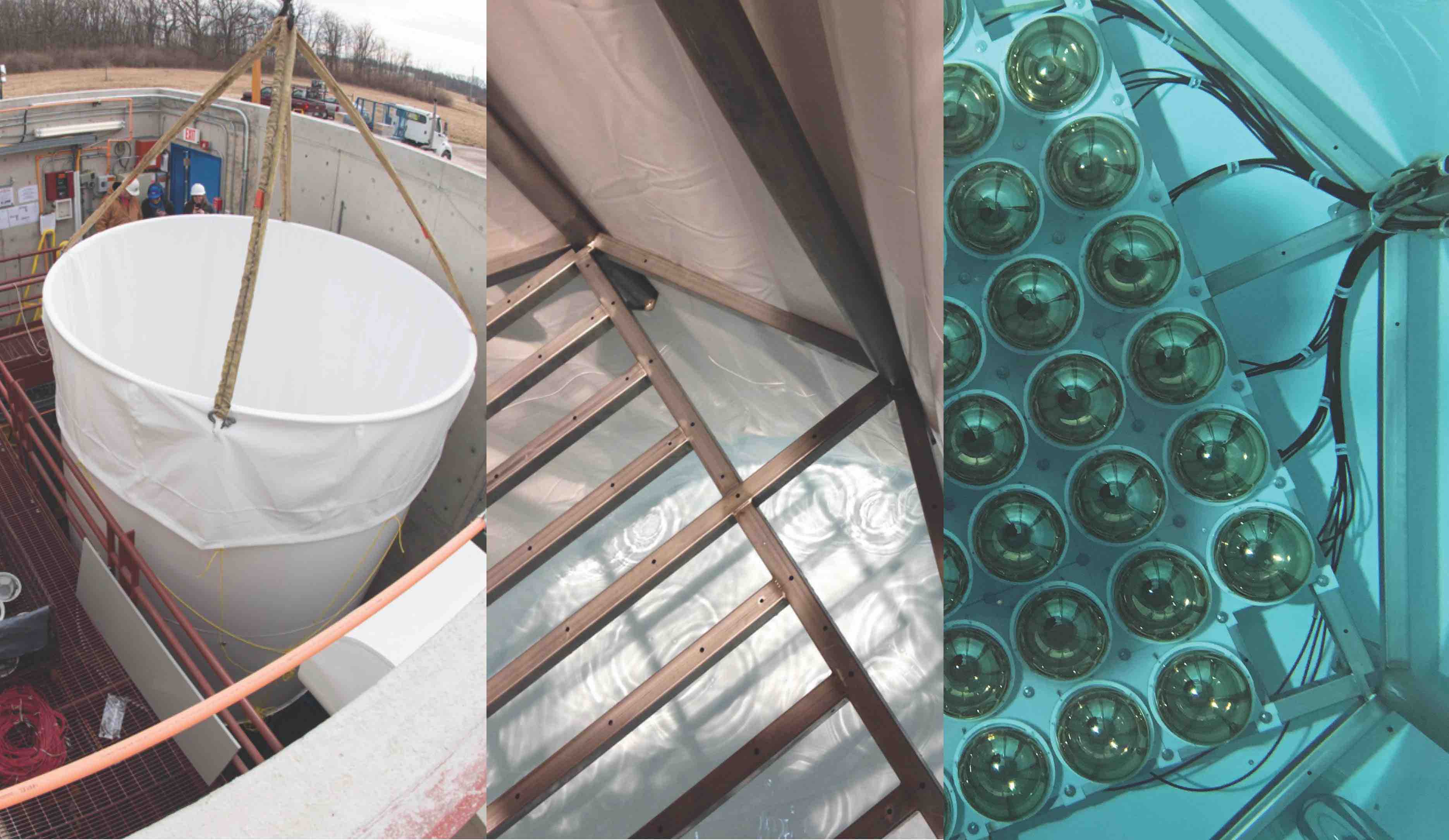}

\end{center}


\newpage 
\includepdf[pages={1}]{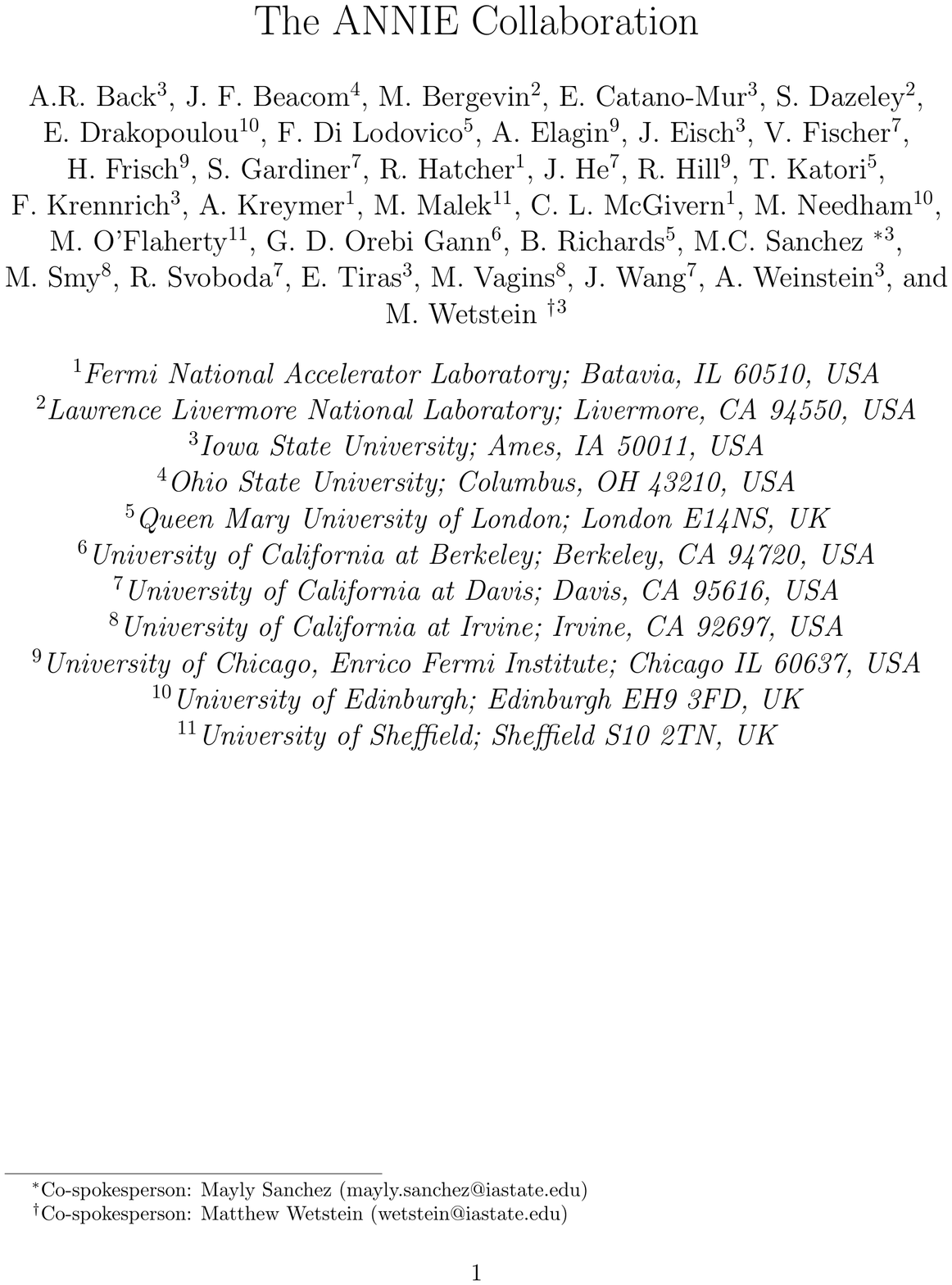}

\newpage

\pagenumbering{roman}

\noindent

\tableofcontents
\setcounter{tocdepth}{3}

\newpage
\pagenumbering{arabic}

\section{Introduction}
\label{sec:Introduction}

This proposal is being submitted to the Fermilab PAC and the Department of Energy to realize the physics and R\&D mission of the Accelerator Neutrino Neutron Interaction Experiment (ANNIE): (1) an important measurement of neutrino-nucleus interactions, focusing specifically on neutron production, and (2) an R\&D effort focused on using new photodetector technology and chemical enhancements to make advanced water-base neutrino detectors. The ANNIE experiment consists of a small Water Cherenkov detector, instrumented with both conventional photomultiplier tubes (PMTs) and Large Area Picosecond Photodetectors (LAPPDs), and deployed on the Booster Neutrino Beam (BNB) at Fermilab.  The experiment was designed to proceed in two stages: a partially-instrumented test-beam run using only PMTs (Phase~I) for the purpose of measuring critical neutron backgrounds to the experiment~\cite{{FNALPAC},{FNALPAC2}}; and a physics run with a fully-instrumented detector (Phase~II).

Phase~I of ANNIE was approved by the Fermilab PAC in late February of 2015 and and has been operating since April 2016, supported in part by funding through the DOE Intermediate Neutrino Program (INP).  The ANNIE collaboration has now measured the neutron backgrounds and shown them to be low enough to proceed.  In addition, the main technological component of Phase~II, LAPPDs, are now being produced by Incom Inc. Thus, the ANNIE collaboration is ready to move forward with the detector modifications to enable Phase~II. 

With most of the infrastructure already built and tested, the bulk of the changes to the detector will consist of the addition of LAPPDs, more conventional PMTs, as well as adding electronics channels to the existing DAQ. Funding for these equipment expenses is being requested from the DOE. \textbf{The Fermilab request is for continued operational support, use of the SciBooNE Hall, and modest FY18 engineering and technical support on a smaller scale than was needed to execute Phase~I.}

\subsection{Physics Motivation}
\label{sub:physics_motivation}

The primary physics goal of ANNIE is to study the multiplicity of final state neutrons from neutrino-nucleus interactions in water. ANNIE provides a unique opportunity to study this physics in a controlled beam experiment. 
Identifying and counting final state neutrons provides a new and critical experimental handle for the systematic uncertainties of the neutrino energy reconstruction in oscillation experiments and for signal-background separation for future neutrino experiments. The ANNIE physics program is in fact complementary to similar measurements of proton multiplicity in the BNB using liquid argon time projection chambers (LAr-TPCs).




\subsubsection{Neutrons and Long Baseline Oscillation Physics}

A key requirement in precision long-baseline physics will be the accurate estimation of the neutrino energy using the final state lepton. Experimentally, charged-current quasi-elastic (CCQE) interactions are often identified by the presence of a high energy lepton and no other particles other than the nuclear recoil. However, a number of intermediate states and three-body interactions can lead to inelastic interactions that mimic CCQE. Of particular interest are the so-called two particle-two hole (2p-2h) interactions, where the neutrino scatters off a correlated pair of nucleons, rather than a single nucleon~\cite{{Benhar},{Martini},{Nieves}}. The reconstructed energy from the lepton momentum in such events underestimates the neutrino energy. 
Increasingly, it has become clear that understanding these multi-nucleon interactions is necessary to explain observed cross sections~\cite{MB1,MB2,Minerva2p2h}. A growing body of literature finds that these contributions have a significant effect in smearing the oscillation spectrum of long-baseline experiments~\cite{{oscillationsE},{oscillationsE2}}. 


The presence of extra final-state nucleons, beyond the one recoil nucleon predicted to first order, is a strong indicator of the inelasticity of events with CCQE-like characteristics. Understanding proton and neutron multiplicity can aid in better understanding neutrino energy reconstruction and in identifying a more robust sample of CCQE events. Moreover, the multiplicity of final state nucleons and the relative abundance of neutrons and protons in two-nucleon final states also provide a strong handle for constraining the widely varying models of the mechanism and the frequency of these interactions. 

\begin{figure}[htb]
       	\begin{center}
		\includegraphics[width=0.5\linewidth]{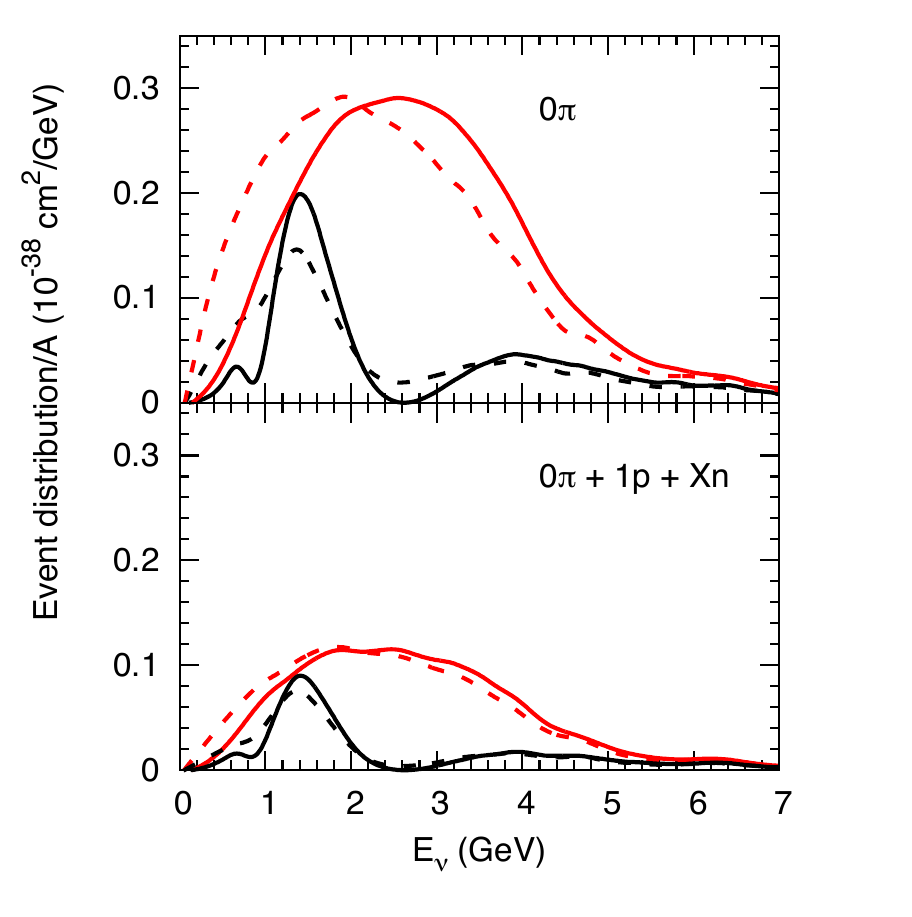} 
	\end{center}
	\caption{A figure taken from Ref~\cite{oscillationsE2}, showing how a more pure CCQE sample selected on the basis of exactly one final state proton (BOTTOM) shows better agreement between the reconstructed (dashed) and true (solid) neutrino energies for oscillated (black) and unoscillated (red) spectra. }
	\label{mosel-lalakulich}
\end{figure}

Figure~\ref{mosel-lalakulich} (from Ref~\cite{oscillationsE2}) uses simulations to illustrate how dramatic this effect might be. For an inclusive sample of CCQE-like events (top panel), defined by a high energy lepton and no visible pions, there is wide discrepancy between the reconstructed and true neutrino energy spectrum.  Using a more restrictive definition of CCQE with the added requirement of exactly one final-state (FS) proton (bottom panel) yields much better agreement between the reconstructed and true energy. This follows from a significant enhancement in true CCQE composition and a reduction of ``stuck-pion" events, where an emitted charged pion is reabsorbed in the nucleus. Such events typically produce more than one neutron~\cite{ransometalk}. 

\begin{figure}[htb]
	\centering

	  	\includegraphics[width=0.7\linewidth]{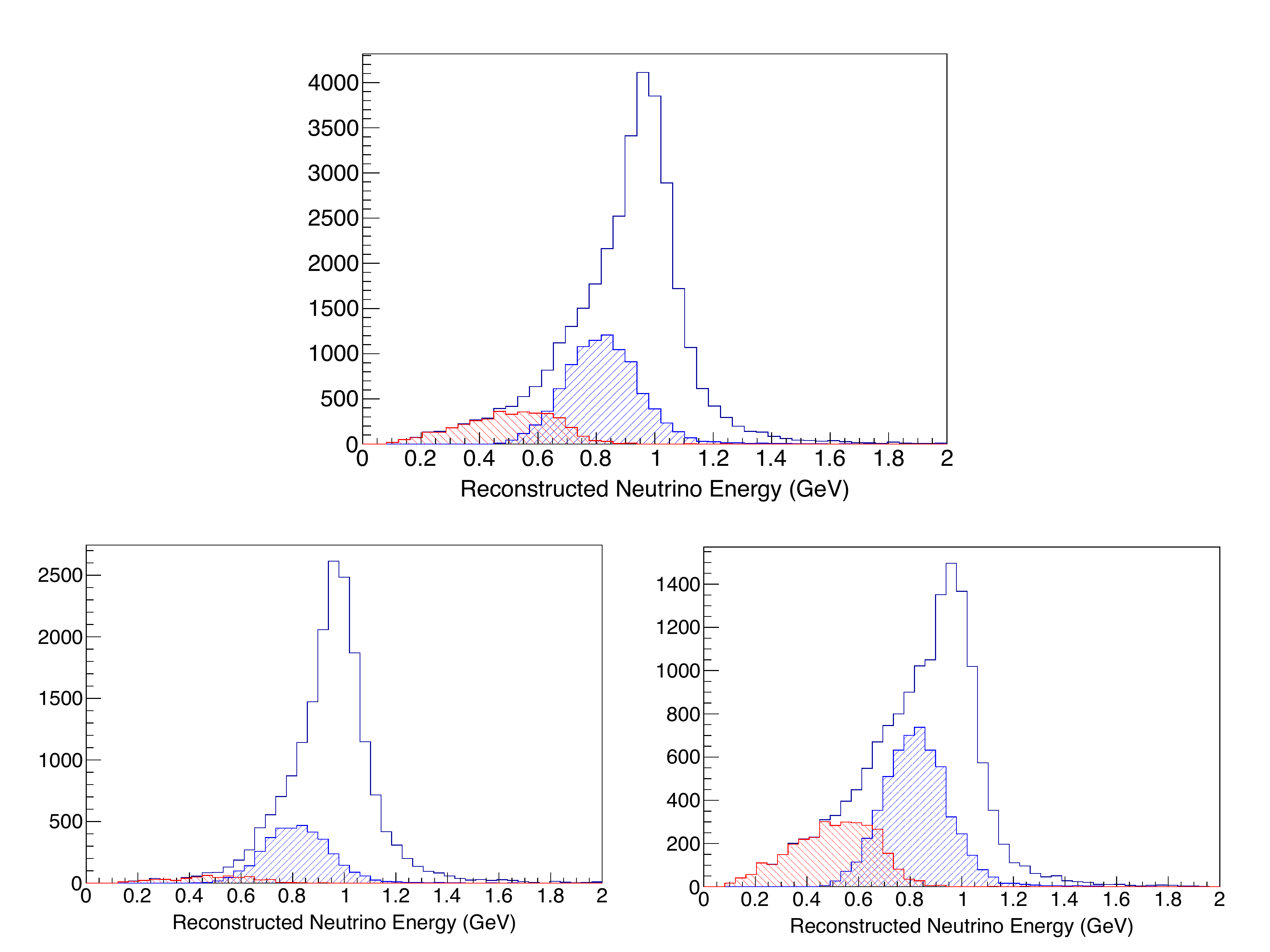}	
        
	\caption{TOP: GENIE \cite{genie1, genie2} Monte Carlo simulation showing the reconstructed energy distribution of 1 $\mu$ + 0$\pi$ events from a 1 GeV monoenergetic neutrino beam, derived from muon kinematics assuming elasticity. The overall shape is shown in black, with the red and blue histograms showing the stuck-pion and 2p-2h contributions, respectively.
    BOTTOM~LEFT:~The analogous distribution for the subset of events with no final state neutrons. 
    BOTTOM~RIGHT:~The same distribution for events with one or more final state neutron.}
	\label{n_shennanigans}
\end{figure}

Another dramatic example is illustrated in Fig.~\ref{n_shennanigans}, which shows the neutrino energy using kinematic reconstruction of a monoenergetic 1 GeV $\nu_{\mu}$ beam~\cite{hill}. 
There is smearing from the non-zero momentum of the struck nucleon(s), as well as a broad low-energy tail from 2p-2h interactions and events with stuck pions. To first order, true CCQE interactions in a pure neutrino beam should produce a single proton and no neutron. The presence of final state neutrons in a neutrino beam is thus an indicator of inelasticity. In this simulated example, selecting events with no final state neutron provides a much purer sample of CCQE events as shown in the bottom left panel of Fig.~\ref{n_shennanigans}.

\begin{figure}[htb]
        	\begin{center}
 		\includegraphics[width=0.85\linewidth]{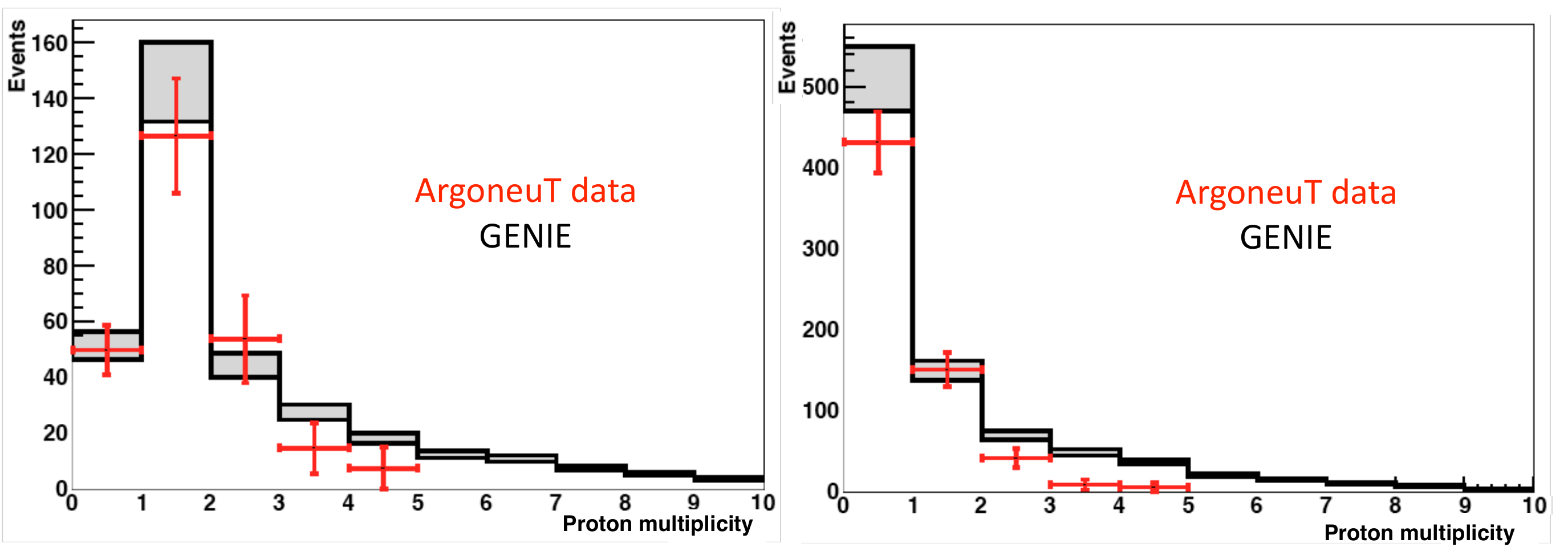} 
 	\end{center}
 	\caption{A figure comparing measured proton multiplicity from data (red points) with predictions derived from GENIE (black lines, systematic uncertainties in grey). The left panel is for a neutrino dominated beam, and the right panel is for anti-neutrino mode where the primary final-state nucleon is typically a neutron. Taken from Ref~\cite{palamara}.}
 	\label{protonmult}
 \end{figure}

Large uncertainties on final state nucleon abundance still remain despite general progress in neutrino interaction physics. Figure~\ref{protonmult} shows the disagreement between measured proton multiplicity in ArgoNeuT and predictions made by GENIE simulations~\cite{genie1, genie2}, which overestimate the proton abundance at high multiplicities~\cite{palamara}. This figure also shows the dramatic difference in proton multiplicity between neutrino and ant-neutrino mode, suggesting that nucleon final states are an effective handle for estimating wrong-sign contamination in neutrino beams. The relative abundance of n-n to n-p pairs varies widely among the different neutrino interaction generators. These wide uncertainties stem, in part, from a lack of data. The ArgoNeuT proton measurements provide an important first step. Given the complexity of the problem and the experimental challenges involved in detecting final state particles, the best strategy is to measure nucleon yields in as many contexts as possible.

\subsubsection{Neutron Tagging for Proton Decay and Supernova Neutrino Searches}

Proton decay remains one of the most generic predictions of Grand Unified Theories~\cite{SNOWMASS_PDK}, and  is also  a feature in  some models  with low-scale gravity and Large Extra Dimensions~\cite{MohapatraLED}. While to date no definitive evidence for proton decay has been seen, experiments are now running into backgrounds from atmospheric neutrino interactions which restrict their future sensitivity. For example, Super-Kamiokande (Super-K) has recently set limits on two important decay modes $\tau/B(p\rightarrow e^{+}\; \pi^{0})>1.6\times 10^{34}\;y$ and $\tau/B(p\rightarrow \mu^{+}\; \pi^{0}) > 7.7\times 10^{33}\;y$ using 0.31 Mton years of data. ~\cite{SKPDK2017}. For these modes the measured (expected) number of background events was 0(0.61) and 2(0.87) respectively. Thus even these relatively ``easy" modes are now starting to run into background issues from atmospheric neutrinos. 

 
In this recent Super-K analysis it was possible to make a neutron tagging cut due to a roughly 20\% efficiency in detecting the 2.2~MeV gamma following a neutron capture. The collaboration estimates that this cut reduced backgrounds by roughly a factor of two. This is due to the assumption that many atmospheric neutrino interactions should be accompanied by one or more neutrons, but less than 10\% of proton decays in oxygen (and obviously none at all in hydrogen) should have associated neutrons.

Neutron  production in atmospheric neutrino interactions is not well understood theoretically. Such neutrons can come directly from the neutrino interaction vertex, from intranuclear scattering of  recoil hadrons,  and  from final state interactions in the target.  Super-K has made a low-efficiency (18\%) measurement of the energy spectrum integrated neutron multiplicity  as a function of visible energy ~\cite{Zhang2011} but these data are not easy to use to model the  neutrons expected from atmospheric neutrino  interactions that are proton decay candidates. This is due to the fact the direction and energy or the parent neutrino is known very poorly, so the momentum transferred to the nucleus ($Q^2$) is not known. The sign of the parent neutrino is also unknown except in those rare cases where  a  $\mu^{-}$ capture can be observed. {\it Since  atmospheric neutrino interactions  that mimic proton decay are not  the``typical"  interaction, it is not possible to  directly  apply the  Super-K measurement to  determine the rate of expected background candidates in a simulation.}

ANNIE will make the first measurement of neutrino induced neutron multiplicity as a function of momentum transfer in the energy range of the atmospheric neutrinos.  Thus ANNIE data will be directly applicable  to estimates of proton decay backgrounds for oxygen, and will also provide guidance for similar background calculations for other nuclei such as carbon and argon.

In addition to  proton decay  backgrounds, ANNIE data is also relevant to the question of neutron  production in neutrino interaction events that could be background for a Diffuse Supernova Neutrino Background (DSNB)~\cite{DSNB} measurement, as planned by future experiments such as SK-Gd (the approved addition of Gd to the Super-Kamiokande detector) and Hyper-K.  In this case, signal events from Inverse Beta Decay (IBD) interactions are tagged by a neutron capture and thus separated from low energy ``stealth" atmospheric muon neutrino interactions. The issue then becomes how often inelastic atmospheric events produce a real neutron and thus avoid rejection. ANNIE will provide the first controlled measurement of this potential background.



\subsection{Technological Goals}
\label{sub:TechnologicalGoals}

\subsubsection{LAPPD Development and Demonstration}

The use of advanced, high resolution photodetectors in place of conventional PMTs could have a transformative impact on future neutrino detectors relying on light collection, ranging from water Cherenkov (WCh) detectors to liquid argon time projection chambers (LAr-TPC). One technology presently undergoing early commercialization is the Large Area Picosecond Photodetector (LAPPD)~\cite{LAPPD}: a $8" \times 8"$ microchannel plate (MCP) based photomultiplier with single photoelectron time resolutions less than 100 picoseconds, and spatial imaging capabilities to better than one centimeter~\cite{8inchtiming}.

\begin{figure}[tb]
        	\begin{center}
 		\includegraphics[width=0.65\linewidth]{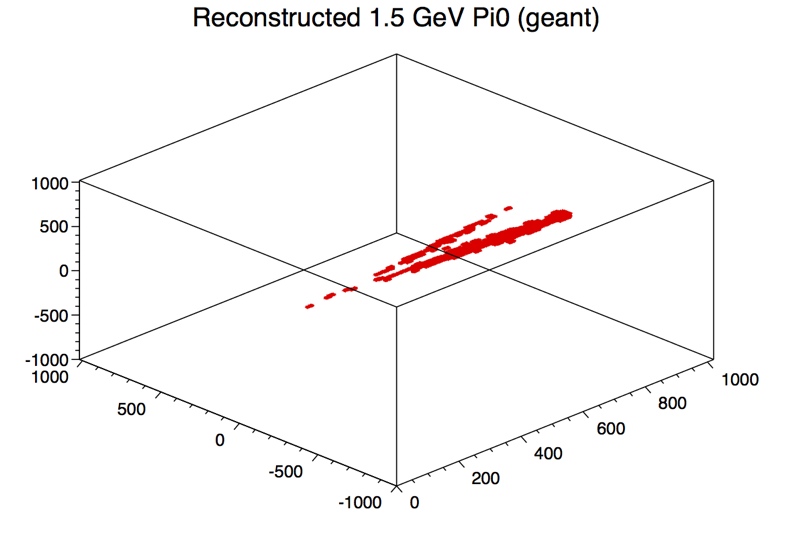} 
 	\end{center}
 	\caption{An example of a reconstructed gamma showers 1.5 GeV $\pi^0$ (simulated), using a time-reversal algorithm in an ANNIE-like detector with 100\% PMT coverage. This figure is illustrative of the capabilities that come with excellent timing and spatial resolution.}
 	\label{TrackReconstruction}
 \end{figure}

LAPPDs offer significant advantages over conventional PMTs. While conventional PMTs are single-pixel detectors, LAPPDs are imaging detectors, able to resolve the position and time of incident single photons within the sensor itself. This provides a much crisper detection of the Cherenkov ring edge and greatly improves the ability to distinguish between closely separated rings. The combination of imaging with the exquisite time resolution of LAPPDs ($\sim$ 50 psec), allow even a small number of LAPPDs to achieve the dramatic improvements in vertex reconstruction necessary to meet ANNIE's physics targets (see Sec~\ref{sec:vertex_fidu}). Ultimately LAPPDs may enable ANNIE to provide even more detailed event reconstruction, as illustrated in Fig~\ref{TrackReconstruction}.
In addition, the small LAPPD thickness (less than 1.5 cm) also allows for more efficient use of the detector volume.
As will be shown in Sec~\ref{sec:lappd_status}, LAPPD capabilities also translate into better energy resolution and better discrimination between dark noise and photons from neutron captures. 




The ANNIE experiment will evaluate the capabilities of LAPPDs in the context of an actual physics measurement. In the process the collaboration will be designing and testing a water-proof housing, developing fast high channel density electronics to take full advantage of the position and timing information available, and writing simulation and analysis software. Thus, ANNIE will take LAPPDs from a prototype technology to an established one available to the entire HEP community.

\subsubsection{First Application of Gd-Loaded Water on a Neutrino Beam}

Over the last decade there has been considerable  interest in the development of the technology to load water with gadolinium as a neutron capture agent~\cite{FSneutrons}. Experimental tests of transparency, cleanliness, and material compatibility have been made at Super-K~\cite{{EGADS},{Xu:2016cfv}} and at LLNL~\cite{GDCLNIM}.  While initial testing was done for Gadolinium Chloride  ($GdCl_{3}$), recent work has centered on Gadolinium Sulfate  (Gd$_{2}$(SO$_{4}$)$_{3}$) in order to improve material compatibility.  

The two large Gd-doped detectors built thus far (EGADS and WATCHBOY) have been installed in deep underground mines in order to measure neutron fluxes and/or perform technical R\&D for the SK-Gd project.  ANNIE Phase~II will be the first use of Gd-doped detector on the surface in a test beam.  Thus ANNIE would be able to demonstrate precision event reconstruction and long-term stability in a surface neutrino beam environment. 

\subsubsection{R\&D for Theia}
\label{TheiaRnD}
\label{theiaphys}

The construction of the Long Baseline Neutrino Facility (LBNF) will enable additional physics experiments in addition to DUNE. There is considerable interest in the international particle physics community in constructing the Theia detector, an advanced optical detector at a deep depth in order to address questions in neutrino physics and astrophysics such as: (1) the possible Majorana nature if the neutrino\footnote{Leptogenesis requires CP violation {\it and} a Majorana component to generate a baryon asymmetry in the universe.}, (2) exploration of ``invisible" modes of nucleon decay, (3) determination of the CNO component of the solar energy cycle, (4) precision measurement of the diffuse supernova neutrino flux, and (5) a measurement of the 2nd oscillation in the LBNF beam as a complimentary measure of CP violation with DUNE. The preliminary concept for Theia has been developed in a series of workshops at Fermilab and in Europe~\cite{Frost1,Frost2,Frost3}. The Theia design calls for a roughly 40-kton (fiducial) detector filled with Lightly-doped Liquid Scintillator (LdLS) or Water-based Liquid  Scintillator (WbLS) and instrumented with both high quantum efficiency PMTs and fast optical sensors (e.g. LAPPD's) to enable separation of Cherenkov and scintillation light~\cite{{Elagin:2016zgp},{ASDC}}. 
In fact, several Theia collaborators have expressed interest in joining ANNIE Phase~II.

The ANNIE detector is also an excellent test platform for the development of technology for Theia. 
The Phase~II  LAPPD deployment described above could be followed by deployment of twenty or more LAPPDs in a future Phase~III, where precision reconstruction of multi-track events could be investigated. In addition, Phase~III ANNIE could be filled with WbLS or LdLS to demonstrate reconstruction combining Cherenkov and scintillation light in a neutrino test beam. Thus, while not part of the Phase~II proposed here, the use of ANNIE for Theia R\&D would be an exciting future use of the facility in the future.




\subsection{Experiment Overview}
\label{sec:overview}

\subsubsection{Physics Measurement Strategy}
\label{sec:physics}

As established in Sec.~\ref{sub:physics_motivation}, detection of final state neutrons from nuclear interactions would have a transformative impact on a wide variety of neutrino physics analyses. 



To first order, the processes to be considered are charged-current (CC) exchange, which produces a single final-state proton or neutron from the neutrino or anti-neutrino interaction respectively, and high-energy neutral current (NC) interactions, which produce either protons or neutrons.
These expectations are modified by higher-order processes and multi-scale nuclear physics, including secondary (p,n) scattering of struck nucleons within the nucleus, charge exchange reactions of energetic hadrons in the nucleus (pion capture), and 2p-2h processes, where the neutrino interacts with a correlated pair of nucleons. The theory describing these interactions is under development and is still weakly constrained by the available data. 

Given the complexity and variety of nuclear interactions, it may not be possible to fully determine the theory. However, by measuring the kinematic dependence of neutron yields on variables such as $q^2$, $E_\mu$, $\theta_\mu$, and visible energy we expect to place strong constraints on neutrino interaction models and generators. 


Data-data comparison studies between CCQE candidates with N=0 and N$>$0 neutrons should provide fairly robust results even considering the large uncertainties in the underlying physics and secondary production mechanisms. In particular, events with one or more neutrons should show a considerably lower average reconstructed energy than 0-neutron events, since the presence of neutrons in neutrino mode is strongly indicative of inelasticity. 
This effect will be modified by processes that occur in true CCQE interactions, such as nucleon-nucleon scatter within the nucleus, emission of de-excitation neutrons, and the production of secondary neutrons by the recoil proton. It is expected that these occur less than 50\% of the time but this has not actually been measured. Thus ANNIE will need to be able to determine the neutron detection efficiency to roughly 10\% and in addition be able to reliably measure muon direction and energy from a well-reconstructed vertex. These requirements drive the experimental design.


\subsubsection{Experimental Design}
\label{sec:exp_design}

The ANNIE detector consists of three major components. In order of position in the beam these are: (1) a Front Veto to reject entering backgrounds, (2) a water tank containing the neutrino target and optical instrumentation, and (3) an iron-scintillator sandwich Muon Range Detector (MRD) in back of the neutrino target. These are installed at the former location of the SciBooNE Experiment~\cite{Hiraide:2008eu} on the BNB at Fermilab. 
This beam is about $93\%$ pure $\nu_{\mu}$ (when running in neutrino mode) and has a spectrum that peaks at about 700~MeV. Using the simulations and event reconstruction packages described in Sec.~\ref{sec:phase2_overview}, we estimate approximately 26,000 charged-current muon neutrino interactions in our fiducial volume per year, roughly 5,000 of which will produce muons that enter and range-out in the MRD. 

The ANNIE target tank was installed in Phase I. It is an upright cylindrical steel tank (10~ft diameter x 13~ft tall) with a plastic liner, filled with 26~tons of Gd-loaded water. The Gd enhances the neutron-capture cross section of the target relative to pure water and produces a detectable (8~MeV) delayed photon signal with a mean time constant of about 30~$\mu$s.
The concentrated solution of gadolinium sulfate will be added after an initial fill of very pure water. Drawing on prior experience with similar-scale Gd-loaded detectors with hermetic seals (Water-SONGS \cite{Reyna:2011qxa} and WATCHBOY \cite{Dazeley:2015uyd}), ANNIE will rely on the plastic liner to prevent ions that might compromise transparency from leaching into the water from the tank walls and maintain only minimal recirculation and nitrogen sparging to suppress biological growth. The tank internal structure will need to be modified for Phase II to accommodate LAPPDs and additional PMTs.

The FACC consists of two layers of overlapping muon paddles in order to reject charged particles produced in the dirt upstream of the hall. Made from components from the CDF detector, it was also installed in Phase I. This will not be modified for Phase II.

The MRD was used in SciBooNE and is already in the hall. It will be used to range out and fit the energy and direction of daughter muons from neutrino interactions in the water target. It will require refurbishment for Phase II, as discussed in section~\ref{sec:mrd_refurb}. 

ANNIE will use early commercial versions of LAPPDs read out by a fast electronics system in order to capture the fast Cherenkov light from the muon track in the water and allow for detailed track and event vertex reconstruction (see Sec.~\ref{sec:lappd_status}). 
Our studies show that vertex reconstruction, in particular, shows a strong dependence on LAPPD coverage (see Sec.~\ref{sec:vertex_fidu}). Thus, LAPPDs will be deployed in a staged fashion as they become available. Conventional PMTs, read out by Analog-to-Digital Converters (ADCs) with a deep buffer, are mostly used to detect the delayed neutron capture signal, but will also contribute to the vertex fit. Section~\ref{sec:conventional_pmts} considers optimal PMT coverage, using the number of photoelectrons produced per capture as a figure of merit, and details the plan to augment the existing stock of Phase~I PMTs using a mix of conventional and high quantum-efficiency tubes. 

\begin{figure}
	\begin{center}
           	\begin{tabular}{c c}	
			\includegraphics[width=0.5 \linewidth]{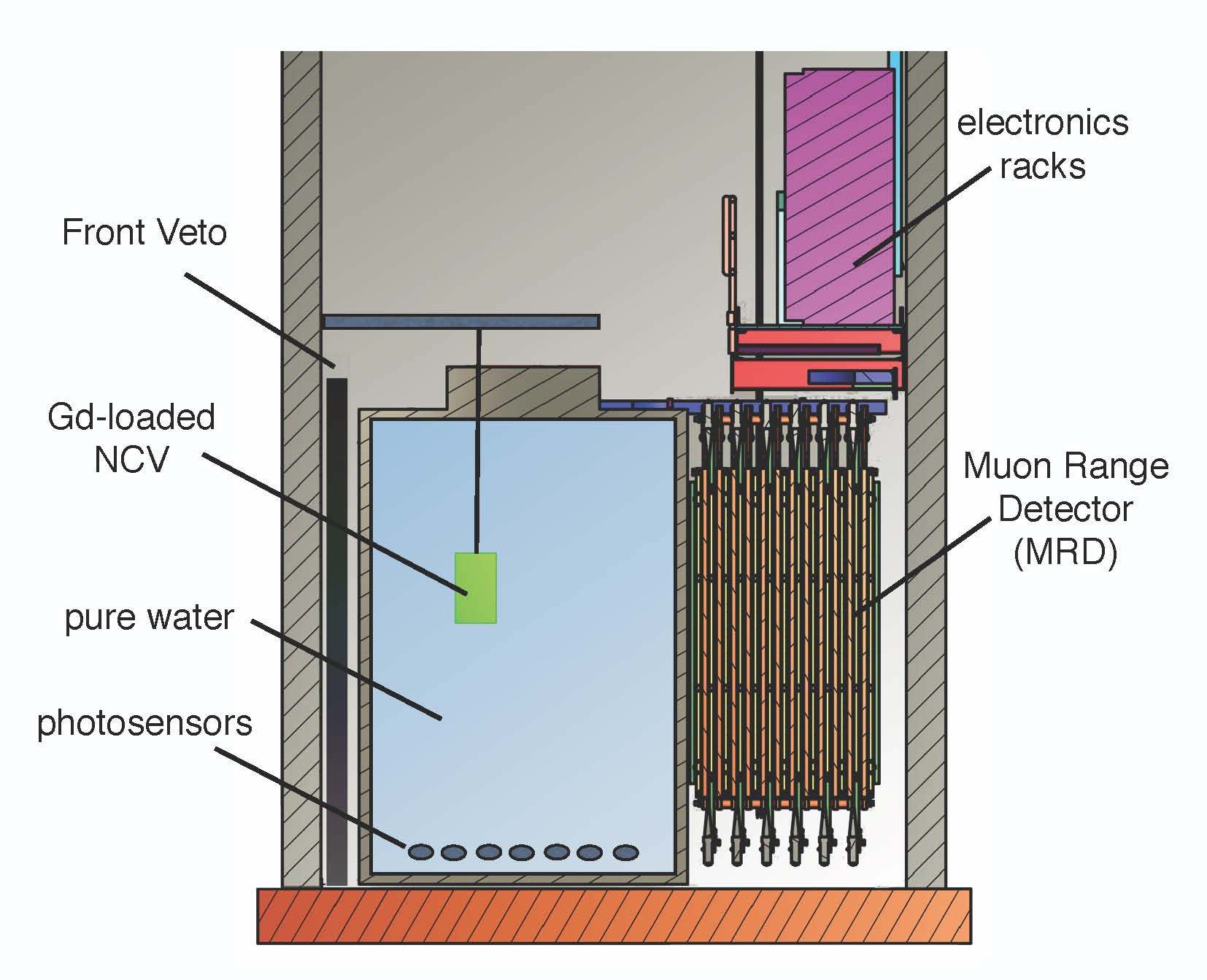} &
            \includegraphics[width=0.5 \linewidth]{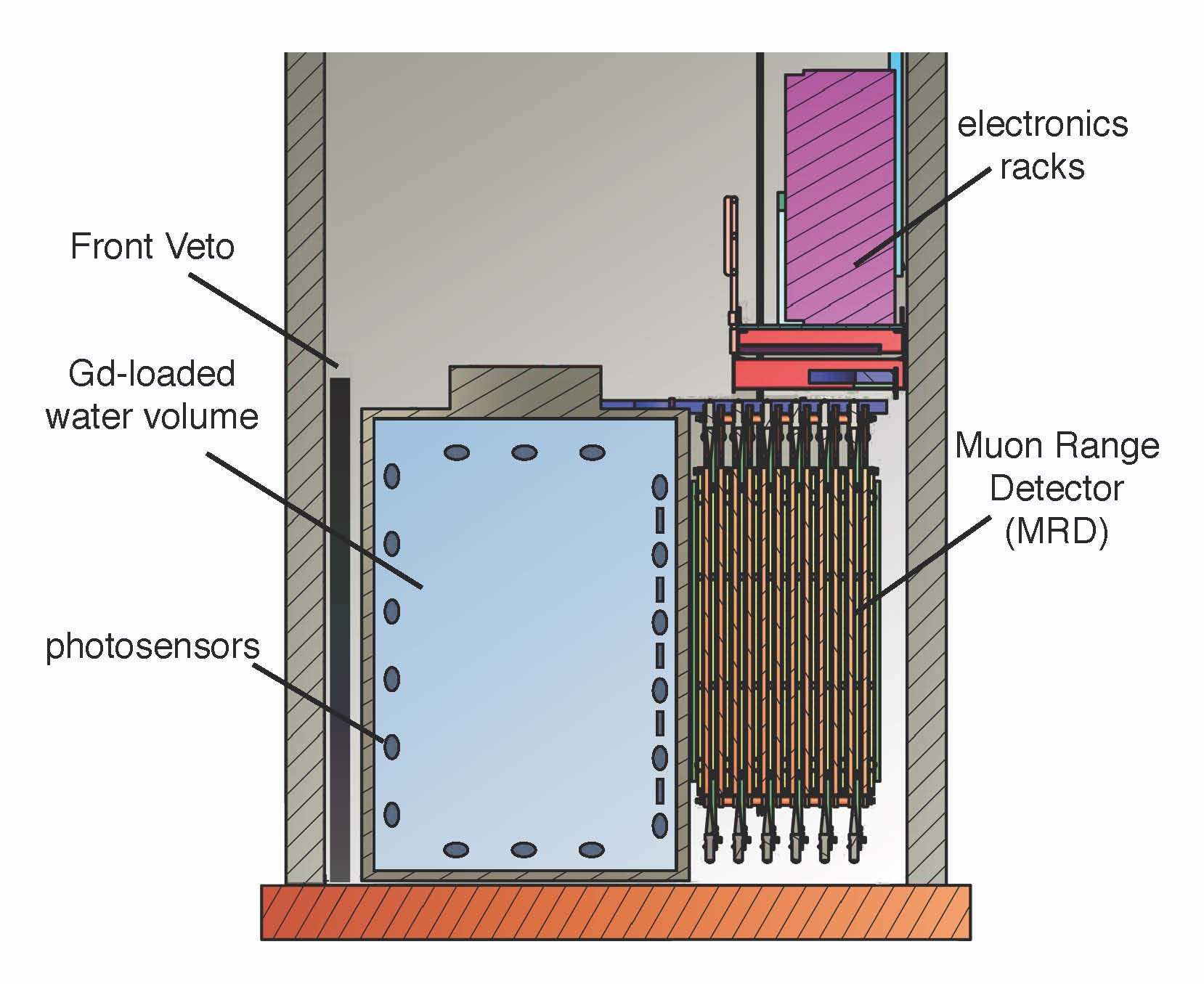} \\
   			    &  \\
                &  \\
            \multicolumn{2}{c}{\includegraphics[width=1.0 \linewidth]{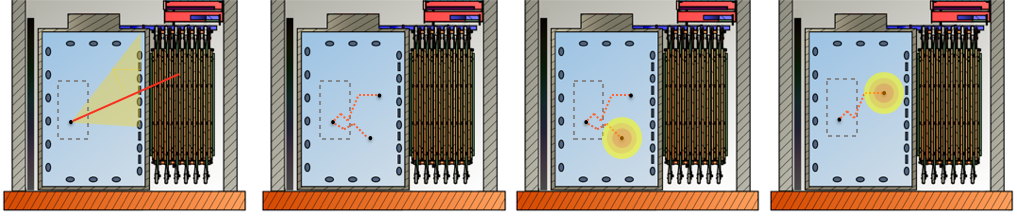}}
			\end{tabular}
	\end{center}
	\caption{ TOP LEFT: A concept drawing of the Phase~I ANNIE detector system. 
    TOP~RIGHT:~A concept drawing of the complete Phase~II detector.
    BOTTOM:~A series of 4 panels showing a typical ANNIE event (from left to right). A prompt muon produced in the fiducial volume is reconstructed using the MRD and photodetectors in the water volume. Neutrons produced by the neutrino interaction scatter around and thermalize. After a few tens of microseconds, each neutron is captured on Gd, producing light from the 8~MeV gamma cascade.}
		\label{fig:anniedetector}
\end{figure}

The trigger and readout design (Sec.~\ref{sec:electronics}) takes into account the timing properties of not only the prompt and delayed signal components but those of the various neutron backgrounds. These include neutrons produced by cosmic-ray interactions in the target, neutrons produced by the interactions of beam neutrinos with rock and dirt (``dirt neutrons''), and neutrons from the beam dump that leak into the atmosphere and then scatter into the detector (``sky-shine''). The primary trigger is based on the beam spill, which rejects the majority of cosmic neutrons.  The ADC readout window is deliberately widened to include a period following the beam spill, in order to characterize the sky-shine background \emph{in situ}. Data from all systems is transferred to a data acquisition system based on the ToolDAQ framework (see Sec.~\ref{sec:daq}) via VME bus transfer. Drawings of the detector system configuration (for both Phase I and Phase II) are shown in Fig.~\ref{fig:anniedetector}.  

\subsubsection{Phase I and II}

ANNIE was designed for staged deployment in two phases. A schematic representation of the Phase~I and Phase~II detectors can be found in Fig.~\ref{fig:anniedetector}. Some of the key differences between the two phases are outlined in Table~\ref{tab:pipii}.

Completion of Phase~I (described in Sec.~\ref{sec:phaseI}) enabled the collaboration to (1) partially build and test each of the main components of the experiment, (2) perform a critical measurement of background neutrons on the physics measurement, and (3) demonstrate control over the key experimental risks.
Phase~I was approved by the PAC in Spring of 2015, commissioned in April 2016. The collaboration has been taking data since May of 2016. More than half of Phase~I data has been processed and analysis is on track towards a publication and PhD thesis. \textbf{Preliminary findings, indicate that the dominant backgrounds can be suppressed by an optically isolated veto region and that background neutron rates in the bulk of the detector are at the few percent level (see Sec.~\ref{sec:neutronbg}).}

ANNIE was partially funded through the DOE Intermediate Neutrino Program at the level of \$150k~\cite{INP}. These funds were used to replace electronics borrowed from the University of Chicago. They are also being used to demonstrate the LAPPD readiness of the experiment, with the following specific goals: (1) to acquire and test LAPPD prototypes from Incom Inc, (2) to complete work on the PSEC-4 based readout system for the LAPPDs, (3) to perform a vertical slice test with an LAPPD working along side the rest of the any readout/DAQ system, (4) to design, build, and test a prototype waterproof housing for the LAPPDs, and (5) to perform Gd compatibility tests.

The key requirement for moving on to Phase~II was the availability of the critical LAPPD sensor. Incom Inc has now produced multiple working detectors, each improving on the previous, as described in Sec.~\ref{sec:lappd_status}. {\bf LAPPDs are now clearly on track to be available for ANNIE Phase~II.} 

Since most of the major components of the ANNIE experiment were built and tested during Phase~I, readying the detector for the Phase~II physics measurement largely consists of procurement of photodetectors and electronics, with the installation of the new equipment being done by collaborator efforts. Further details on the Phase~II implementation can be found in Sections~\ref{sec:phase2_overview},~\ref{sec:phase2_mechupgrade}, and \ref{sec:phase2_readoutupgrade}. Funds to purchase the needed equipment is being requested from the DOE through a proposal to be submitted in early Fall 2017.

ANNIE has secured additional conventional PMTs for use in Phase II, and now is requesting DOE funding for the addition of just 40 HQE 8" tubes (for a total of 127) to replace a stock of borrowed Irvine phototubes and meet the neutron detection requirements (see Sec~\ref{sec:ncap_efficiency}). ANNIE will also request funding from the DOE to purchase 5 LAPPDs sufficient to meet the minimal Phase II physics goals, with the intent of expanding this coverage to 20 photosensors in FY19, for further vertex and tracking resolution improvements (see Sec~\ref{sec:vertex_fidu}). 

From Fermilab, we are requesting modest engineering support. The scale of the requested activity is considerably smaller than that of Phase~I, and is detailed in Sections~\ref{sec:phase2_mechupgrade} and~\ref{sec:FNALrequest}. Labor estimates, in Sec.~\ref{sec:FNALrequest}, are based on comparable prior work. Any upgrades in FY19 will not require removal and redeployment of the detector. By design, the LAPPD system can be deployed and removed from the already filled tank (see Sec~\ref{sec:lappds}).\\ 

\begin{table}[]
  \centering
  \caption[foobar]{Essential differences between ANNIE Phase~I and Phase~II.}
  \label{tab:pipii}
  \begin{tabular}{lcc}
    \toprule
    detector component & Phase~I & Phase~II \\
    \midrule
	MRD & 2 layers & fully active \\
	Gd in water & No & Yes\\
	NCV & Yes & No \\
   	Front Veto & fully active & fully active \\
    \midrule
	Conventional PMTs & 60 & 127 \\
    LAPPDs & 0 & 5-20\\
    \midrule
	ADC readout cards  &  16  &  30 - 50 \\
    PSEC readout cards &  0   &  10-40 \\
	CAMAC TDC cards  &  3 & 12 \\
    CAMAC discriminator cards  &  3 & 12 \\
	Positive-HV channels  &  60 & 127 \\
	Negative-HV channels  &  71 & 361 \\
	\bottomrule   
  \end{tabular}
\end{table}

\noindent The main tasks requested directly from Fermilab for ANNIE Phase II in FY18 are:

\begin{itemize}
\item Modifying the inner structure of the detector to accommodate side- and top- mounted PMTs. 
\item Several modifications to the top of the tank in order to add several ports and feedthroughs. 
\item Minor engineering consultation to interface the various PMTs with the inner structure design. 
\item Minor technical support on power distribution and rack protection for the addition of electronics. 
\item Transportation and redeployment of the detector into the experimental hall.
\item Operational support at the level of current Phase~I operations. 

\end{itemize}


\section{ANNIE Phase~I Results}
\label{sec:phaseI}

ANNIE Phase~I 
 enabled the collaboration to build and test all of the major infrastructure for the full experiment, including the target tank and water system. The goal of Phase~I was to measure and understand beam-induced neutron backgrounds to the physics measurement to be conducted in Phase~II.

\subsection{Phase~I Motivation}

\begin{figure}[htb]
  \begin{center}
    \begin{tabular}{c  c}
      \includegraphics[width=0.35\linewidth]{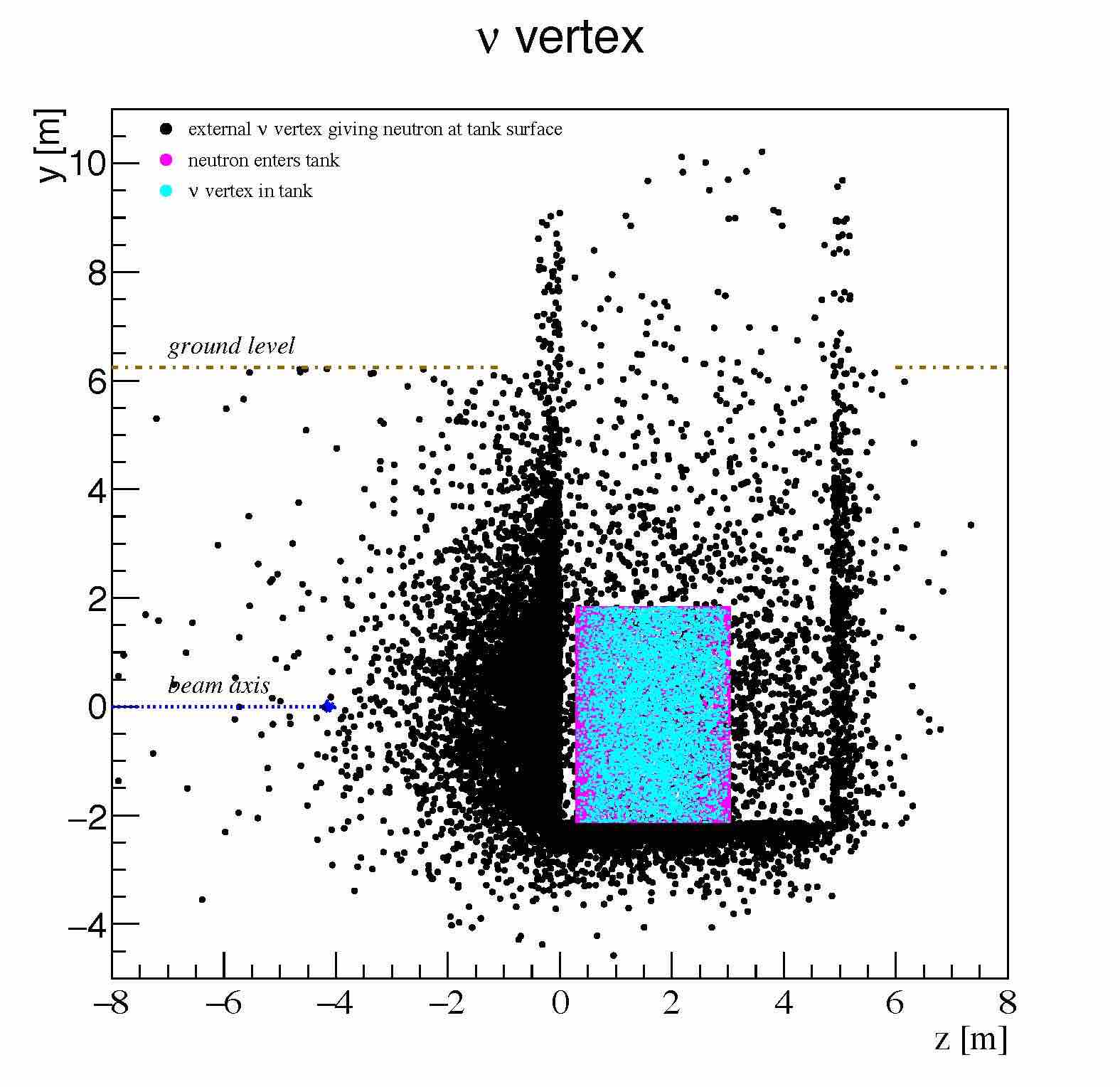} &
      \includegraphics[width=0.50\linewidth]{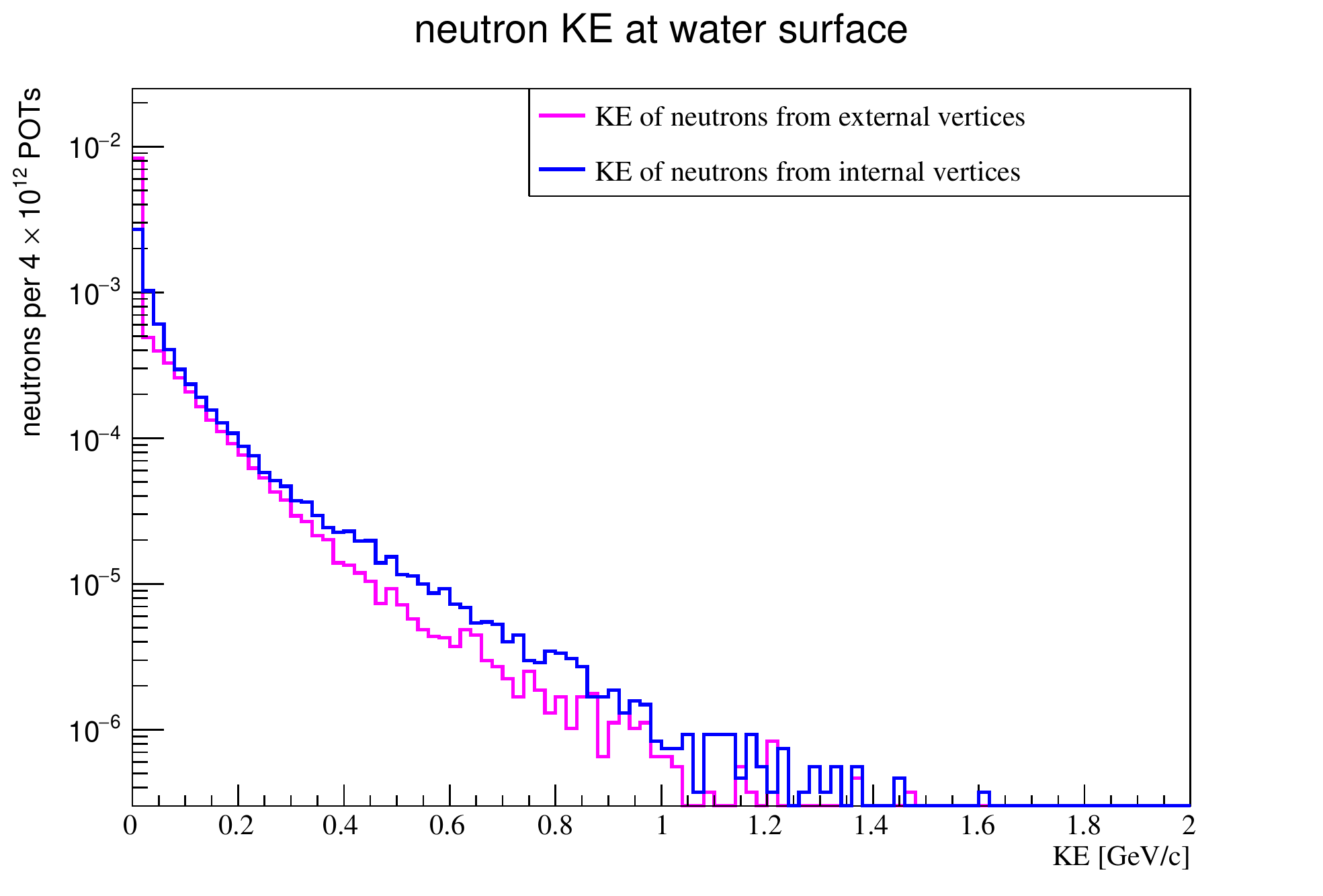}
    \end{tabular}
  \end{center}
  \caption{LEFT: The distribution of neutrino vertices that contribute neutrons
           that reach the tank (black points).  Magenta points are where
           the neutrons enter the tank. Neutrino interactions inside the tank
           are cyan. 
           RIGHT:~Kinetic Energy spectrum of the neutrons reaching the water
           in the tank from outside (magenta) and those originating
           from neutrino vertices within the tank (blue). 
           }
  \label{fig:nke}

  \label{fig:nsrcpos}
\end{figure}

Several sources introduce neutron backgrounds in the ANNIE detector. A continuum of ambient neutrons from cosmic radiation and long-lived isotopes will be present, but can be largely suppressed by strict time cuts around the beam window, and characterized with data from an off-beam trigger. 
Neutrinos from the BNB can interact with dirt and rock upstream of the experimental hall, producing a beam-correlated neutron background. This background arrives later with respect to the prompt component of a neutrino interaction and will capture on Gd after thermalizing. The vertices of these interactions that give rise to neutrons in the tank are shown in the left panel of Fig.~\ref{fig:nsrcpos}. The kinetic energy of neutrons created by interactions in and out of the tanks are very similar, as can be seen in the right panel of Fig.~\ref{fig:nsrcpos}.

An additional neutron background is that of sky-shine, namely secondary neutrons produced in the beam dump that leak into the atmosphere and make it into the detector after undergoing multiple scatterings~\cite{{LeeSkyshine},{BartolNeutronMonitor},{msu-skyshine}}. Preliminary results from SciBooNE indicate an observable excess of events after the beam time window with a clear dependency on the height in the detector hall~\cite{Takeiskyshine}.
The $y$-dependence of the event count in the detector hall suggests that having a fiducial volume away from the top of the detector significantly reduces sky-shine and cosmic backgrounds.


ANNIE Phase~I was built to confirm this rapid falloff of neutron backgrounds by performing a direct measurement of the neutron backgrounds in our target water volume as a function of distance from the front wall and the top of the detector. The
variation with position of the neutron capture rate is being characterized by using a movable neutron capture target. Measurements taken using this target at several positions provide the background neutron flux information needed for the physics measurements in Phase~II.

\subsection{Phase~I Overview}
\label{sec:run1_overview}

\begin{figure}
	\centering
              	\includegraphics[width=0.8\linewidth]{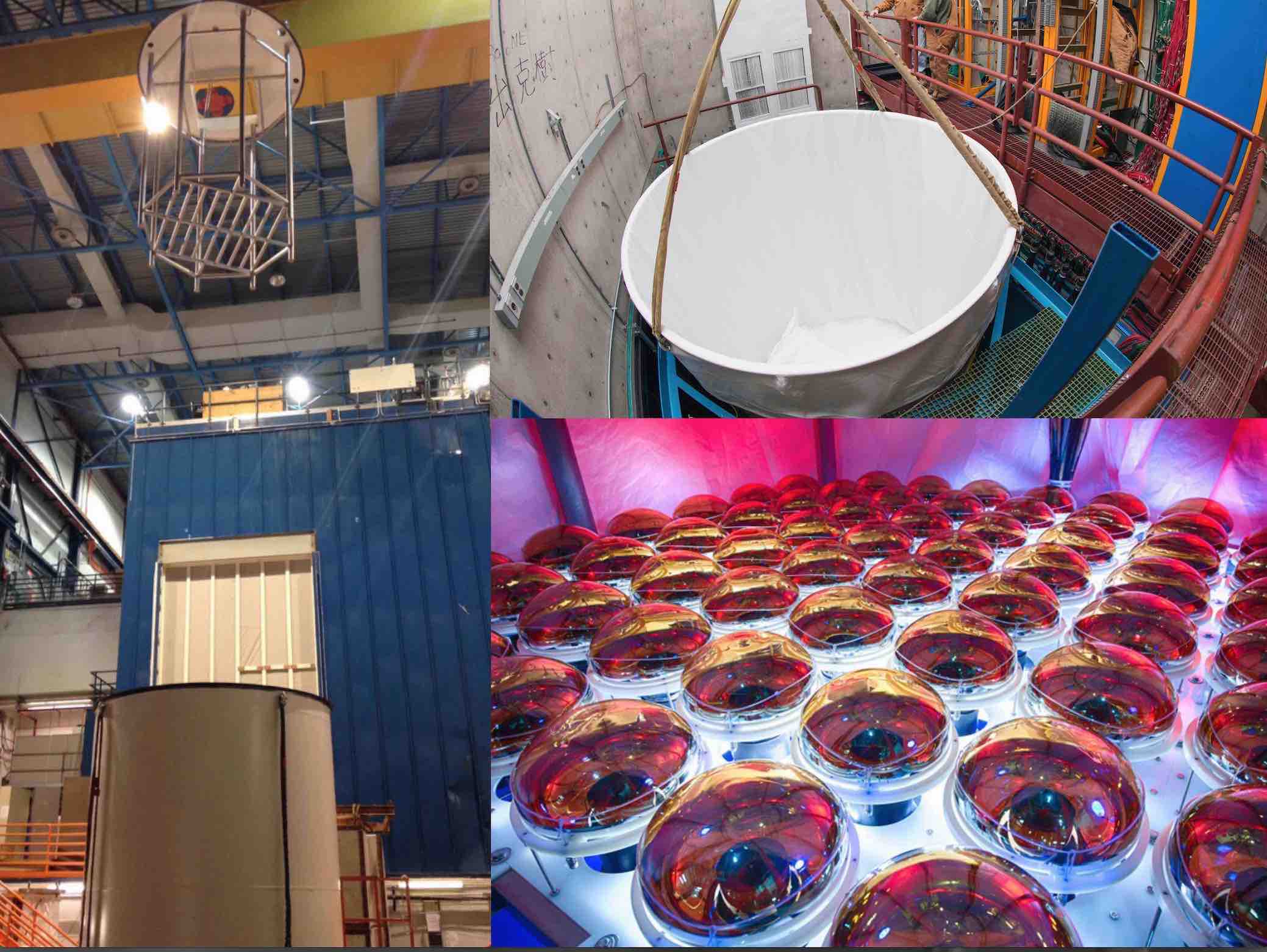}
	\caption{LEFT: The inner structure being lowered in the ANNIE tank. RIGHT TOP: The ANNIE tank being lowered into the experiment hall. RIGHT BOTTOM: The installed ANNIE Phase~I PMTs.}
	\label{fig:tank_pics}
\end{figure}

Phase~I, shown schematically in the left panel of Fig.~\ref{fig:anniedetector} and pictured in Fig.~\ref{fig:tank_pics}, is a partially instrumented implementation of the ANNIE experiment with all of the major components in place. The steel tank, common to both phases, is covered with a white reflective PVC liner in order to maximize light collection and is filled with 26~tons of ultrapure (unloaded) water.  The front veto and two orthogonal layers of the MRD are instrumented to tag muons entering and leaving the water volume.

\begin{figure}
	\centering
    \begin{tabular}{c c}
    	    \includegraphics[width=0.39\linewidth]{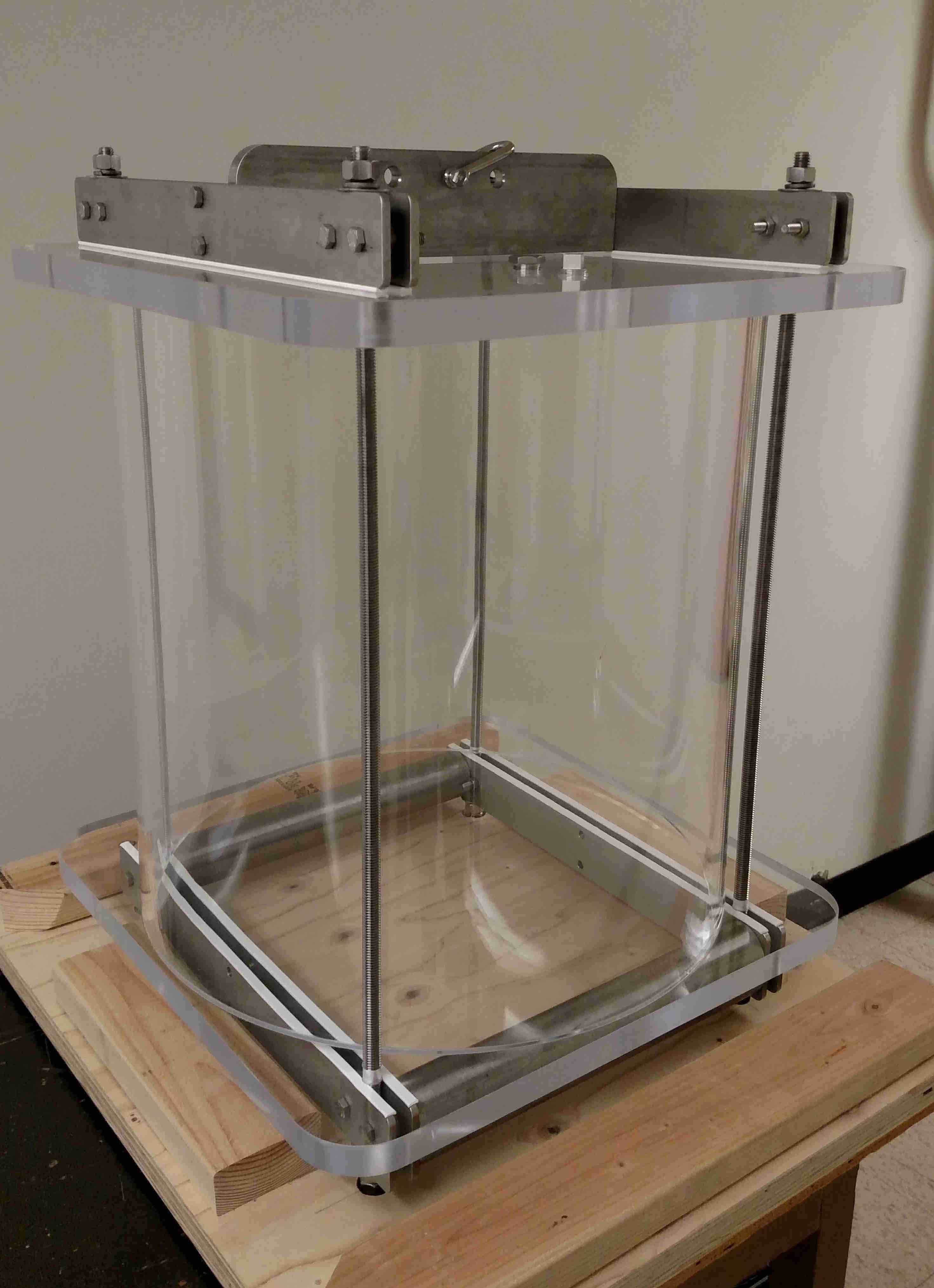}
	       	\includegraphics[width=0.585\linewidth]{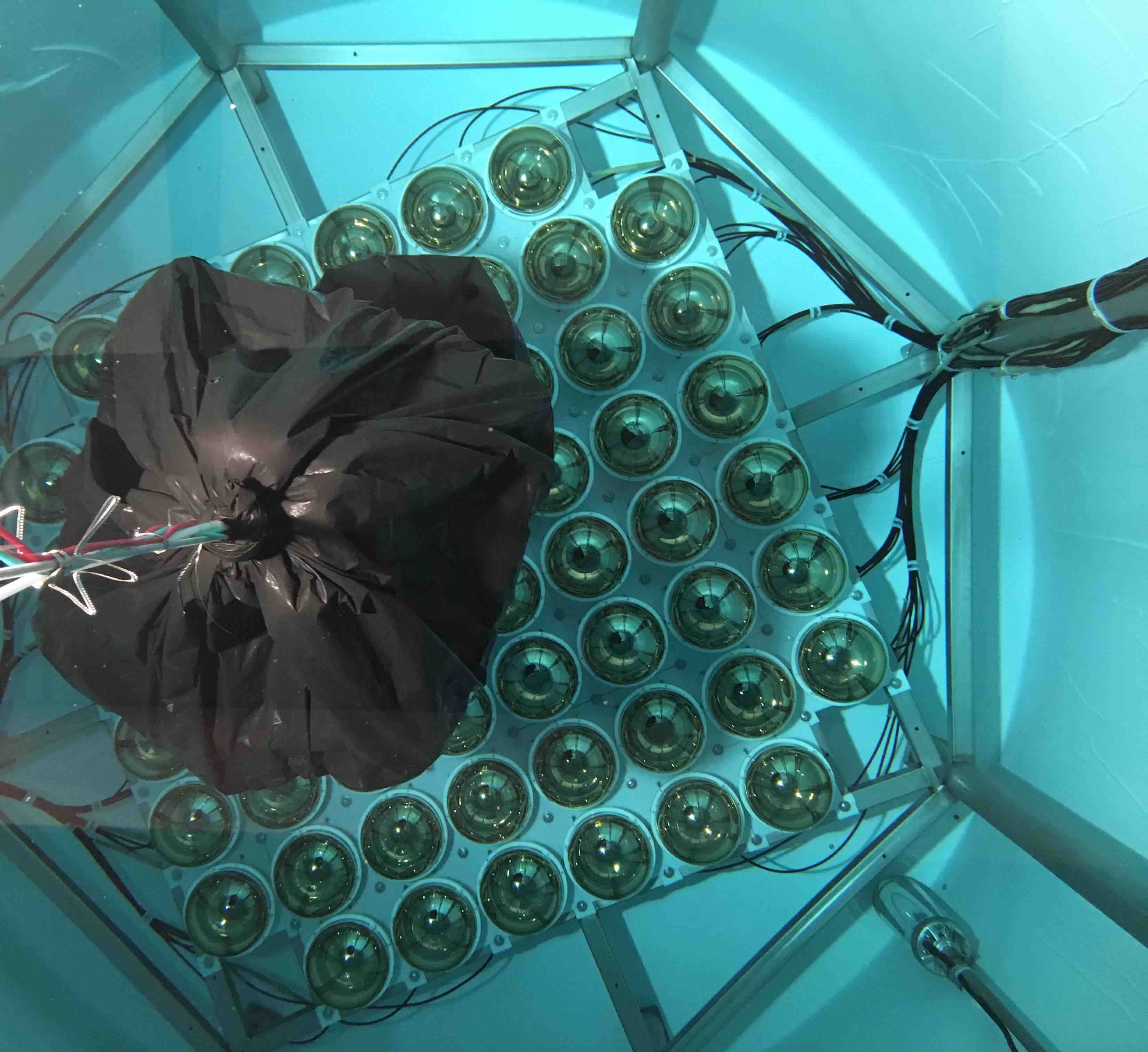}
    \end{tabular}
	\caption{LEFT: The Neutron Capture Volume, before deployment. RIGHT: Inside the installed and filled ANNIE Phase~I tank.  The NCV can be seen enclosed in black plastic.  Below the NCV is the array of 60 PMTs installed on the bottom of the tank.  The stainless steel inner structure, PMT cables and the white plastic tank liner can also be clearly seen.}
	\label{fig:tank_pics2}
\end{figure}

The unique feature present in Phase~I is the deployment in the tank of the Neutron Capture Volume (NCV), a 50x50~cm acrylic cylinder filled with EJ-335, a Gd-loaded (0.25\% w/w) liquid scintillator manufactured by the Eljen Technology company~\cite{ej-335}. The NCV can be moved within the water volume using a winch, thus allowing a neutron rate measurement at different locations in the tank. Two 3-inch photomultiplier tubes are installed on top of the NCV in order to tag energy depositions in the liquid scintillator. As shown in Fig.~\ref{fig:tank_pics2}, the NCV is enclosed in black plastic to optically isolate it from the rest of the tank and allow the use of the bottom PMT grid as a veto to tag muons from beam neutrinos and cosmic rays.


\subsubsection{Data Taking Progress}
\label{subsub:data_progress}
\begin{figure}
  \centering
          \includegraphics[width=0.5\linewidth]{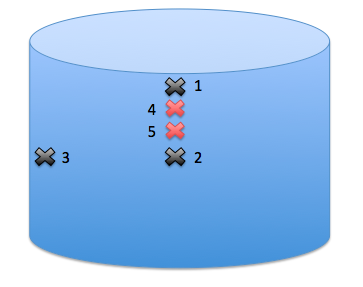}
  \caption{Planned Phase~I NCV positions within the water volume. Black crosses represent NCV positions where data taking has been completed. Red crosses represent positions where data are currently being taken or will be taken before the summer shutdown. The positions are numbered in chronological order beginning with position \#1 at the top center of the tank, just below the water line. The beam travels from left to right, and all of the positions lie in a vertical plane that passes through the center of the tank in the beam direction.}
  \label{fig:ncv_positions}
\end{figure}

Phase~I data taking will continue until the summer shutdown of the BNB in early July. A large fraction of the neutron background data has already been taken, and early analyses are in
progress. Figure~\ref{fig:ncv_positions} shows a diagram of the locations within the tank where the
background neutron rate is being measured in Phase~I. Data has already been taken in positions \#1, \#2 and \#3. 
Data are currently being taken at an intermediate position (position \#4), with one or two more NCV positions expected before the beam turns off.

During data taking at NCV position \#2, the ANNIE DAQ was upgraded to allow NCV events to be collected at a much higher rate. These DAQ settings, referred to as ``Hefty mode,'' were subsequently used for all beam data taking after the upgrade was complete. The Phase~I analysis results shown here are restricted to data taken at positions \#1 and \#2 before the DAQ upgrade. Updates to the ANNIE reconstruction software for Hefty mode are nearly complete, and analyses using the newer data are expected soon. Table~\ref{tab:phase_I_data_progress} summarizes the Phase~I data taking progress as of this writing.

\begin{table}[b]
  \centering
  \caption[foobar]{Summary of Phase~I data taken as of 12 June 2017. The triggering modes are
    \begin{inparadesc}
    \item[beam] for IRM triggers from the BNB, \item[source] for \isotope[252]{Cf} calibration source triggers (see
    Sec.~\ref{sub:source_cal}),
    \item[cosmic] for cosmic muon triggers, and 
    \item[hefty] for beam data taken in the ``Hefty mode''.
    \end{inparadesc}}
  \label{tab:phase_I_data_progress}
  \begin{tabular}{ccccccc}
    \toprule
     & \multicolumn{4}{c}{DAQ triggers by type} &
     & Approximate \# of \\
    NCV position & Beam & Source & Cosmic & Hefty & Total DAQ triggers & recorded beam spills \\
    \midrule
    1 & 1.96$\times$10$^6$ & 2.58$\times$10$^5$ & 1.72$\times$10$^4$ & 5.19$\times$10$^3$
    & 2.24$\times$10$^6$ & 2.13$\times$10$^6$ \\
    2 & 9.25$\times$10$^5$ & 0.00 & 2.25$\times$10$^3$ & 2.91$\times$10$^5$ & 1.22$\times$10$^6$ & 11.98$\times$10$^6$ \\
    3 & 0.00 & 0.00 & 0.00 & 1.62$\times$10$^5$ & 1.62$\times$10$^5$ & 6.16$\times$10$^6$ \\
    4 & 0.00 & 0.00 & 0.00 & 3.80$\times$10$^4$ & 3.80$\times$10$^4$ & 1.44$\times$10$^6$ \\
    \bottomrule
  \end{tabular}
\end{table}

\subsection{Source Calibration}
\label{sub:source_cal}

In order to understand the NCV neutron capture efficiency a calibration apparatus was built using a \isotope[252]{Cf} neutron source\footnote{Californium-252 (\isotope[252]{Cf}) undergoes spontaneous fission with a branching ratio of 3.1\% thus emitting an average of 20~gammas, 80\% with an energy below 1~MeV, and 3.8 neutrons whose energies approximately follow a Maxwell-Boltzmann distribution averaged at 2.14~MeV~\cite{PDG2016}.} and a LYSO (Lutetium Yttrium Oxyorthosilicate) scintillation crystal coupled to a 1-inch PMT. The \isotope[252]{Cf} source used in this apparatus had an initial activity of 5.4~mCi in January~1988, and its estimated current activity is about 2.7~$\mu$Ci.
The relatively high density and effective atomic number of the LYSO crystals allows the efficient detection of the gammas originating from the californium fissions thus providing a trigger signal to the data acquisition system. By triggering on fission events from the calibration source and counting the number of subsequent neutron captures observed, we can obtain a measurement of the detection efficiency of the NCV. Determining the efficiency in this way is needed so that measurements of the background neutron event rates at different positions in the tank can be converted into absolute fluxes.

\begin{figure}
	\centering
       	 \begin{tabular}{c c}
          	\includegraphics[width=0.47\linewidth]{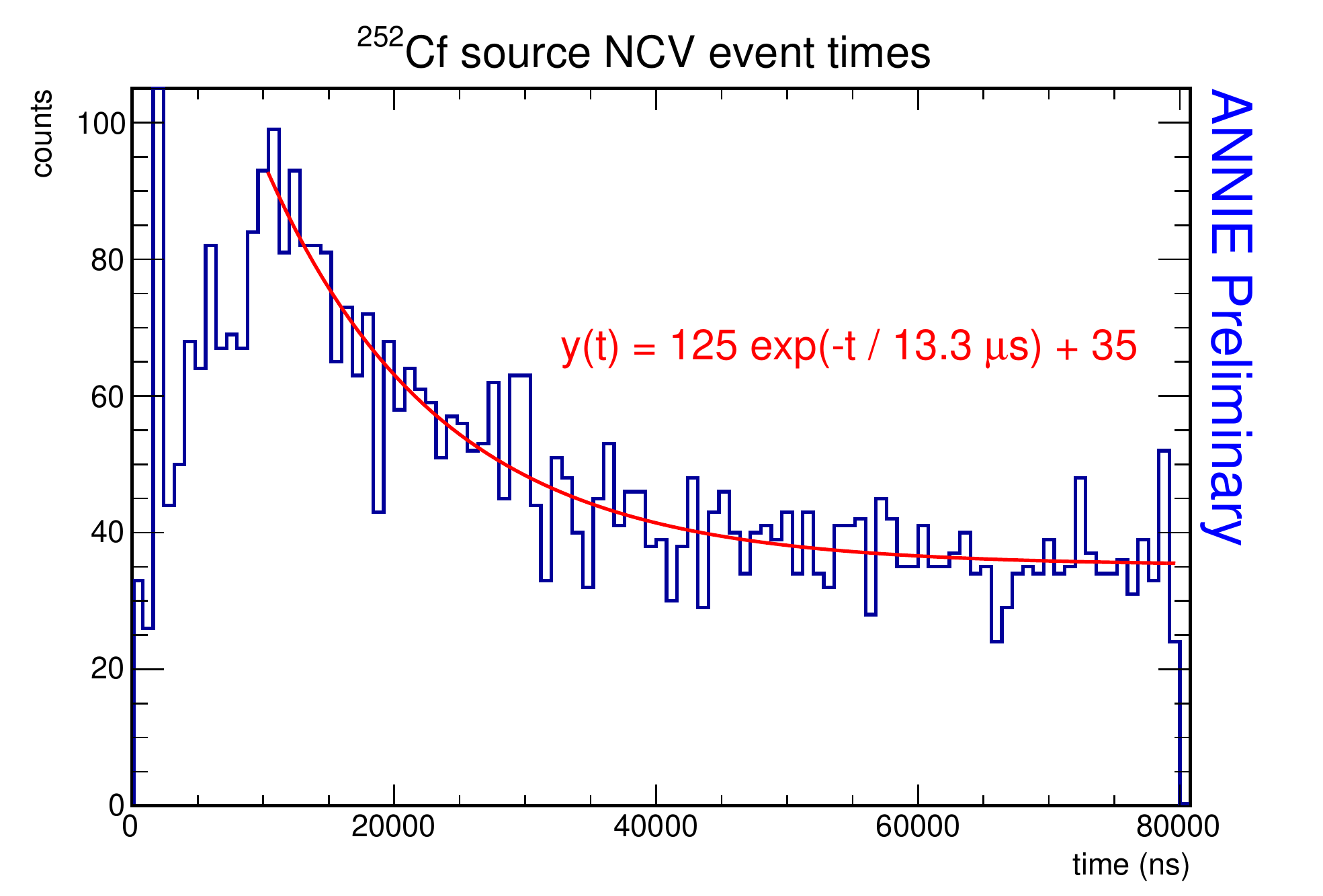}
	       	\includegraphics[width=0.45\linewidth]{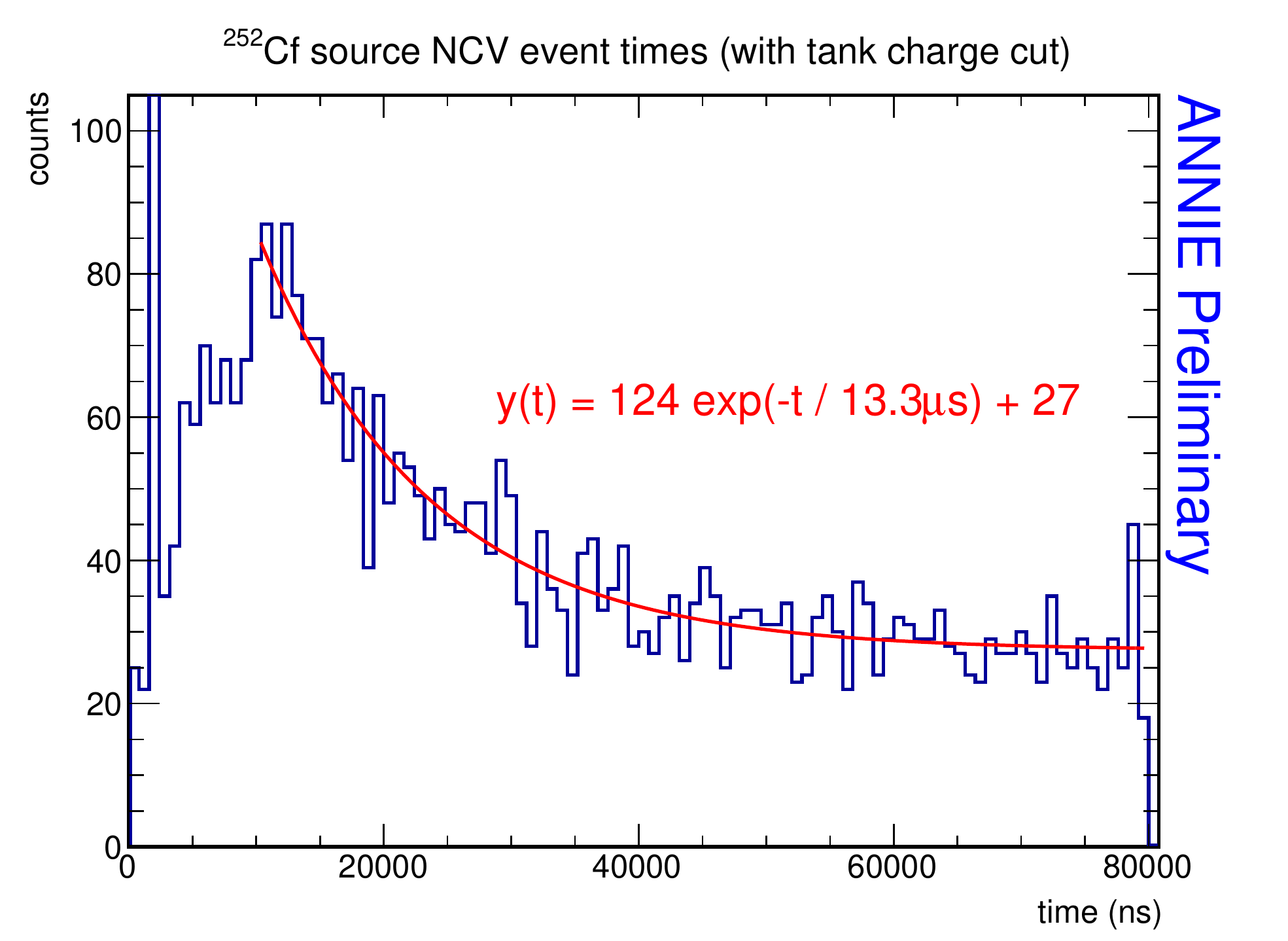}
         \end{tabular}         
	\caption{LEFT:~Time distribution of NCV events (coincidences of both NCV PMTs) observed using the \isotope[252]{Cf} calibration source trigger with the NCV at position \#1 with the trigger occurring at 2~$\mu$s.
    RIGHT:~Time distribution of NCV events from the same dataset after applying an analysis cut on the total integrated charge observed on the water tank photomultiplier tubes.
    }
	\label{fig:source_data}
\end{figure}

The left-hand panel in Fig.~\ref{fig:source_data} shows the data acquired using the Phase~I calibration apparatus with the NCV in position \#1.
The distribution of times when both NCV PMTs fired simultaneously may be divided into several regions: 1~$\mu$s of pre-trigger background followed by (1) a sharp peak at about 2~$\mu$s, corresponding to the arrival of fission $\gamma$-rays from the source in the NCV; (2) a gradual rise over the next 10~$\mu$s or so as neutrons from the source arrive in the NCV, thermalize, and begin to capture; and (3) an exponential decay over the remainder of the data acquisition window as more neutron captures occur. A fit to the exponential decay region yields a time constant of 13.3~$\mu$s, which is in good agreement with the expected value of about 10~$\mu$s for thermal neutron captures in EJ-335 liquid scintillator. Neutrons may also capture within the water surrounding the NCV (see Sec.~\ref{sec:gd_loading}) with an expected time constant of about 200~$\mu$s. When a capture event occurs close to the NCV, the 2.2~MeV capture $\gamma$-ray may sometimes enter the acrylic vessel and be detected. This effect will raise the NCV capture time constant a little above the 10~$\mu$s value expected for the liquid scintillator alone.

While both cosmic-ray muons and PMT dark pulses are expected to contribute flat backgrounds (of 3.6$\times$10$^{-3}$ muons\footnote{This estimate is based on the standard value\cite{PDG2016} of 70 muons~m$^{-2}$~s$^{-1}$~steradian$^{-1}$ integrated over the NCV cylinder.} and 9.9$\times$10$^{-4}$ dark pulse coincidences per trigger, respectively) to the timing distribution shown in Fig.~\ref{fig:source_data}, simply subtracting off the constant part of the timing distribution will eliminate many neutron capture events. 

\begin{figure}
  \centering
  \includegraphics[width=0.60\linewidth]{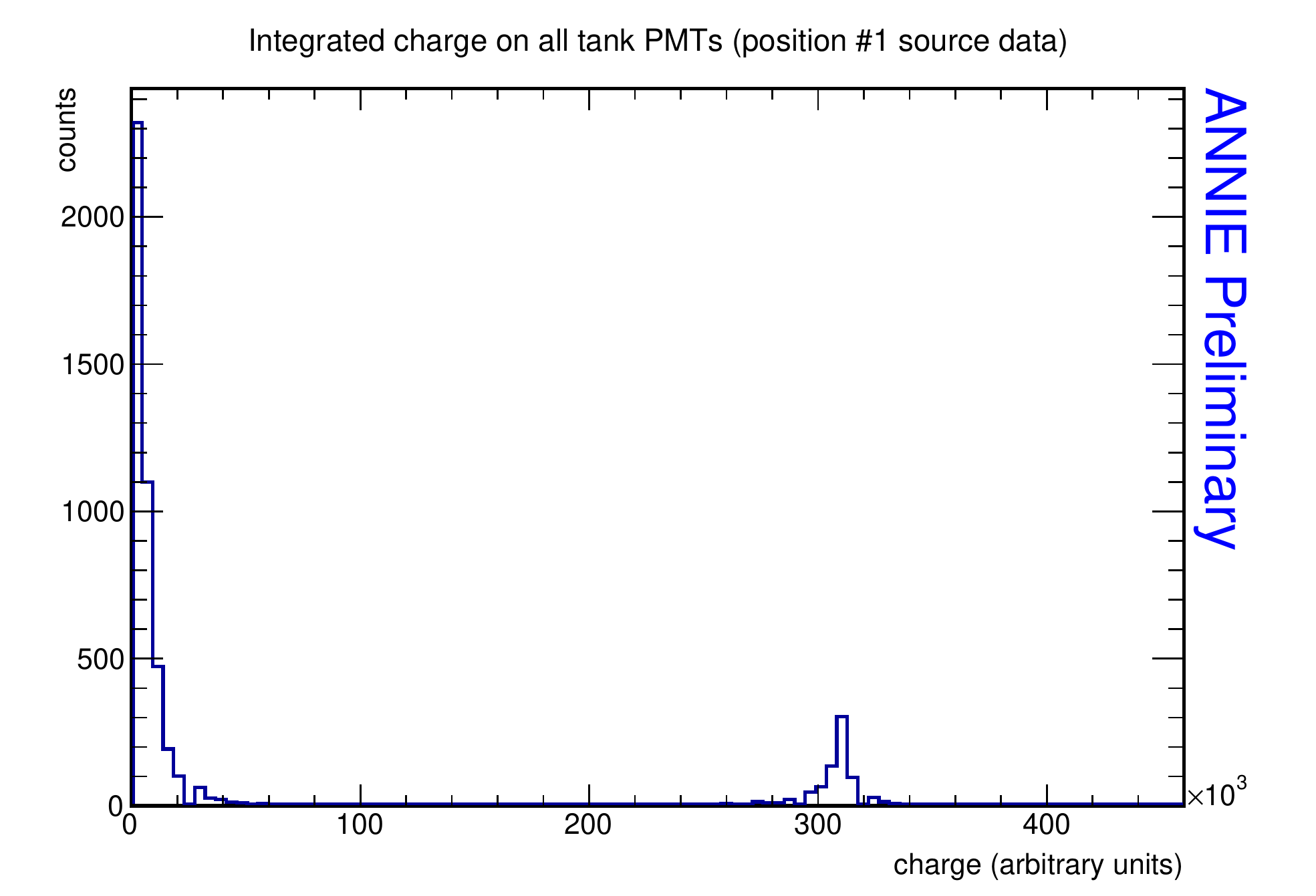}
  \caption{Distribution of integrated charge on all tank PMTs for all NCV events from the position \#1 calibration source dataset. The bimodal shape seen here is consistent with low-charge neutron capture events and high-charge muon events.}
  \label{fig:tank_cut}
\end{figure}

The optical isolation of the NCV from the rest of the detector allows for an efficient veto of cosmic-ray muons using the water tank PMTs. While muons above the Cherenkov threshold\footnote{About 55 MeV for muons in water.} will produce a large amount of light in the water tank, neutron captures in the NCV will be nearly invisible to the tank PMTs.\footnote{Neutron capture gammas that escape from the NCV and Compton scatter within the water volume may produce a comparatively small amount of Cherenkov light.} Applying a cut on the total charge observed on all of the tank PMTs (``tank charge'') allows one to discriminate between muon and neutron-capture events.
Figure~\ref{fig:tank_cut} shows the distribution of tank charge observed in the neutron calibration source data. The tank charge distribution consists of two well-separated peaks, one at high charge and the other at low charge, with very few events occurring in the intermediate
region. The right-hand panel in Fig.~\ref{fig:source_data} shows the calibration source data after applying an analysis cut\footnote{Specifically, integrated total tank PMT charge $<$ 10$^5$ (arbitrary units).} that excludes the entire high tank charge peak. The time constant of the fitted decaying exponential (cf. Fig.~\ref{fig:source_data} left panel) remains essentially unchanged, while the constant term is noticeably reduced, suggesting that the tank charge cut indeed reduces the flat background we expect from cosmic-ray muons. 

\begin{figure}
  \centering
  \includegraphics[width=0.60\linewidth]{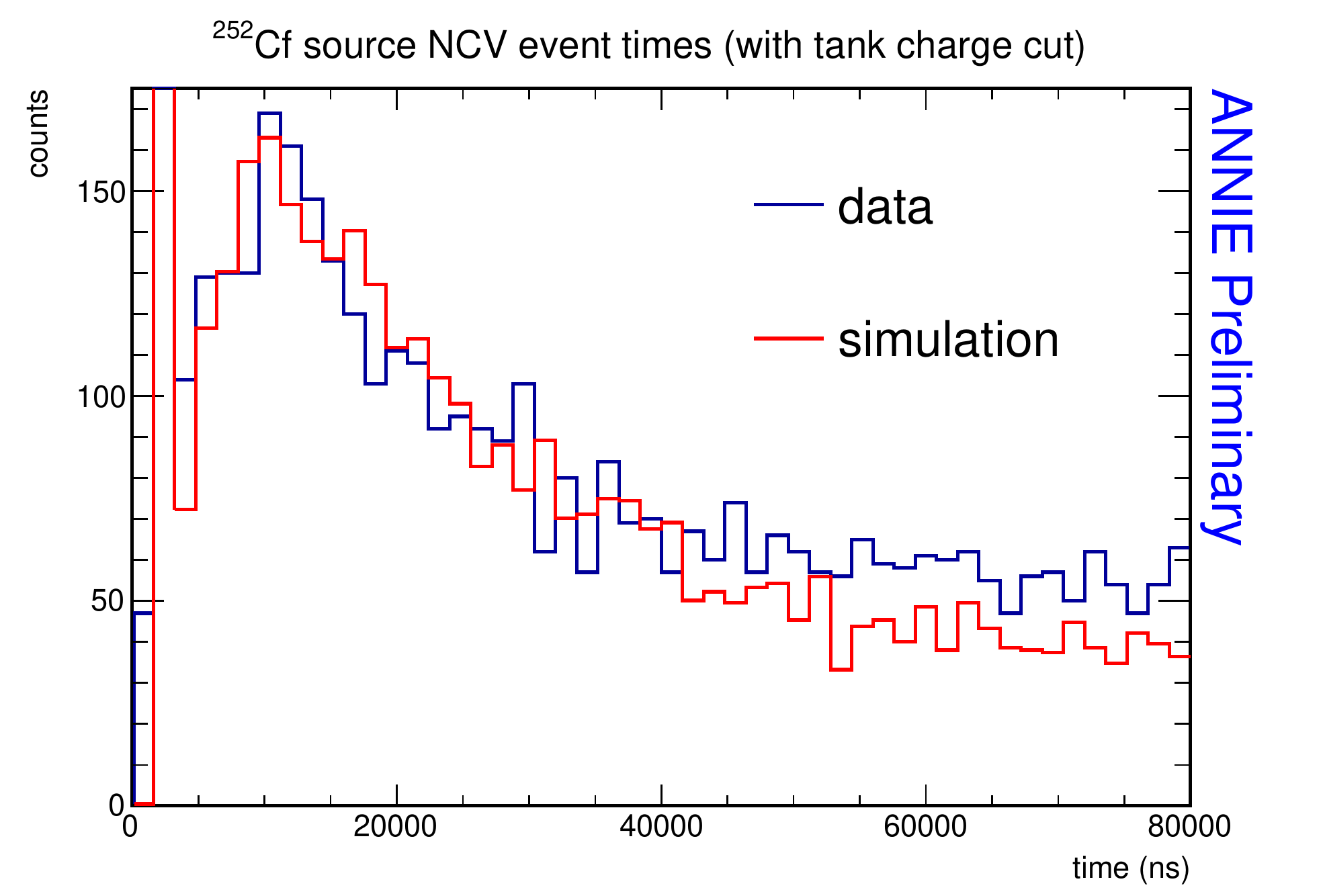}
  \caption{Comparison of the calibration source data (blue) with the results of a RAT-PAC simulation (red). The data histogram contains the same events as the right-hand panel
    in Fig.~\ref{fig:source_data}
    (the same analysis cut has been applied), but it has been rebinned. The simulation histogram contains zero events in the first bin because the pre-trigger time region was not modeled.}
  \label{fig:source_mc_comp}
\end{figure}

Monte Carlo simulations performed using RAT-PAC \cite{ratpac}
predict that about 44\% of neutrons which capture within the NCV in the first 80~$\mu$s after a \isotope[252]{Cf} fission event will take at least 10~$\mu$s to enter the NCV. 
The distribution of the arrival times of these neutrons at the NCV has a long tail that decreases slowly over a broad range of the 80~$\mu$s data acquisition window, making the subsequent neutron captures difficult to distinguish from background. Additionally, because roughly half of the neutron captures will occur at times later than 80~$\mu$s, pileup neutrons from fissions other than the one detected by the trigger will contribute to the flat part of the timing distribution. A direct comparison between a Monte Carlo simulation and the calibration source data that passed the tank charge cut is shown in Fig.~\ref{fig:source_mc_comp}. The features seen in the data are well reproduced by the simulation. Work is ongoing to understand
the small difference seen in the exponential decay region between the data and the simulation.

The calibration source data taken to date confirm that the NCV is sensitive to both neutrons and $\gamma$-rays produced from \isotope[252]{Cf} fissions. Work is still required to
determine the NCV's neutron detection efficiency. One
aspect of this measurement that is currently being addressed is the effect of the age of the \isotope[252]Cf source on the expected $\gamma$-ray and neutron fluxes. Because \isotope[252]{Cf} sources are manufactured with a variety of Cf isotopes present, the contributions from nuclides other than \isotope[252]{Cf} become increasingly important on
tens-of-year timescales~\cite{radev2014neutron}. The activity of the 
\isotope[252]{Cf} source used for these calibration measurements was last measured in January 1988, so a precise analysis of the neutron source data must account for the source's initial
composition and age.

The collaboration is currently considering additional measurements that would involve either characterizing the \isotope[252]{Cf} source or taking new data with a neutron source whose activity is already precisely known. These
measurements can easily be performed after the BNB shutdown in July either at Fermilab or at one of our collaborating institutions. The results of these new measurements will allow a precise determination of the NCV efficiency and will thus inform the final determination of the background neutron fluxes observed in the Phase~I data.

\subsection{Status of the Background Neutron Rate Measurement}
\label{sec:neutronbg}

\begin{figure}
  \centering
  \begin{tabular}{c c}
    \includegraphics[width=0.45\linewidth]{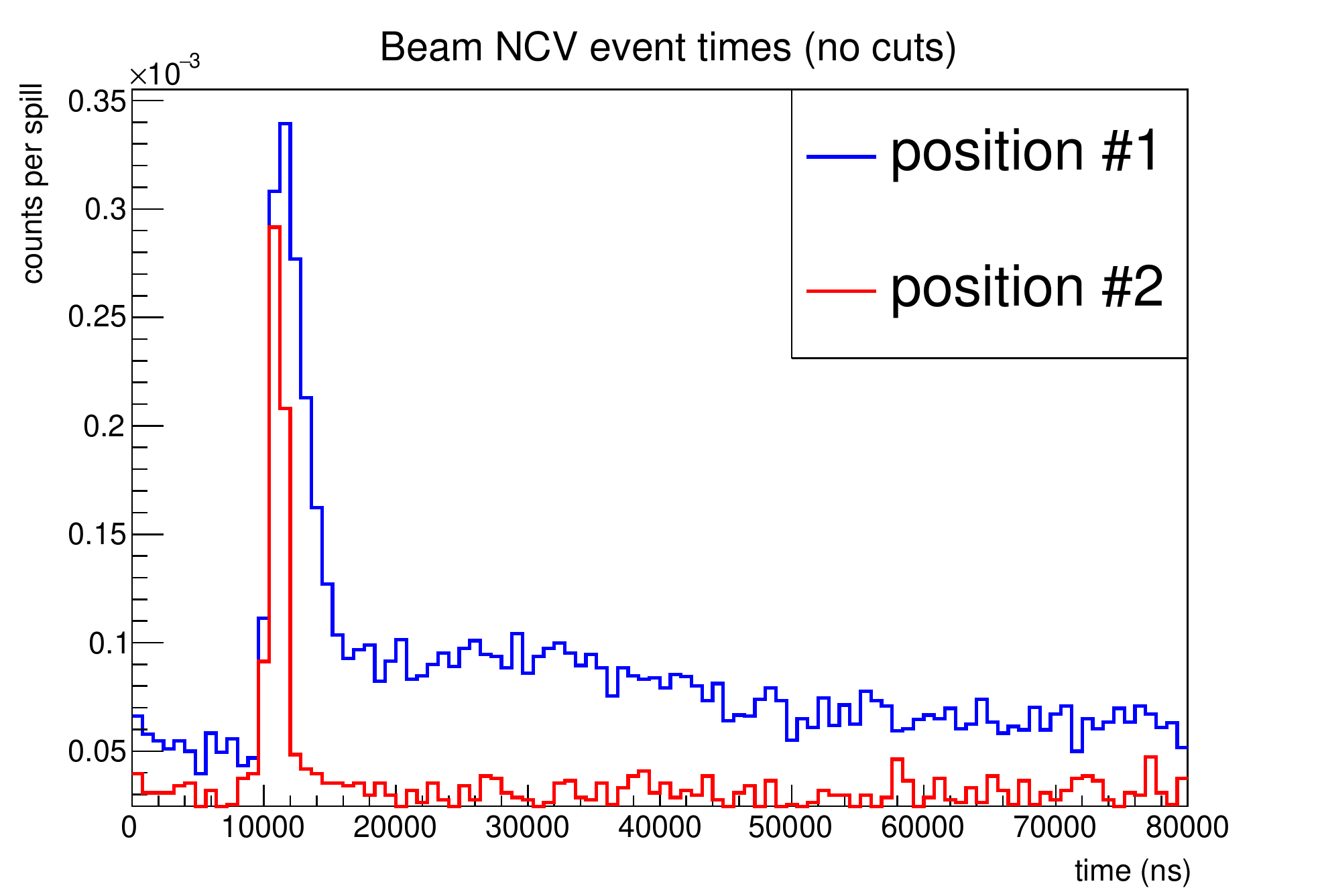}
    \includegraphics[width=0.45\linewidth]{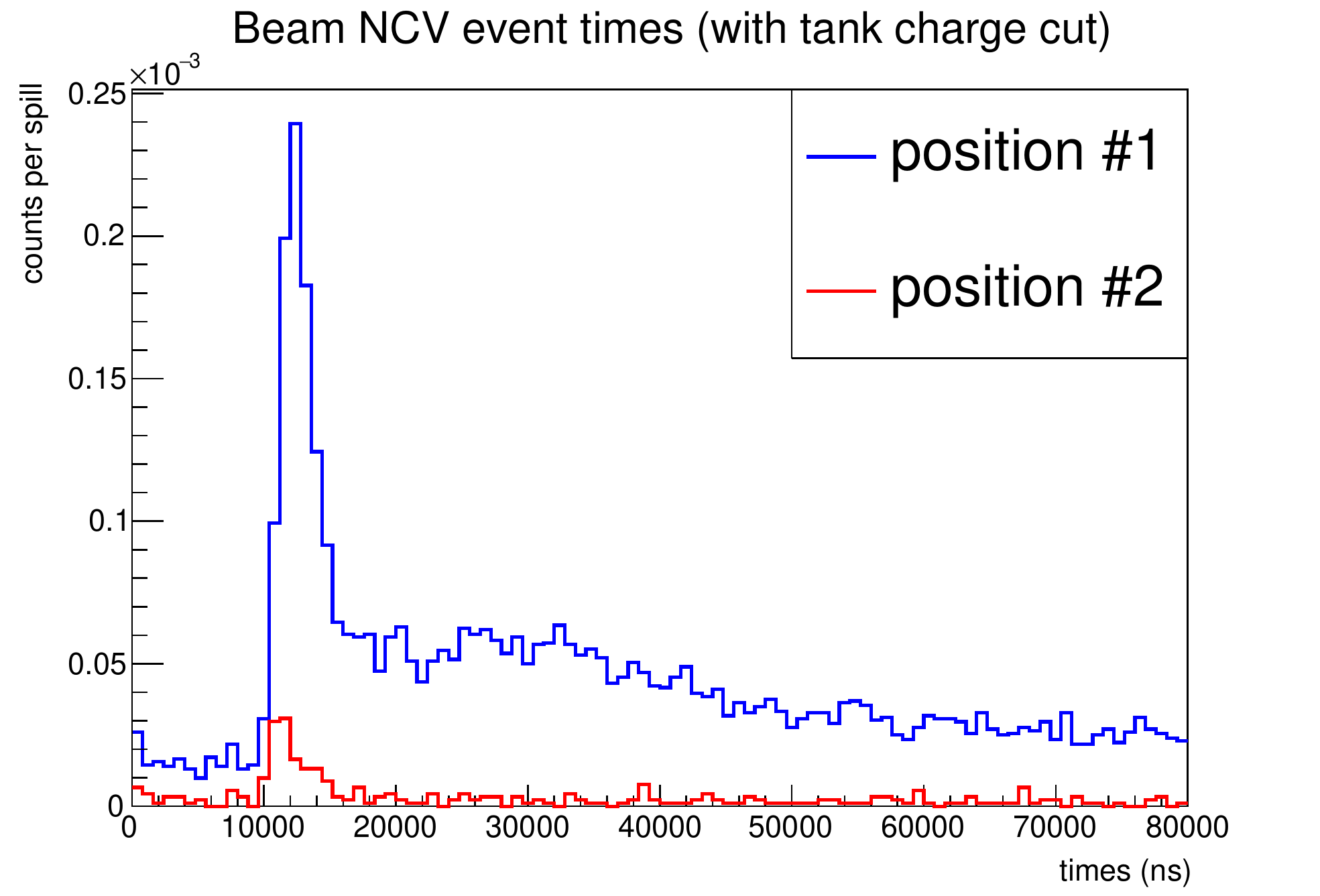}
  \end{tabular}
  \caption{LEFT:~NCV coincidence times for beam data taken at position \#1 (blue) and position \#2 (red). No analysis cuts have been applied. 
  RIGHT:~The same data after applying the tank charge cut described in the text.}
  \label{fig:beam_comparison}
\end{figure}

As was discussed in Sec.~\ref{subsub:data_progress}, beam data taking at NCV positions \#1 and \#2 has been completed, and progress is being made on a detailed analysis of the neutron background rates. The left panel of Fig.~\ref{fig:beam_comparison} shows event time distributions at positions \#1 and \#2. The position \#1 data show a long-lived excess of events in the time region after the beam crossing, which begins at 10~$\mu$s. This excess strongly suggests the presence of a significant flux of sky-shine neutrons at position \#1, particularly when one notes that the event
rate at position \#2 (which is near the center of the tank) returns back to its pre-beam level within less than 10~$\mu$s. Except in the immediate vicinity of the beam crossing, the timing distribution for position \#2 is consistent with a flat background, suggesting that the observed events at that location are almost entirely due to cosmic rays and muons associated with the beam. Further evidence for this conclusion is provided in the right panel of Fig.~\ref{fig:beam_comparison}, in which the
tank charge cut introduced in Sec.~\ref{sub:source_cal} is applied to the beam data. The dramatic suppression seen in the post-cut position \#2 data is consistent with a strong contribution of muons above the Cherenkov threshold to the pre-cut data.

Although a precise determination of the background neutron rate at each NCV position is still in progress, we may already obtain conservative estimates for the rates at positions \#1 and \#2. To do so, we first estimate the competing backgrounds that are not associated with the beam (e.g., cosmic muons, dark
noise) by fitting a constant to the pre-beam portion of the NCV event timing distribution.\footnote{In the data taken so far, this corresponds to the first 10~$\mu$s at each position. For the position \#2 data, a scaled integral of the pre-beam region was used rather than a fit due to poor statistics.} We may then subtract the expected contribution of the fitted constant background from the total number of events in the remainder of the distribution. To ensure that we obtain a conservative upper limit for the background neutron rates, we may further assume an NCV efficiency of 50\% and apply only the tank charge analysis cut to the beam data at each position (this helps suppress events caused by beam-associated muons). Following this procedure provides upper limits of about 6.0$\times$10$^{-3}$ background neutron events per beam spill at position \#1 and 7.7$\times$10$^{-5}$ events per spill at position \#2. Because these estimates include contributions from signal neutrons (those
produced by neutrino interactions in the tank) as well as background neutrons, we expect the final rates to be somewhat lower than the values given here. If we assume that the rate at position \#2 (the center of the tank) is representative of the average background neutron rate in the water tank as a whole, then we may scale up our position \#2 estimate and conclude that no more than 2\% of beam spills will give rise to a background neutron capture in the Phase~II Gd-loaded water tank.

The large difference that we observe between our estimated background neutron rates at NCV positions \#1 and \#2 suggests that the thickness of water shielding the top of the NCV strongly influences the observed background neutron rate. An important priority in the remainder of Phase~I data taking will therefore be measurements at intermediate positions between \#1 and \#2 in order to characterize how the background neutron rate scales with depth below the water line. This will determine the position of the optical isolation near the tank top in order to suppress sky-shine neutrons.



\section{LAPPD Status}
\label{sec:lappd_status}

\begin{figure}[h]
	\centering
           \includegraphics[width=0.95\linewidth]{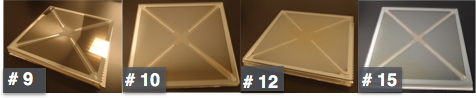}\\       
	\caption{Working LAPPD prototypes from Incom.}
	\label{fig:LAPPD_prototypes}
\end{figure}

Since ANNIE last reported to the Fermilab PAC, Incom Inc has succeeded in making several fully functional sealed LAPPD prototypes. Progress has been rapid. LAPPD-9 was the first working detector with an aluminum photocathode. LAPPD-10, with a multi-alkali photocathode has a quantum efficiency (QE) of 5\%. LAPPD-12 has a QE of 15\%. LAPPD-15 has a uniform photocathode in excess of 25\% QE. LAPPDs are now clearly on track to be available for ANNIE Phase~II. Figure~\ref{fig:LAPPD_prototypes} shows photographs of working LAPPD prototypes from Incom Inc.  

\begin{figure}
	\centering
           \includegraphics[width=0.70\linewidth]{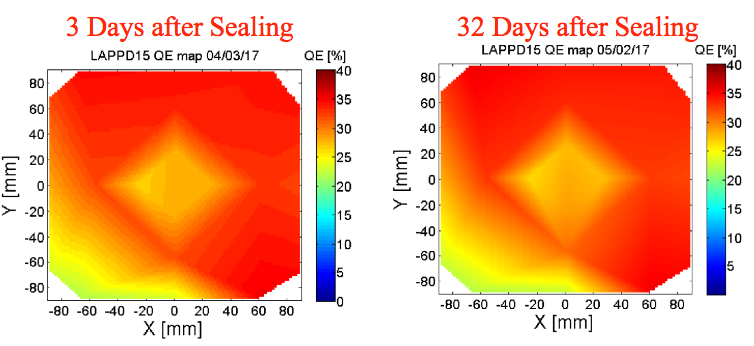}\\ 
           \includegraphics[width=0.70\linewidth]{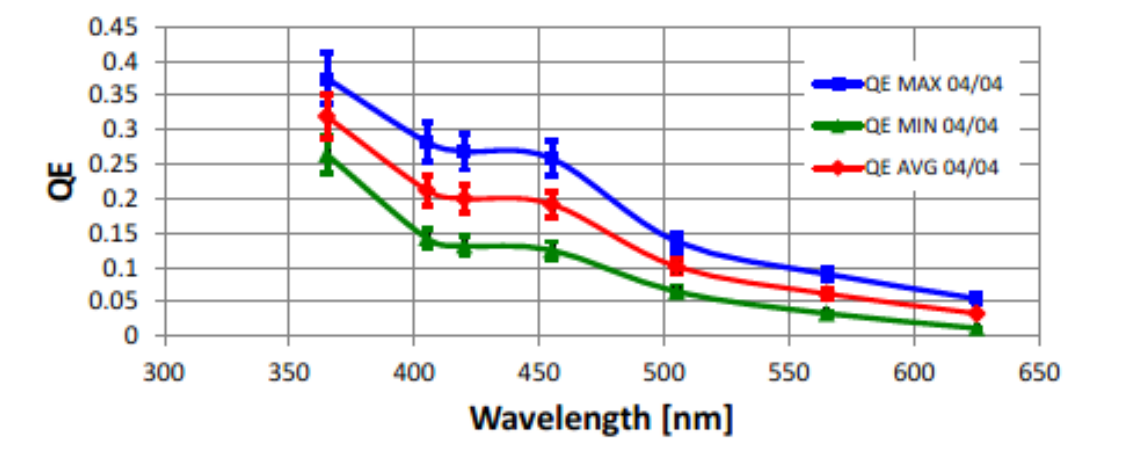}\\ 	
	\caption{TOP: LAPPD-15 QE map at 3 days (LEFT) and 32 days (RIGHT) after sealing. BOTTOM: The average QE at 375 nm remains at 30\%, with a maximum 35\% and minimum of 22\%.}
	\label{fig:LAPPD_15_maps}
\end{figure}

Figure~\ref{fig:LAPPD_15_maps} shows the QE map for the most recent tile, LAPPD-15 at 3 days and 32 days after sealing. There was no visible change over that period and the detector has continued to operate at the QE for several months since. The QE varies from 35\% to 22\% spatially, averaging 30\% at 375 nm. The slightly diminished QE in the lower left corner is due to a line-of-sight deposition issue that has since been resolved and will not be present in future iterations. The lower panel of Fig~\ref{fig:LAPPD_15_maps} shows the wavelength dependence.

Incom is presently installing added tile making capacity (4 tiles per month), which is expected to be in place well in advance to meet ANNIE needs (November 2017). A collaboration between Incom and U Chicago on Gen II LAPPDs has potential to increase yields, improve performance, and reduce costs. Incom’s development costs are currently being partially offset by DOE SBIR support which has enabled them to provide a quote for 20 LAPPDs within the modest target budget for ANNIE. Successful early adoption will go a long way in providing momentum for the technology and in developing the tools and techniques for future LAPPD deployments.

\subsection{Iowa State Test Facility and Preliminary Results}

\begin{figure}
	\centering
           \includegraphics[width=0.8\linewidth]{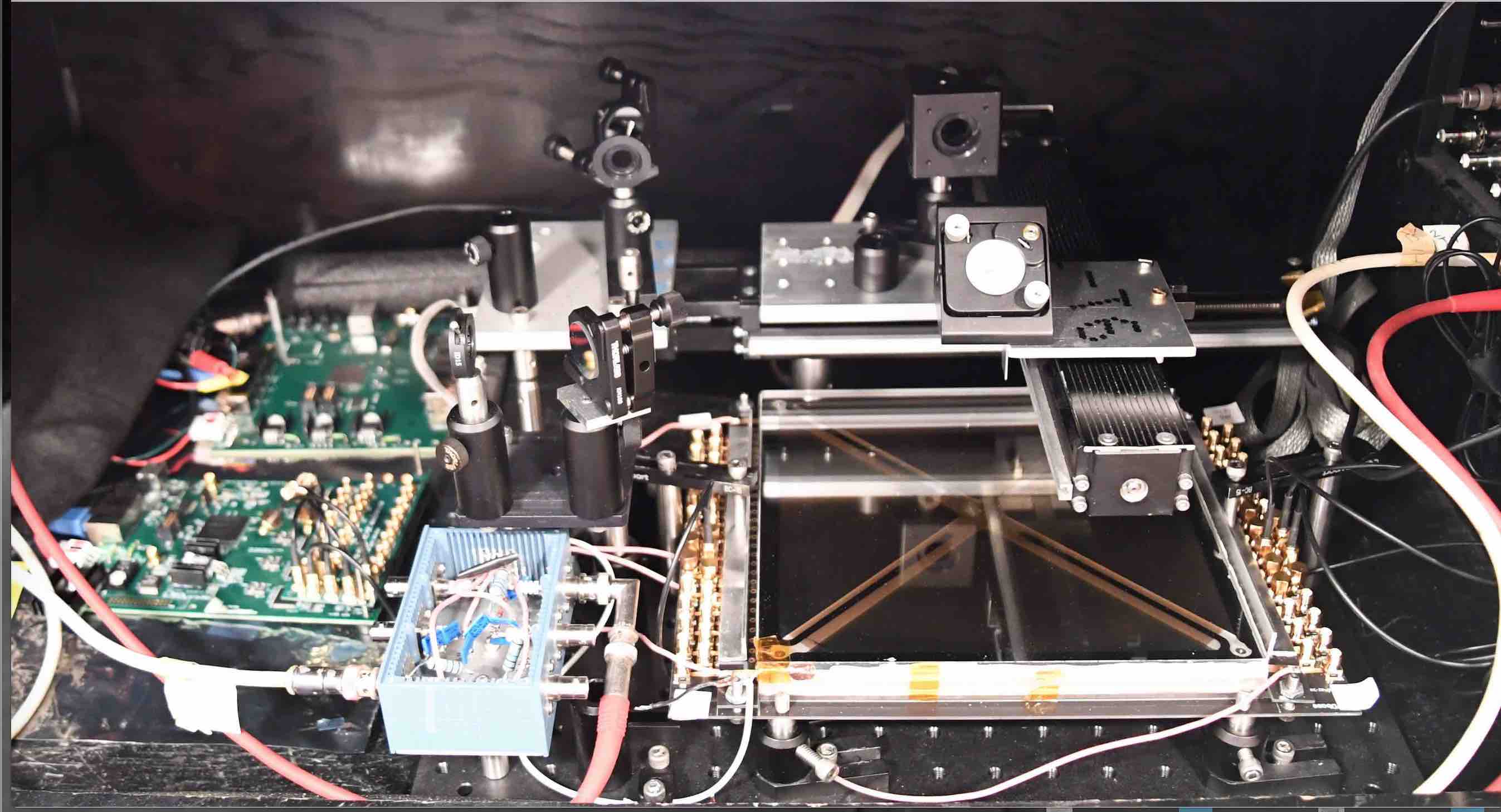}\\       
	\caption{LAPPD-12 installed in the ISU test stand.}
	\label{fig:ISU_teststand}
\end{figure}

A test facility was built at Iowa State University (ISU) to characterize these LAPPDs and to develop a test stand for vertical integration tests of the ANNIE electronics, shown in Fig.~\ref{fig:ISU_teststand}. The facility consists of a dark box with a PiLas laser capable of delivering pulsed blue laser light with rough 30 picosecond pulse duration. The optics allow for beam characterization, 2D translation, and possible triggering on a photodiode. Readout options include a fast, 40 Gsample/second oscilloscope and the custom designed PSEC-4 electronics to be used in ANNIE. Commissioning of the ISU test-stand included the development of a printed circuit board and mounting structure to enable easy connection between the detector, readout, and HV.

\begin{figure}
	\centering
	 \begin{tabular}{c c}   
                \includegraphics[width=0.5\linewidth]{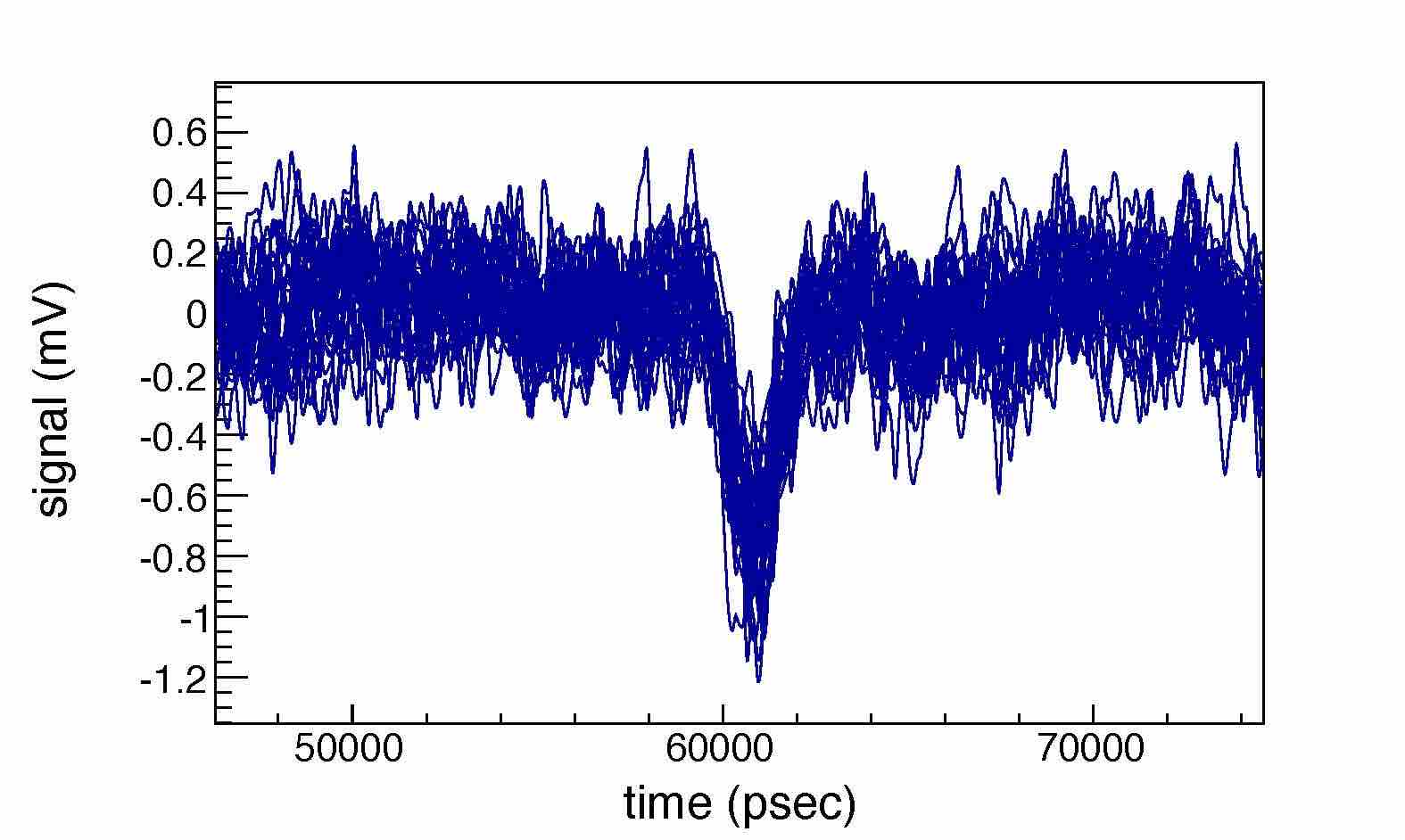}&
                \includegraphics[width=0.5\linewidth]{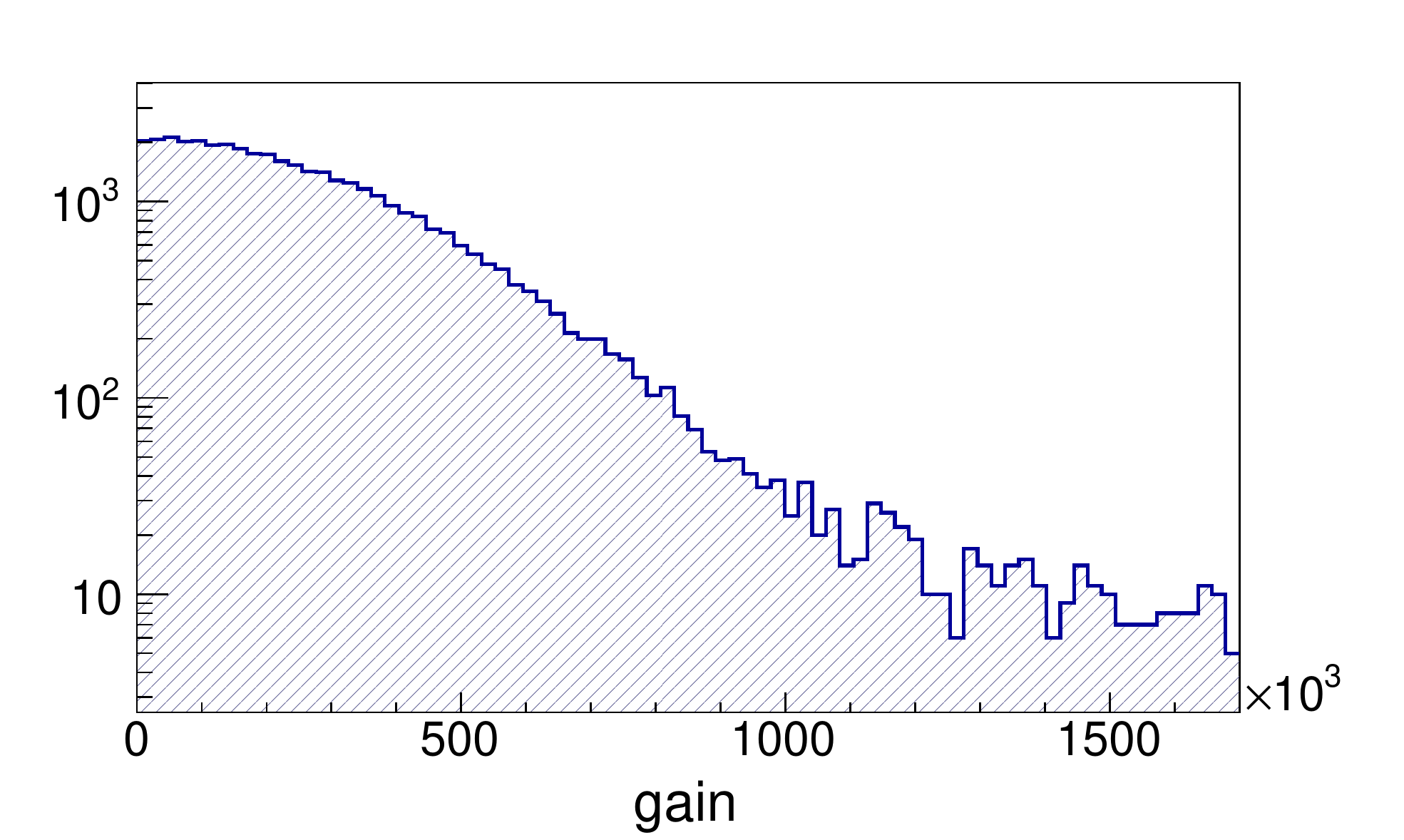}\\
                \includegraphics[width=0.5\linewidth]{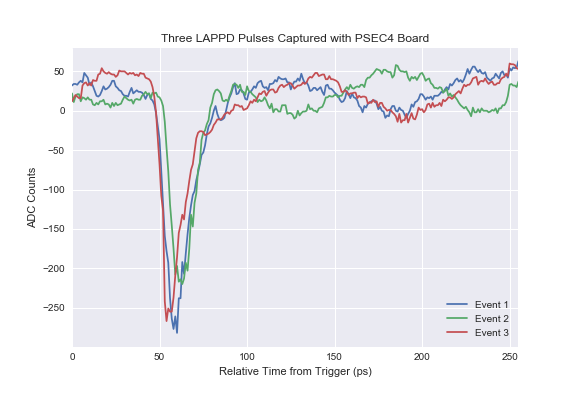}&
                \includegraphics[width=0.5\linewidth]{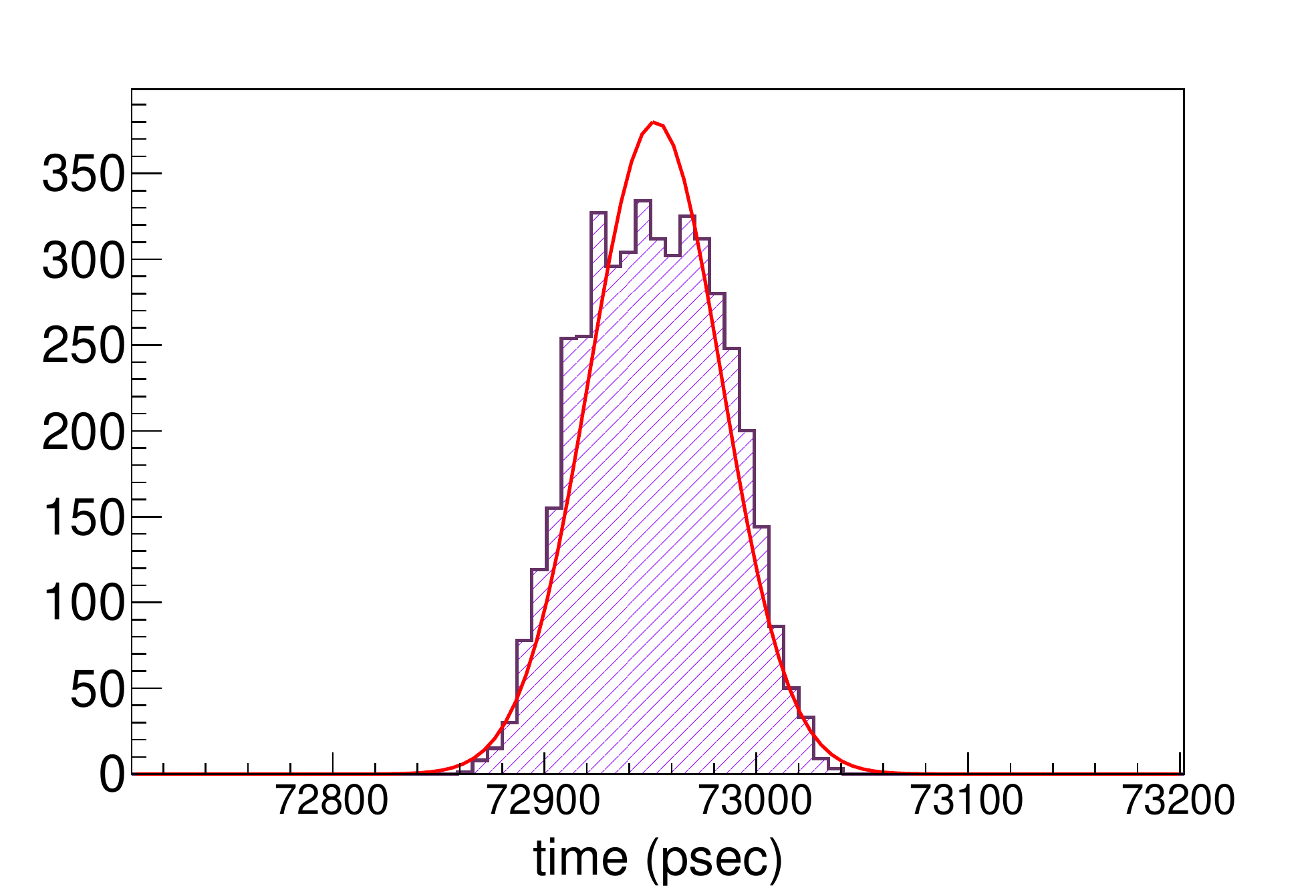}\\
     \end{tabular}         
	\caption{TOP~LEFT:~Example of single photoelectron pulses from LAPPD-9. 
    TOP~RIGHT:~The single-PE gain distribution of LAPPD-9. 
    BOTTOM~LEFT:~Several example multi-PE pulses from LAPPD-12, acquired using the PSEC front end readout. 
    BOTTOM~RIGHT:~The multi-PE TTS distribution measured using the ISU test stand. The 30 psec sigma and non-Gaussian shape is due to the limitations of the laser, which should be sufficient for characterizing 50 psec photosensors.}
	\label{fig:LAPPD_Characterization}
\end{figure}

LAPPD-9 was delivered to Iowa State in January of 2017 and used to commission the Iowa State LAPPD characterization setup. In early March 2017, LAPPD-9 was exchanged for LAPPD-12 which is still being tested at ISU and being used for development work of the PSEC-4 electronics and triggering. Various characteristics of LAPPD-9 were measured, including the gain and single photoelectron (PE) time resolution, shown in Fig.~\ref{fig:LAPPD_Characterization}.  Gains of these two tiles are currently below the target of 10$^6$. However, of all of the characteristics of these MCP-based detectors, this parameter should be the easiest to address. Early test modules have shown that these detectors will be able to achieve peak gains as high as 10$^7$~\cite{LAPPDtiming}. Moreover, gains of order 10$^5$ are acceptable with electronic pre-amplification and the printed circuit to which the LAPPDs are mounted is being designed to accommodate amplifiers.

Early tests of the timing characteristics focused on establishing the sensitivity of the test setup. At high laser intensities we have measured transit time spreads (TTS) of 30 picoseconds, limited by the characteristics of the laser. This should be more than sufficient for characterization of baseline LAPPDs, which are expected to have an intrinsic resolution of 50 picoseconds driven by the geometry of the detector~\cite{LAPPDtiming}. At lower light levels, we have observed resolutions approaching 100 picoseconds, limited by signal-to-noise due to the low gain. We expect to see considerable improvements with higher gain prototypes, although 100 picoseconds is sufficient for ANNIE. Efforts at the ISU test stand are focused on optimizing the operational voltages of the detector and measuring the single PE time resolution.

\subsection{Vertical Integration of the PSEC Electronics}

A major objective the Iowa State LAPPD facility, in addition to characterizing the LAPPDs, is the vertical integration of LAPPDs into the complete ANNIE readout and DAQ. LAPPDs can be connected to the PSEC4-electronics designed by UChicago for LAPPD readout (further described in Sec.~\ref{sec:electronics}). The lower left panel of Fig.~\ref{fig:LAPPD_Characterization} shows an example signal readout by the PSEC system.

The ISU test stand has a vertical slice of the ANNIE DAQ complete with one module from each system. Once work on the PSEC-4 triggering is complete, the next step will be to incorporate the readout of an Incom LAPPD into the vertical slice. This work is expected to take place over the summer and fall of 2017.

ANNIE PSEC system development is planned to proceed as follows:
(1) development of the trigger scheme with ANNIE's central card; (2) transfer of firmware and triggering scheme to a new central card; (3) vertical integration of the PSEC system with a slice of the ANNIE readout, and (4) operation of a cosmic ray test stand with scintillator paddles and LAPPD, using the integrated ANNIE readout and DAQ.

\section{Phase~II Overview and Physics Optimization}
\label{sec:phase2_overview}

The main physics goal of Phase~II is the detection of neutrons originating from neutrino interactions in the water volume, with a secondary objective of making high-statistics measurements of neutrino cross-sections on water.
The realization of these goals will require reconstruction of muons and neutrons in the tank using PMTs and LAPPDs, reconstructing the energy and momentum of outgoing muons in the MRD, and vetoing neutrino interactions in the rock with the front veto.

As explained in Sections~\ref{sec:phase2_mechupgrade} and~\ref{sec:phase2_readoutupgrade}, the detector design will undergo several changes to fully exploit the capabilities of the sub-detectors already present in the hall. The most significant Phase~II modifications will be made to the ANNIE tank:  the introduction of LAPPDs, an increase in conventional PMT coverage, and loading of the water with gadolinium sulfate. Also important will be the operation of the fully-instrument MRD.

In preparation for Phase~II, the collaboration has developed a suite of reconstruction tools and a set of complementary detector simulations. This work will allow the collaboration to quickly analyze Phase II data. The results presented in this section are based on these tools.

The baseline goal for the ANNIE Phase II physics analysis is to provide a measurement of neutron yields from experimentally CCQE-like interactions as a function of muon kinematics. Muons are reconstructed both in the water volume and matched to tracks in the MRD. The kinematic effects of selecting interactions with muons in the MRD are described in Sec~\ref{sec:kinematics_mrd}. Efficient neutron detection is enabled by the delayed capture on dissolved Gd. This requires more photocathode coverage than can be provided by the number of LAPPDs in the scope of this proposal. ANNIE therefore relies on a significant number of conventional PMTs to enable efficient neutron tagging, as demonstrated in Sec~\ref{sec:ncap_efficiency}. Finally, the analysis will require precision reconstruction of the interaction point by the LAPPD system in order to select a sample of fiducial events far enough from the edges of the detector to fully contain any neutrons produced (Sections~\ref{sec:ncap_containment} and ~\ref{sec:vertex_fidu}). Sufficient LAPPDs may also permit the reconstruction NC events, resonant pion final states, and interactions where the muon stops in the water volume without entering the MRD. Future studies would be needed to demonstrate these capabilities.




\subsection{Kinematics and MRD}
\label{sec:kinematics_mrd}

\begin{figure}
	\centering
	 \begin{tabular}{c c}
                \includegraphics[width=0.55\linewidth]{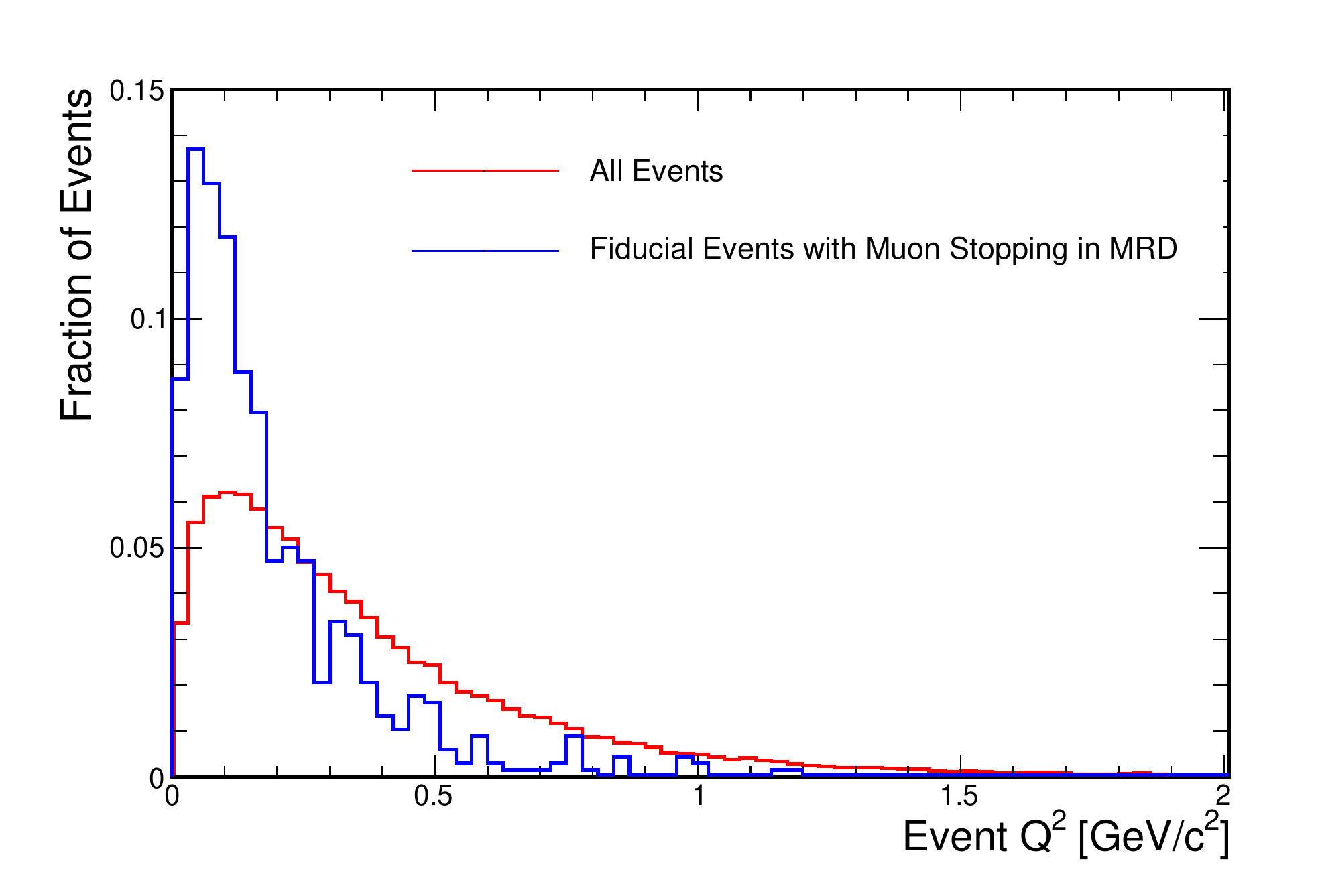}&
                \includegraphics[width=0.55\linewidth]{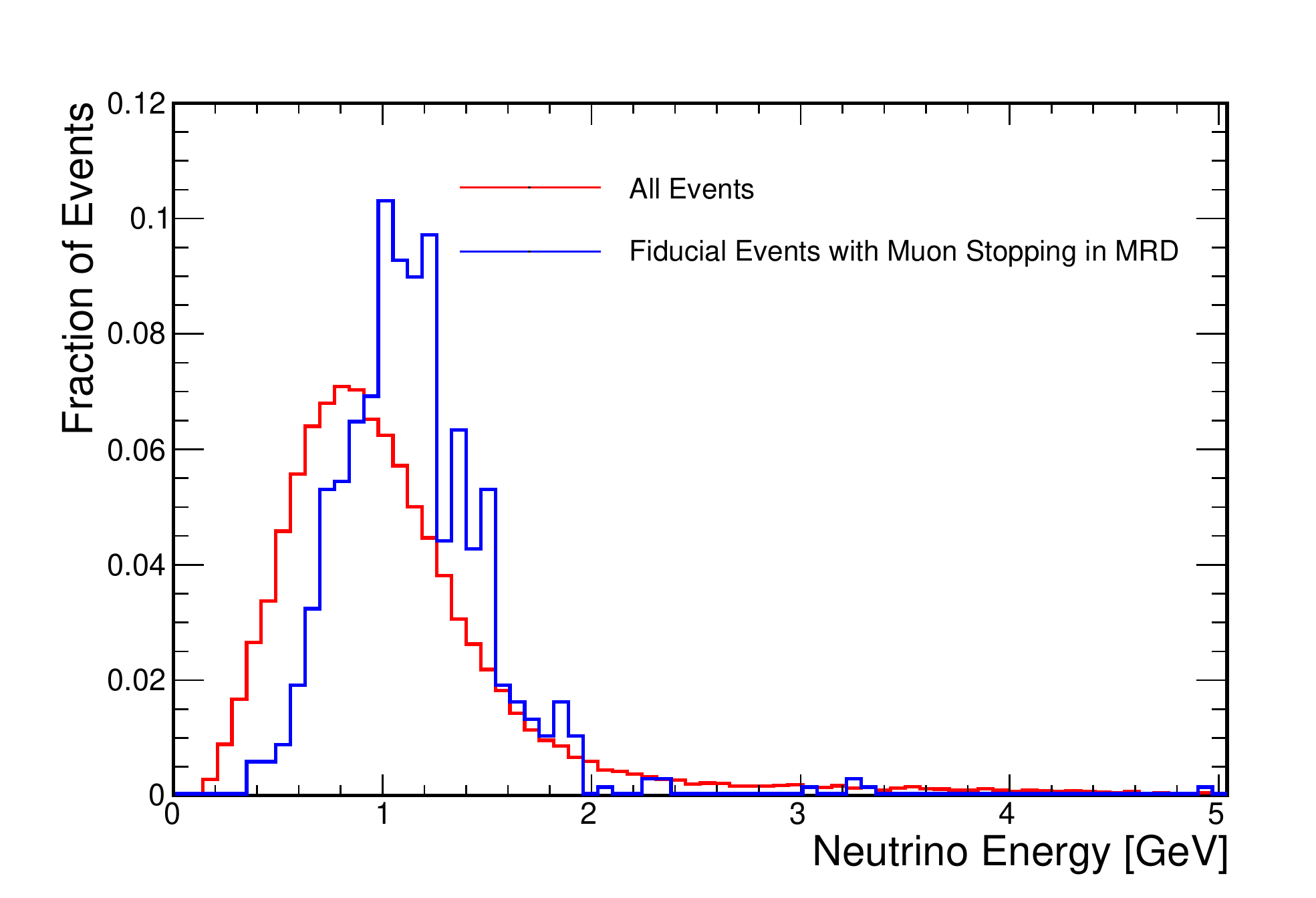}\\
         \end{tabular}
	\caption{LEFT: The normalized $Q^2$ distribution for all events (red line) and for 2.5-ton fiducial events with muons ranging out in the MRD (blue line). 
    RIGHT:~The normalized $E_{\nu}$ distribution for all events (red line) and for 2.5-ton fiducial events with muons ranging out in the MRD (blue line).}
	\label{fig:Q2andEnuDists}
\end{figure}



The MRD is used to measure the final state muon and is central to determining the kinematics of the neutrino event. The kinematic distributions of incident neutrinos from the Booster Neutrino Beam, alongside the distribution of neutrino interactions with a muon stopping within the MRD are shown in Fig.~\ref{fig:Q2andEnuDists} using simulations of the Phase~II detector. The distributions for the incident beam and the MRD acceptance neutrino events present only a small, expected skew toward higher neutrino energies and lower $Q^{2}$ on the latter. 

Table~\ref{tab:Tbl_Event_Counts} shows the number of expected fiducial NC, CC, CCQE and CC-Other interactions that enter, stop within, fully penetrate, or exit through the sides of the MRD.
These numbers represent neutrino interactions within a 2.5-ton central fiducial volume, integrated over 1 year of running ($2\times10^{20}$ POT). For 64\% of events with a muon entering the MRD (18\% of all fiducial CC events), the muon is fully stopped, allowing complete track reconstruction and accurate energy estimation. 

These studies show that a significant number of neutrino interactions will have complete track reconstruction and accurate energy estimation. Furthermore, the kinematics of these interactions are reasonably representative of all incident beam neutrinos. 

\begin{table}[t]
\centering
\caption{Fiducial Event Counts for 1 Year of Running}
\label{tab:Tbl_Event_Counts}
\begin{tabular}{lllll}
                      & NC    & CC    & CCQE  & CC-Other \\
All                   & 11323 & 26239 & 13674 & 12565    \\
Entering MRD          & 2     & 7466  & 4279  & 3187     \\
Stopping in MRD       & 2     & 4830  & 2792  & 2038     \\
Fully Penetrating MRD & 0     & 1454  & 761   & 693      \\
Exiting Side of MRD   & 0     & 1181  & 726   & 455     
\end{tabular}
\end{table}

\subsection{Neutron Tagging Efficiency and PMT Coverage}
\label{sec:ncap_efficiency}

The baseline PMT coverage for ANNIE Phase II is set by the need to efficiently detect neutron captures on Gd. An important early goal of the simulations effort was to establish that ANNIE had a sufficient number of PMTs among the borrowed stock to meet this goal. Since the UC Irvine tubes are to be returned before Phase II, simulations have been done to establish the minimum number of replacement tubes that would have to be purchased. 

Studies were conducted using the light 5~MeV electrons as a proxy for the light yield of Gd captures\footnote{Super-K data and GEANT simulations show that gadolinium capture gamma cascades result in about the same light output as a 5~MeV electron}. These simulations were run in two configurations:  one including 60 of the Phase~I Hamamatsu 8-inch R5912 PMTs and a configuration including 40 new Hamamatsu 8-inch HQE R5912 PMTs instead. In both cases, the light collection efficiencies are quite similar, with efficiencies above 10 PE of nearly 100\% at the center of the tank.


\begin{figure}
	\centering
                \includegraphics[width=0.5\linewidth]{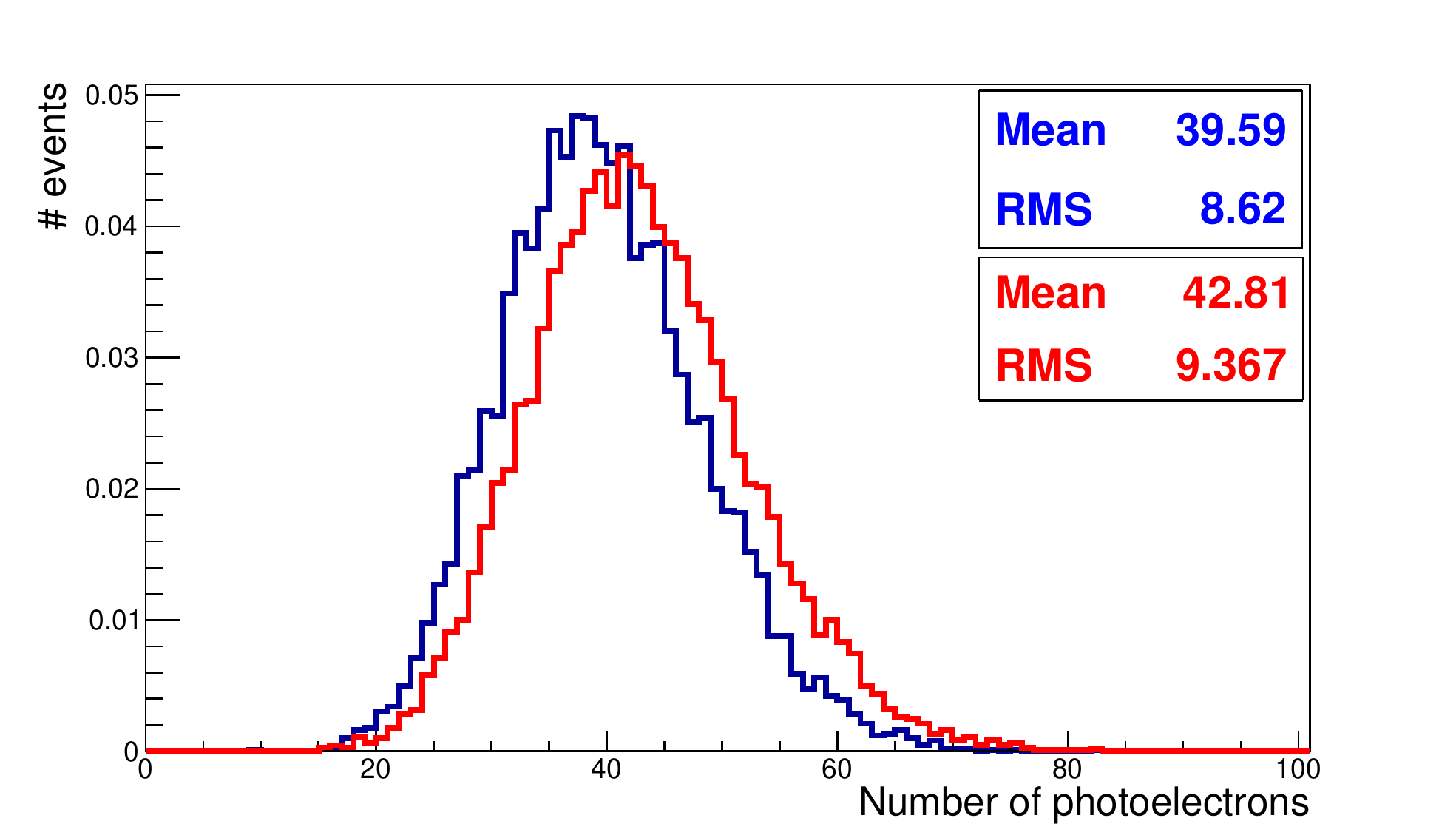}       
	\caption{Total number of PE detected for 2 different Phase~II PMT configurations for 5~MeV electrons generated isotropically at the center of the tank. In both configurations, 22 11-inch PMTs and 20 10-inch PMTs are installed respectively at the top and bottom of the tank and 45 10-inch PMTs are installed on the barrel of the tank. To those barrel PMTs 60 existing 8-inch PMTs (red curve) or 40 new 8-inch PMTs (blue curve) are added.}
	\label{fig:NPE_2PMTconfigs}
\end{figure}

\begin{figure}
	\centering
	 \begin{tabular}{c c}
		\includegraphics[width=0.5\linewidth]{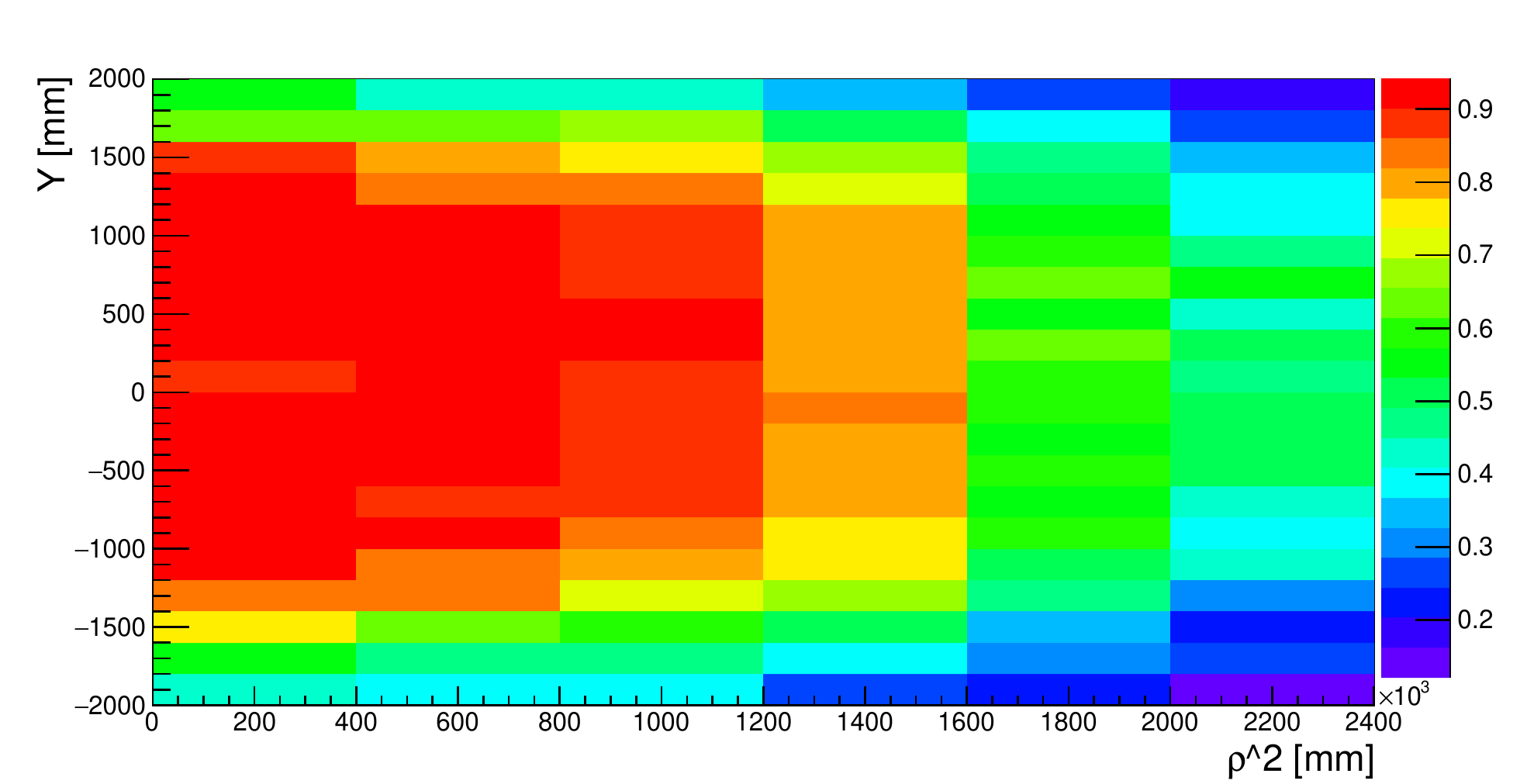}
        \includegraphics[width=0.5\linewidth]{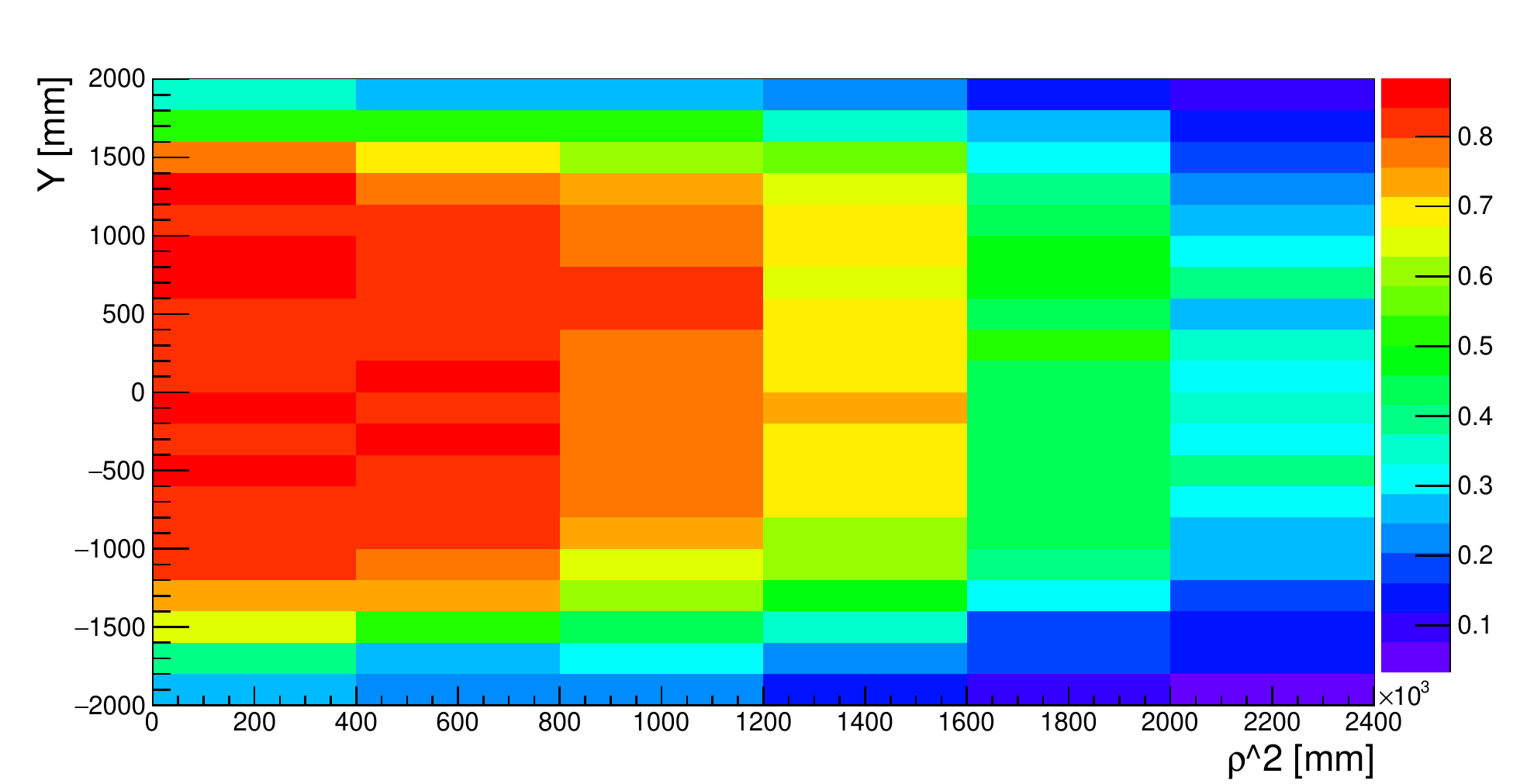}
	\end{tabular}
	\caption{ Temperature plots showing the thermal neutron detection efficiency as a function of the neutron generation position in $\rho^2$ and y, the detector vertical axis, for captures leading to more than 5 (LEFT) or 10 (RIGHT) photoelectrons detected by the PMTs.}
		\label{fig:ThermNeutron_eff}
\end{figure}

In order to assess the neutron capture detection efficiency as a function of the location in the tank for the proposed PMT coverage, we can look at the probability for thermal neutrons generated in the tank to lead to a capture that will exceed a certain threshold of photoelectrons. Figure~\ref{fig:ThermNeutron_eff} shows this probability for both 5 and 10 PE thresholds. The maximum probability at the center of the tank is higher than the ratio of gadolinium and hydrogen captures (about 0.9) since, in some cases, some neutron captures on hydrogen can produce a non-negligible amount of photoelectrons.

These studies show that the planned PMT coverage (with either 60 existing 8-inch PMTs or 40 new 8-inch PMTs) is sufficient for neutron tagging in the ANNIE detector with a reasonably uniform efficiency at 90\% limited by the Gd concentration. 


\subsection{Neutron Containment and Optimization of Fiducial Volume}
\label{sec:ncap_containment}

\begin{figure}
	\centering
           \includegraphics[width=0.6\linewidth]{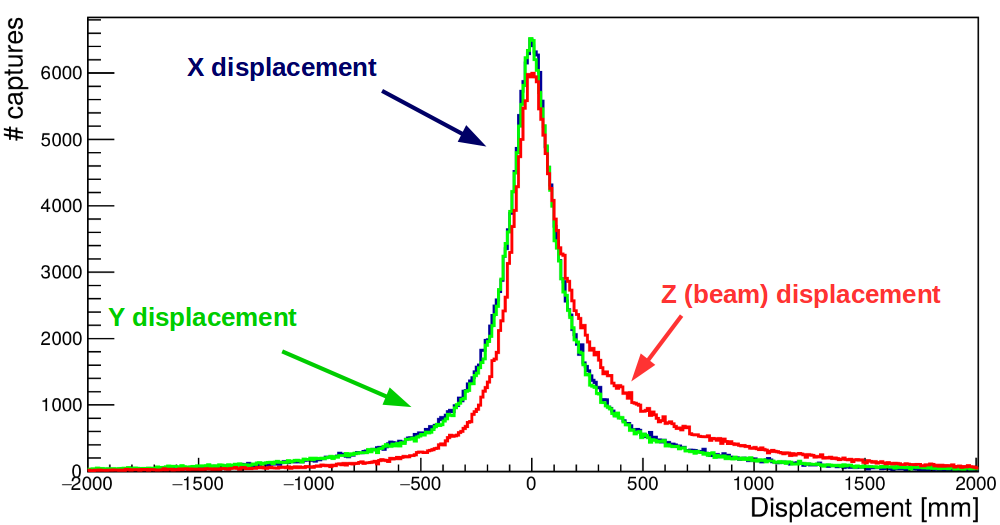}\\       
	\caption{Projections in x, y, and z of the distances between the creation and capture vertices of neutrons resulting from neutrino interactions.}
	\label{fig:NeutronCap_Proj}
\end{figure}

Neutrino interactions in the ANNIE tank create neutrons with energies ranging up to tens of MeV. Emitted on average in the forward direction, those neutrons will travel some distance 
from the initial interaction point before reaching thermal energies through successive elastic scatterings. Figure~\ref{fig:NeutronCap_Proj} shows projections of the distances between the neutron creation vertices and their capture vertices. While the projections in x and y, the directions transverse to the beam, are symmetric as expected, the projection in z, the beam direction, is forward with a mean shift with respect to the center of about 18~cm. This necessitates a fiducial volume set slightly back from the center of the tank in the beam direction in order to maximize acceptance.

\begin{figure}
	\centering
	 \begin{tabular}{c c}
                \includegraphics[width=0.5\linewidth]{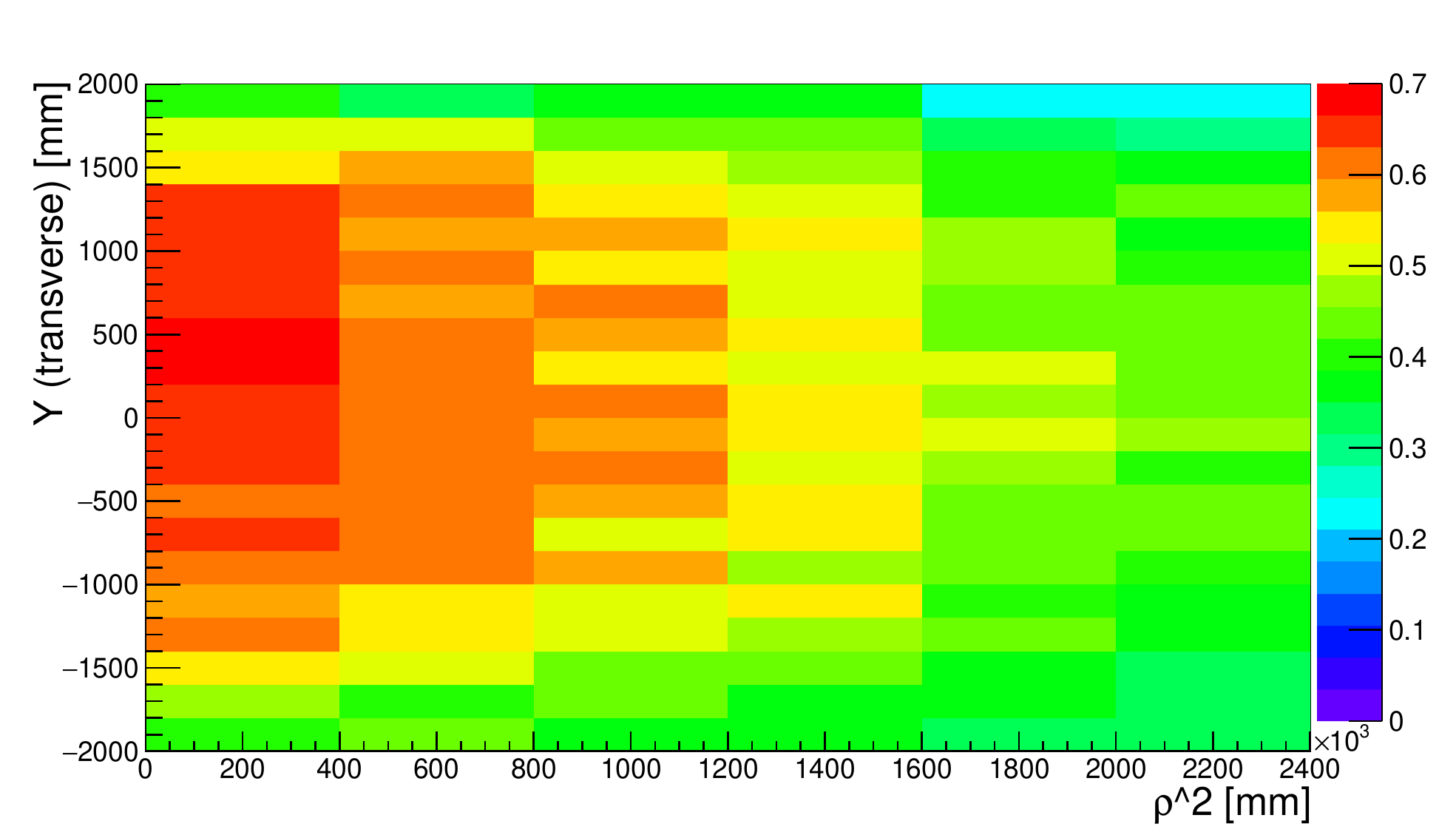}&
                \includegraphics[width=0.5\linewidth]{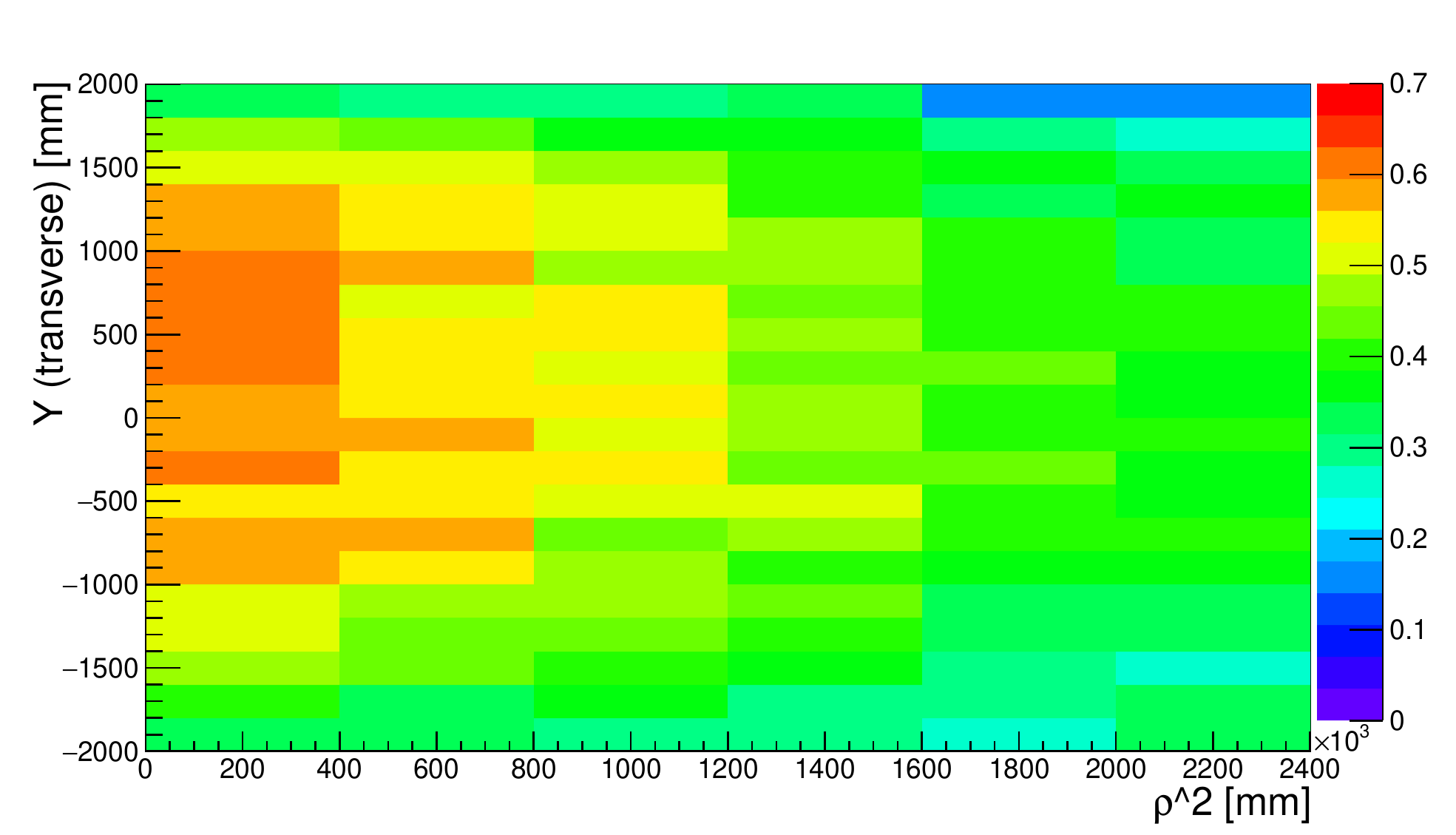}\\
                \includegraphics[width=0.5\linewidth]{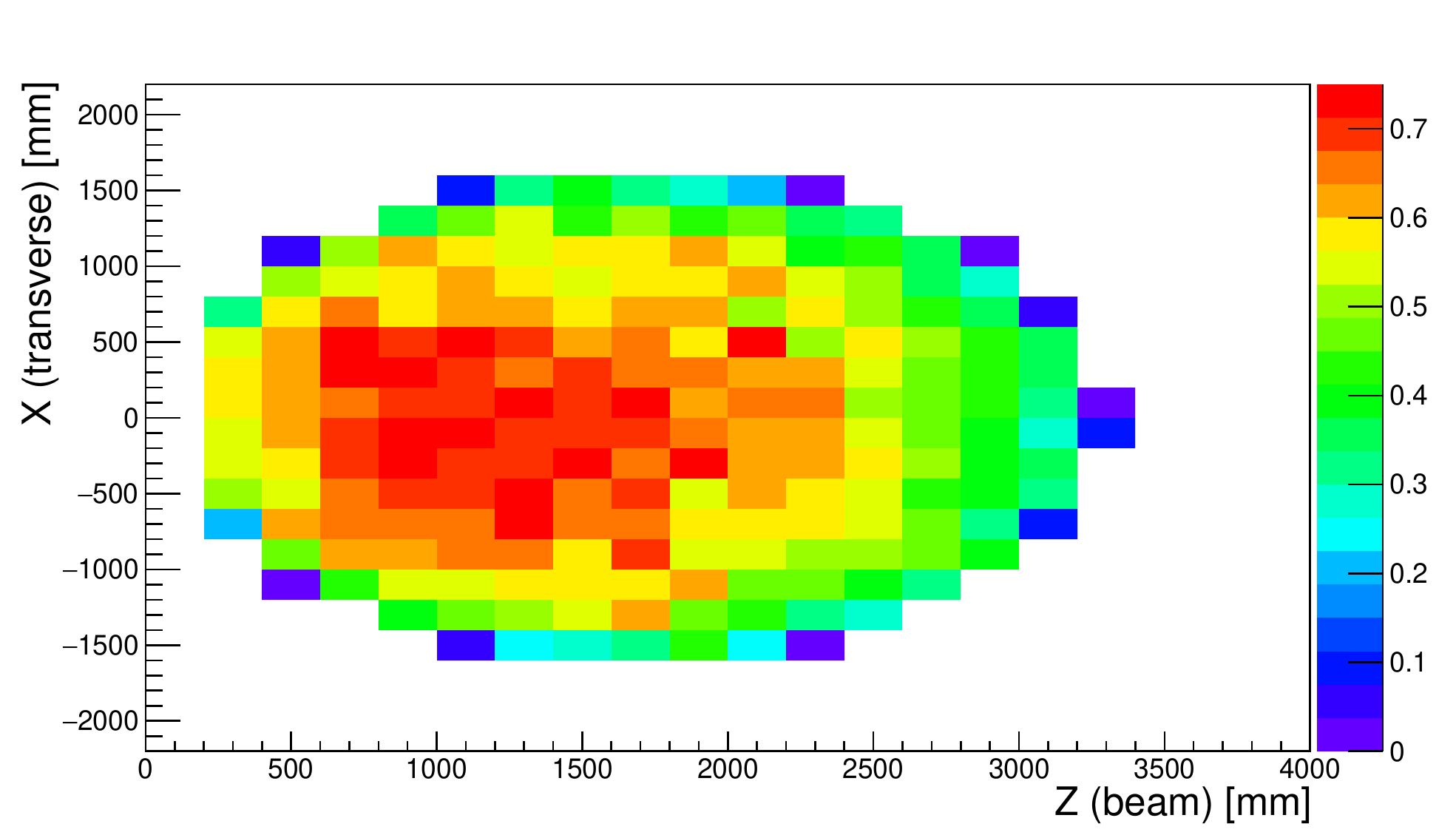}&
                \includegraphics[width=0.5\linewidth]{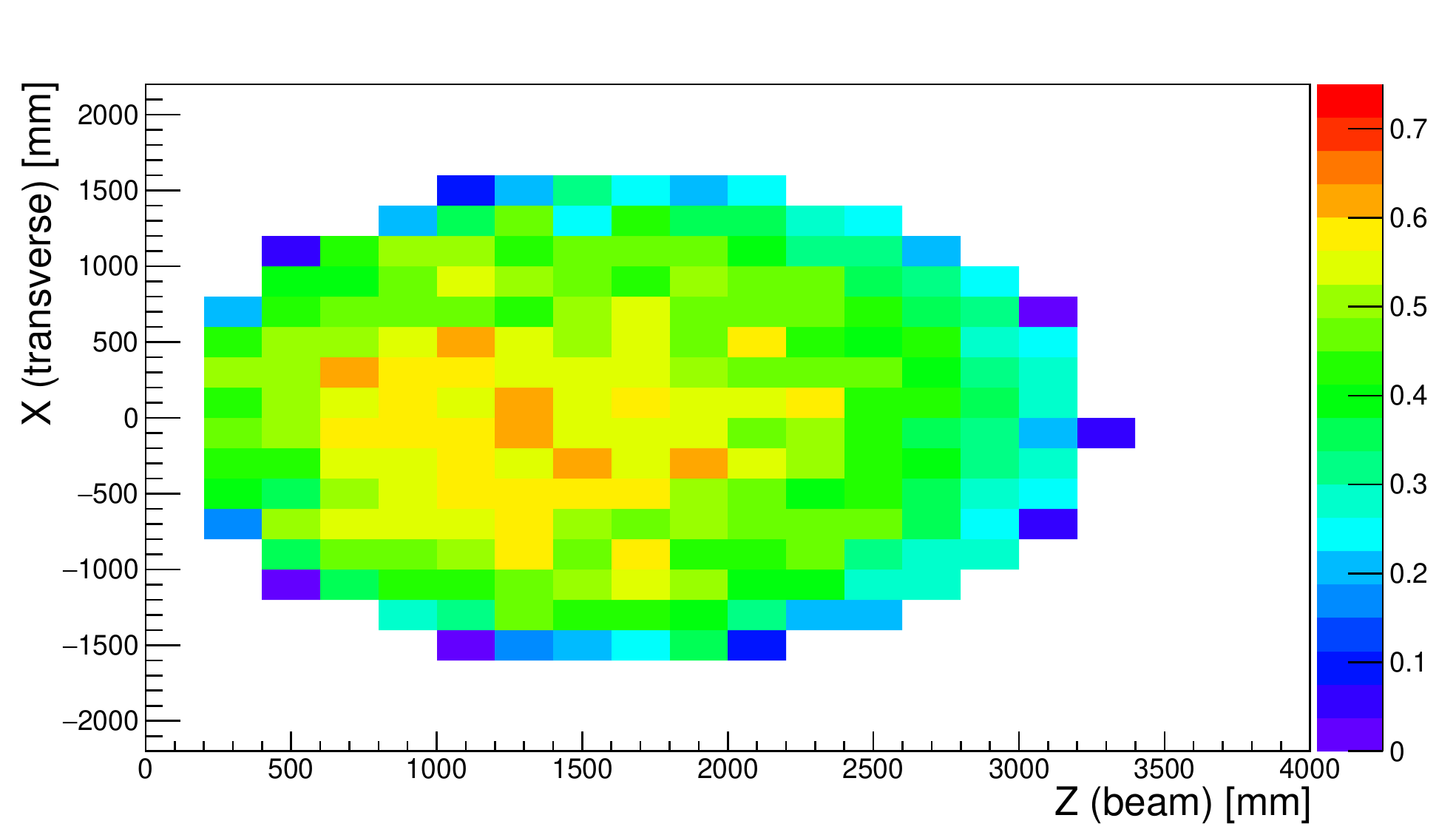}\\
         \end{tabular}         
	\caption{TOP: Temperature plot showing the neutron detection efficiency as a function of the {\it neutrino} interaction position in $\rho^{2}$ and y, the vertical axis, for a 5 photoelectron (LEFT) and 10 photoelectron (RIGHT) threshold. BOTTOM: Probability density plot showing the neutron detection efficiency as a function of the neutrino interaction position in x, the transverse direction, and z, the beam direction, (integrated between~-1~and +1~meters in the vertical direction) for a 5 photoelectron (LEFT) and 10 photoelectron (RIGHT) threshold.}
	\label{fig:NeutronDist_Eff}
\end{figure}

Fig.~\ref{fig:NeutronDist_Eff} shows the product of neutron acceptance and detection efficiency as a function of the position of the neutrino interactions from which the neutrons originate for two different light collection thresholds (5 and 10 photoelectrons per neutron capture). The position for the top two panels is expressed in $\rho^{2}=x^{2}+z^{2}$, with x and z the transverse and beam direction respectively, and y, the vertical direction. While this coordinate system underestimates the detection efficiency in $\rho^{2}$ due to the forward neutron emission and the subsequent higher neutron detection efficiency for neutrino interactions closer to the upstream side of the tank, one can notice that the efficiency is maximized when only considering a volume between -1~and +1~meters in the vertical axis.

When only selecting neutrino interactions occurring in this [-1;+1]~meters vertical volume, the neutron detection efficiency integrated over the vertical axis dramatically improves as displayed on the two bottom panels of Fig.~\ref{fig:NeutronDist_Eff}.

Based on these studies, we choose a fiducial volume where the detection efficiency of neutron captures is maximized. To allow both an efficient containment of neutrons and a good reconstruction of muon tracks and vertices using the LAPPDs on the downstream wall of the detector, this volume must be placed upstream of the detector center in the beam direction while being centered in both transverse directions. Its size has been optimized to 2.5~tons in a volume ranging from [-0.6;+0.6]~meters in the transverse direction, [-1;+1]~meters in the vertical direction, and [+0.6;+1.6]~meters in the beam direction.

Reconstruction studies described in Sec.~\ref{sec:vertex_fidu} have been performed to confirm the optimal fiducial volume size and its position in the tank.

\subsection{Neutrino Vertex Fiducialization and Resolution}
\label{sec:vertex_fidu}

Selecting fiducial events, as described in Sec~\ref{sec:ncap_containment}, requires accurate vertex reconstruction at a level better than the O(100) cm dimensions of the fiducial volume itself. As noted in Sec.~\ref{sub:TechnologicalGoals}, ANNIE is designed to exploit the excellent timing resolution provided by LAPPDs to meet this need. This section presents the results of a first Monte Carlo study demonstrating the vertex reconstruction capabilities of the ANNIE detector.


To reconstruct a beam neutrino interaction vertex, we employ an algorithm based on the techniques used in previous WCh detectors, such as Super-Kamiokande. The track parameters of a charged particle are determined by a maximum-likelihood method. Three spatial parameters specify the vertex position, one time parameter reflects when the interaction took place and two angular parameters specify the direction of the primary lepton track. These six parameters are varied in the fit to maximize the figure-of-merit (FOM). For each hit we calculate a time residual, the difference between the actual hit time and the predicted hit time as derived from its relationship to the hypothesized track. Given that the primary lepton propagates at the speed of light ($c$) and emits a Cherenkov photon at $42\degree$, traveling at $c/n$ with $n$ being the refractive index of water, we define the time residual ($\Delta t$) as:
\begin{equation}\label{eq:timing_residual}
	\Delta t = t_{hit} - \frac{L_t}{c} - \frac{L_p}{c/n},
\end{equation}
where $t_{hit}$ is the hit time, $L_t$ is the length of the primary lepton track from the neutrino vertex to the point where the photon is emitted and $L_p$ is the distance the photon travels. 
The FOM takes its maximum value when the width of the time residual distribution is minimized. The vertex position which maximizes the FOM is defined to be the best fit vertex position. 



A sample of 780 muons matched to ANNIE event kinematics was generated using a custom Geant4-based simulations package developed for the experiment. We calculate the radial displacement of each reconstructed vertex from the true interaction point ($\Delta r$) on an event by event basis. The vertex resolution is then defined as the value of $\Delta r$ at the 68th percentile of all successfully reconstructed events from the sample. Using this method, vertex resolutions were calculated for three different photodetector configurations:

\begin{itemize}
\item 20\% coverage of the inner surface of the tank by conventional PMTs (approximately 120 10-inch PMTs).
\item 5 LAPPDs on the downstream wall of the tank with no PMTs.
\item 21 LAPPDs on the downstream wall of the tank with no PMTs.
\end{itemize}
\begin{figure}
  \begin{center}
    \begin{tabular}{c  c}
      \includegraphics[width=0.5\linewidth]{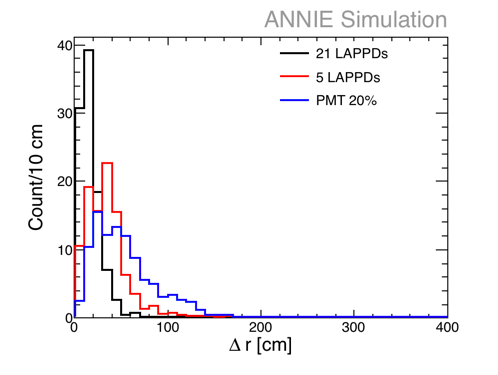} &
      \includegraphics[width=0.5\linewidth]{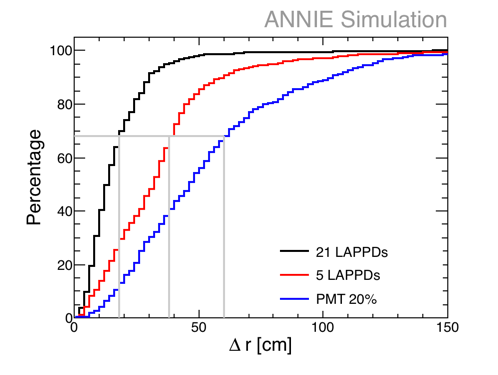} 
    \end{tabular}
  \end{center}
  \caption{Raw $\Delta r$ distributions (LEFT) and cumulative $\Delta r$ distributions (RIGHT), normalized to a percentage of successfully reconstructed events. The panels compare configurations using five LAPPDs (red), 21 LAPPDs (black), and 20\% PMT coverage (blue, about 120 tubes). Calculations for each configuration used the same sample of 780 events.}
  \label{fig:5vs21vsPMTs}
\end{figure}

Figure~\ref{fig:5vs21vsPMTs} compares the vertex reconstruction performance achieved in each of the three configurations. The left panel shows the raw distributions of $\Delta r$, while the right panel shows the cumulative distributions, expressed using percentages of successfully reconstructed events, to highlight the bound at 68\%. The vertex resolution from 20\% coverage with conventional PMTs is only 60 cm.
A configuration with five LAPPDs, and no PMTs, achieves an improved resolution of 38 cm. Increasing the number of LAPPDs to 21 would bring an additional factor of two improvement over the projected vertex resolution with 5 LAPPDs of 18 cm. Fits combining both the LAPPD and PMT systems and more sophisticated reconstruction algorithms are expected to improve further on these results.

These studies confirm that instrumenting the ANNIE detector with at least five LAPPDs will meet the baseline physics needs of the experiment.  The combination of ANNIE PMT coverage with 21 LAPPDs would bring even more dramatic improvements.

\section{Phase~II Mechanical Modifications}
\label{sec:phase2_mechupgrade}

As mentioned in Sec.~\ref{sec:phase2_overview}, the modifications the detector will undergo in preparation for Phase~II are focused on the Muon Range Detector (c.f Sec.~\ref{sec:mrd_refurb}) and the water tank (c.f Sec.~\ref{sec:gd_loading} and~\ref{sec:conventional_pmts}). Updates to the electronics and HV systems will be further discussed in Sec.~\ref{sec:electronics}.


\subsection{Modifications to the Tank and Inner-structure }
\begin{figure}[h]
	\centering
	 \begin{tabular}{c c}
                \includegraphics[width=0.45\linewidth]{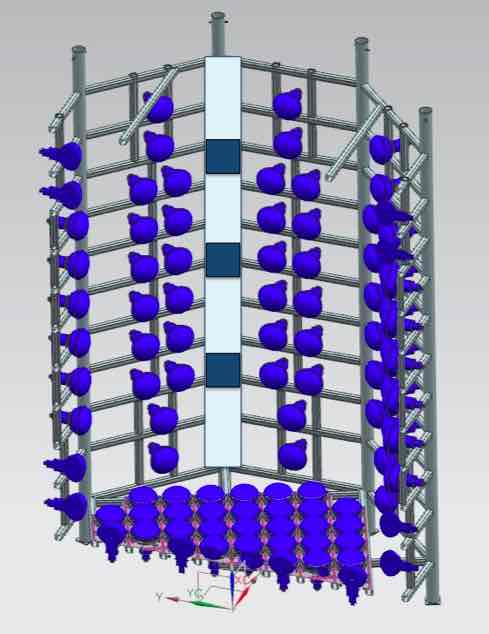}\\
         \end{tabular}         
	\caption{A model of the fully-populated Phase~II inner structure.}
	\label{fig:Inner_structure}
\end{figure}

\begin{figure}[h]
	\centering
	 \begin{tabular}{c c}
                \includegraphics[width=0.5\linewidth]{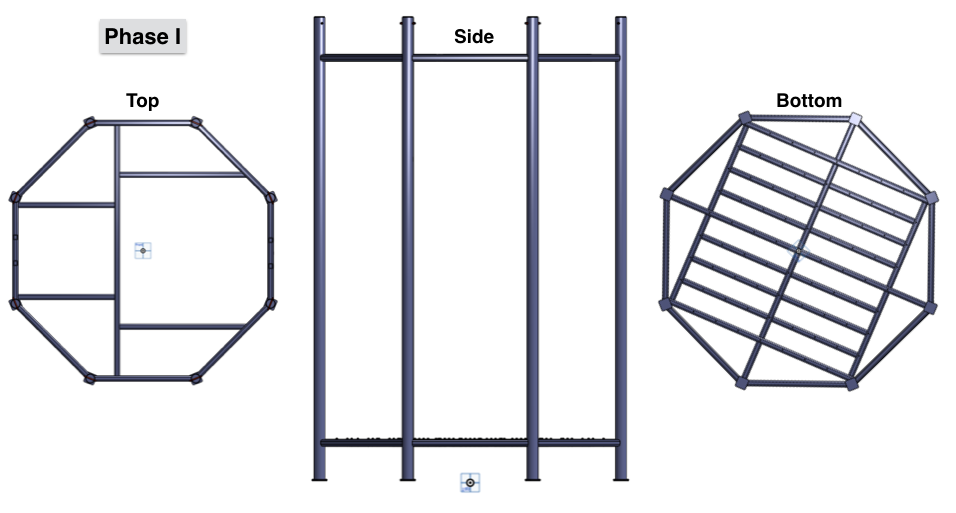}\\
                \includegraphics[width=0.5\linewidth]{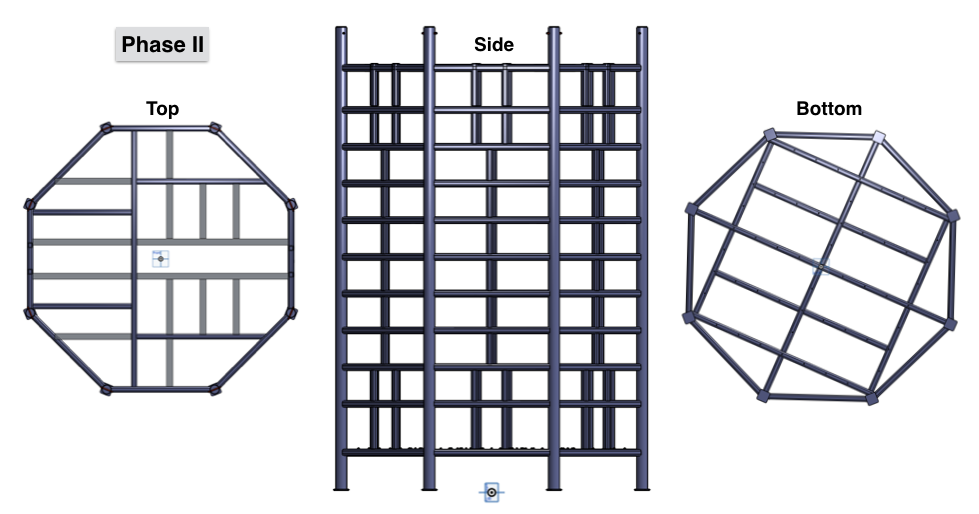}\\

         \end{tabular}         
	\caption{A comparison between the Phase~I (TOP) and Phase~II (BOTTOM) inner structures.}
	\label{fig:Inner_structure-Comp}
\end{figure}

An exploded view of the ANNIE tank is shown in Fig.~\ref{fig:Inner_structure}. As mentioned in Sec.~\ref{sec:run1_overview}, the ANNIE tank is currently filled with ultra-pure water for Phase~I operation, and will be loaded with gadolinium sulfate at a concentration of 0.2\% by weight, corresponding to a 0.1\% concentration of pure Gd by weight. The inner PMT supporting structure has an octagonal shape (see Fig.~\ref{fig:Inner_structure-Comp}) and holds sixty 8-inch photomultiplier tubes (PMTs) at the base of the tank (top panel). It will be modified to hold more PMTs ($\sim$~120) on all sides (bottom panel). In addition to these PMTs, several LAPPDs will be mounted at the sides of the tank. 
 
The sixty 8-inch PMTs at the bottom of the tank will be replaced with twenty 10-inch PMTs. The bottom grid structure was welded specifically for 8-inch PMTs but a slight modification will be sufficient to hold new 10-inch PMTs by  removing every other cross layer tube at the bottom structure. 

The inner structure was previously prepared for holding PMTs on the side. What remains to be done is the mechanical design and welding of each side of the inner structure for mounting PMTs. Two possible side designs are being considered. For the first option, the sides will be designed separately and fixed up on the structure after the PMTs are mounted. In the second design option, the cross bars are directly welded on the structure and the PMTs are mounted on them, shown in Fig.~\ref{fig:Inner_structure-Comp}. 

\subsection{Muon Range Detector (MRD) Refurbishment}
\label{sec:mrd_refurb}
\begin{figure}
	\centering
	 \begin{tabular}{c c}
                \includegraphics[width=0.4\linewidth]{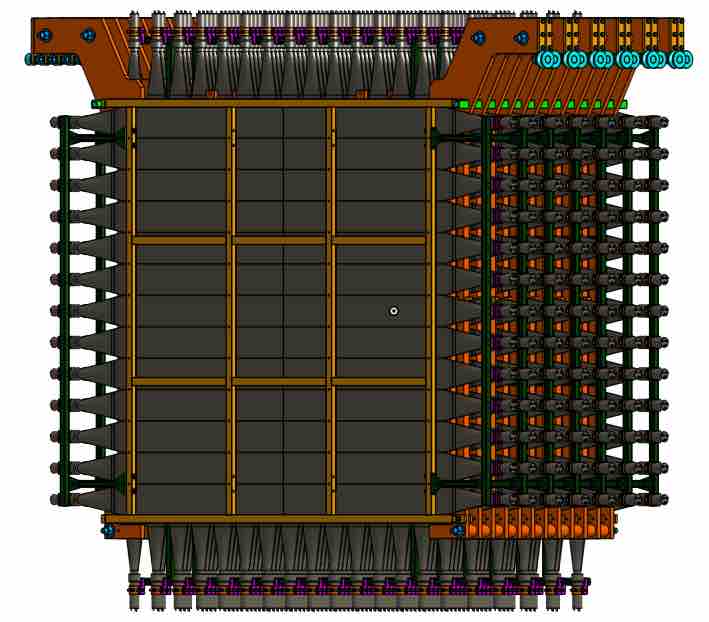}&
                \includegraphics[width=0.35\linewidth]{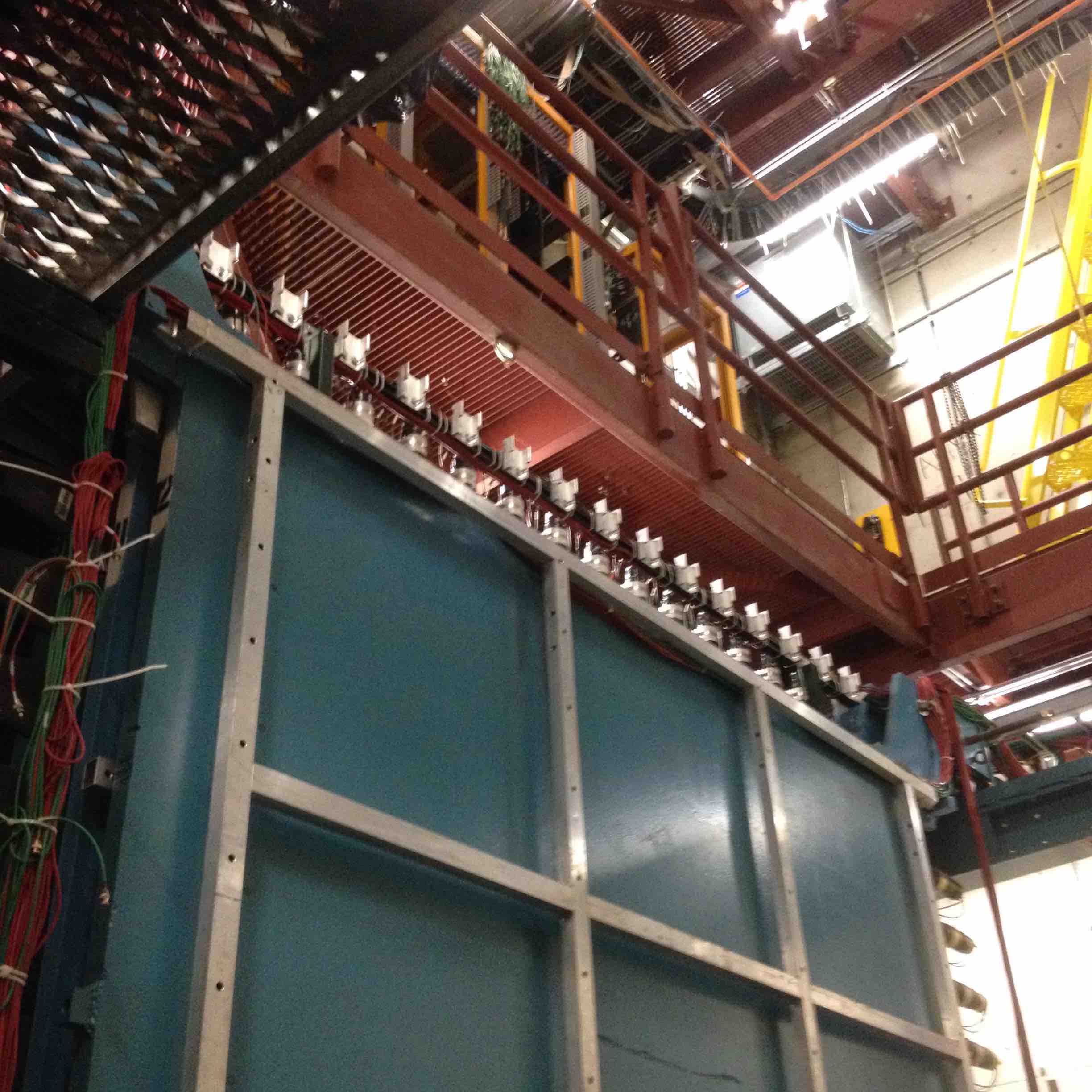}\\
 
         \end{tabular}         
	\caption{A diagram of the MRD (LEFT) and a photograph of the installed MRD in the hall (RIGHT).}
	\label{fig:MRD_pics}
\end{figure}

The ANNIE physics analysis relies on a ready-built Muon Range Detector (MRD), already installed and cabled for the SciBooNE experiment. The MRD (pictured in Fig.~\ref{fig:MRD_pics}) is a steel-scintillator sandwich detector designed to range out muons, the majority of which will not fully stop  within the water volume. Complete reconstruction of the muon energy provided by the MRD will be necessary for the kinematic reconstruction of charged current neutrino events in ANNIE. 


The MRD is currently missing 71 of the total 362 paddles previously existing for the SciBooNE experiment. The ANNIE experiment in Phase~II will require the refurbishing and re-installation of these. The latter can only be accomplished while the water tank is out of the hall. In order to begin commissioning the Phase~II detector in summer 2018, the MRD refurbishment must be completed this year in summer 2017 when collaborators (including undergraduate labor) are most available.

Fermilab has already secured the necessary EMI 9954KB 2” PMTs for these muon paddles. They have also committed to identifying the missing paddles (or a similar stock of scintillator) and providing some training for collaborators to learn how to re-assemble them. However, the light guides and acrylic “cookies” used to couple the scintillators to the PMTs have been damaged in the SciBooNE decommissioning process and are missing (see Fig.~\ref{fig:MRD_pics}). Funding has been requested to the DOE in order to purchase a new stock of light guides with attached cookies, as well as the materials needed to glue and tape the paddles back together. 
The MRD refurbishment work will be led by ISU postdoc Emrah Tiras who is based at Fermilab, and performed with labor from ISU undergraduate students who will be at the laboratory during the summer. 

\subsection{Water System and Gadolinium loading}
\label{sec:gd_loading}

Neutrons in pure water can be captured on either a free proton or an oxygen nucleus, with a thermal neutron radiative cross section of 0.33~barns and 0.19~millibarns respectively. The vast majority of captures will thus occur on a free proton through the following reaction: n + p $\rightarrow$ $^{2}_{1}\text{H}$ + $\gamma$, thus leading to the emission of a single 2.2~MeV gamma that will deposit its energy in the medium through successive Compton scatters. With a Cherenkov threshold of 0.26 MeV in water, a decent fraction of this energy will not be converted into Cherenkov light thus making the detection of this gamma-ray difficult. In addition, with such a relatively low capture cross section, thermal neutrons can diffuse for distances up to several meters in water, and their capture time profile follows an exponential law with a decay constant of about 200~$\mu$s.

Gadolinium (Gd) has a thermal neutron radiative capture cross section of 49,000~barns\footnote{This is the average of all the isotopes found in natural gadolinium. The two main contributors are $^{155}\text{Gd}$ and $^{157}\text{Gd}$, having respective abundances of 14.8\% and 15.65\% and with respective cross sections of 60,800 and 254,000~barns} and those captures lead to a cascade of gammas with a summed energy of about 8~MeV\footnote{The released energy following captures on $^{155}\text{Gd}$ and $^{157}\text{Gd}$ is respectively 8.54 and 7.94~MeV.}, thus yielding a higher detection efficiency. Gadolinium loading is performed by adding water-soluble compounds, such as gadolinium chloride (GdCl$_{3}$) or gadolinium sulfate (Gd$_{2}$(SO$_{4}$)$_{3}$) to ultrapure water. The latter compound, gadolinium sulfate, has been identified as a suitable loading candidate by the ANNIE collaboration, for availability and material compatibility reasons.
The time profile of neutron captures on Gd obeys an exponential decay as well, but with a constant of about 30~$\mu$s in case of a Gd compound concentration of 0.2\% by weight, equivalent to a concentration of pure Gd of about 0.1\% by weight.

While gadolinium sulfate is less reactive than gadolinium chloride, compatibility tests must be performed to ensure all materials in contact with the Gd-loaded water will not undergo degradation nor lower the water light transparency over the time of the experiment. Such tests will be performed by UC Davis collaborators, who have experience dealing with Gd-loaded water. 

The purification system used during Phase~I (c.f. Sec.~\ref{sec:run1_overview}) is not suitable for Gd-loaded water. Indeed, the de-ionizing filters would remove all impurities present in the water, including gadolinium ions. This issue can be circumvented by not circulating water through the filtration for longer periods, typically between 6 months and a year. 
Alternatively the collaboration will consider the acquisition of specialized ion exchange resins which remove only monovalent and divalent, leaving the trivalent gadolinium ions in solution. While not possible on the scale of Super-K due to the high cost of the resin (about \$100/liter of resin), the small size of ANNIE makes this feasible at a small cost. The custom resin would be compatible with our existing Phase~1 purification system.

\subsection{Conventional PMTs}
\label{sec:conventional_pmts}


During the first phase of ANNIE, neutron captures were observed within the optically isolated NCV and sixty 8-inch photomultiplier tubes were used for the water volume to identify potential beam neutrinos or beam and cosmic muons in the tank. 

\begin{figure}
	\centering
	 \begin{tabular}{c c c}
                \includegraphics[width=0.3\linewidth]{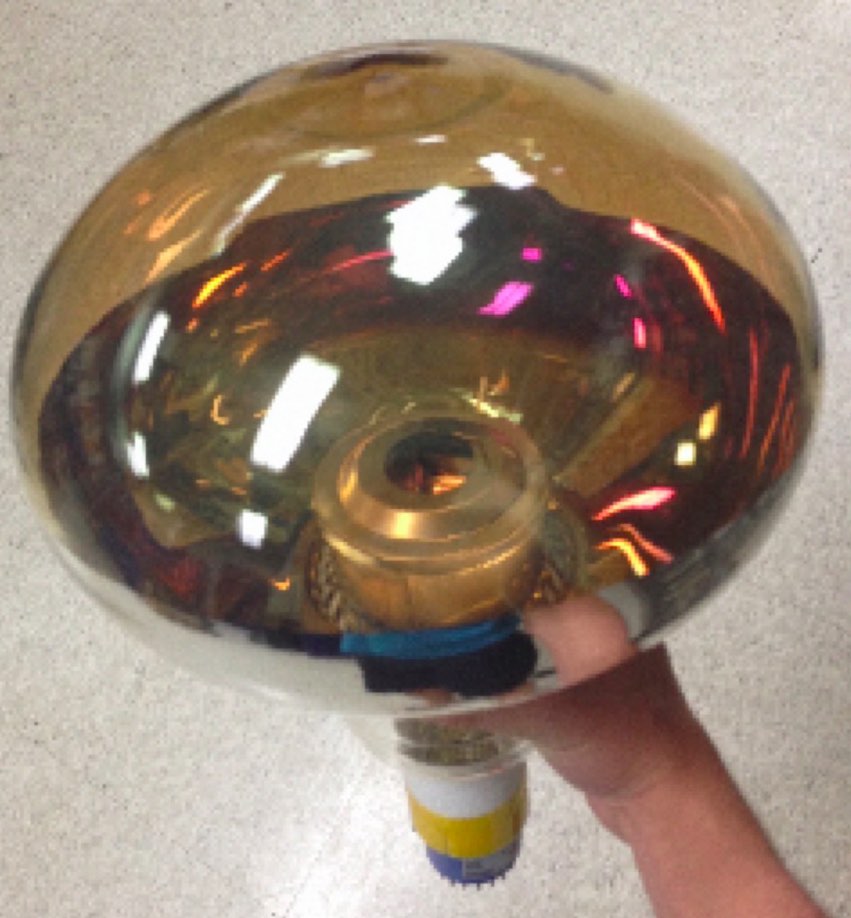}&
                \includegraphics[width=0.3\linewidth]{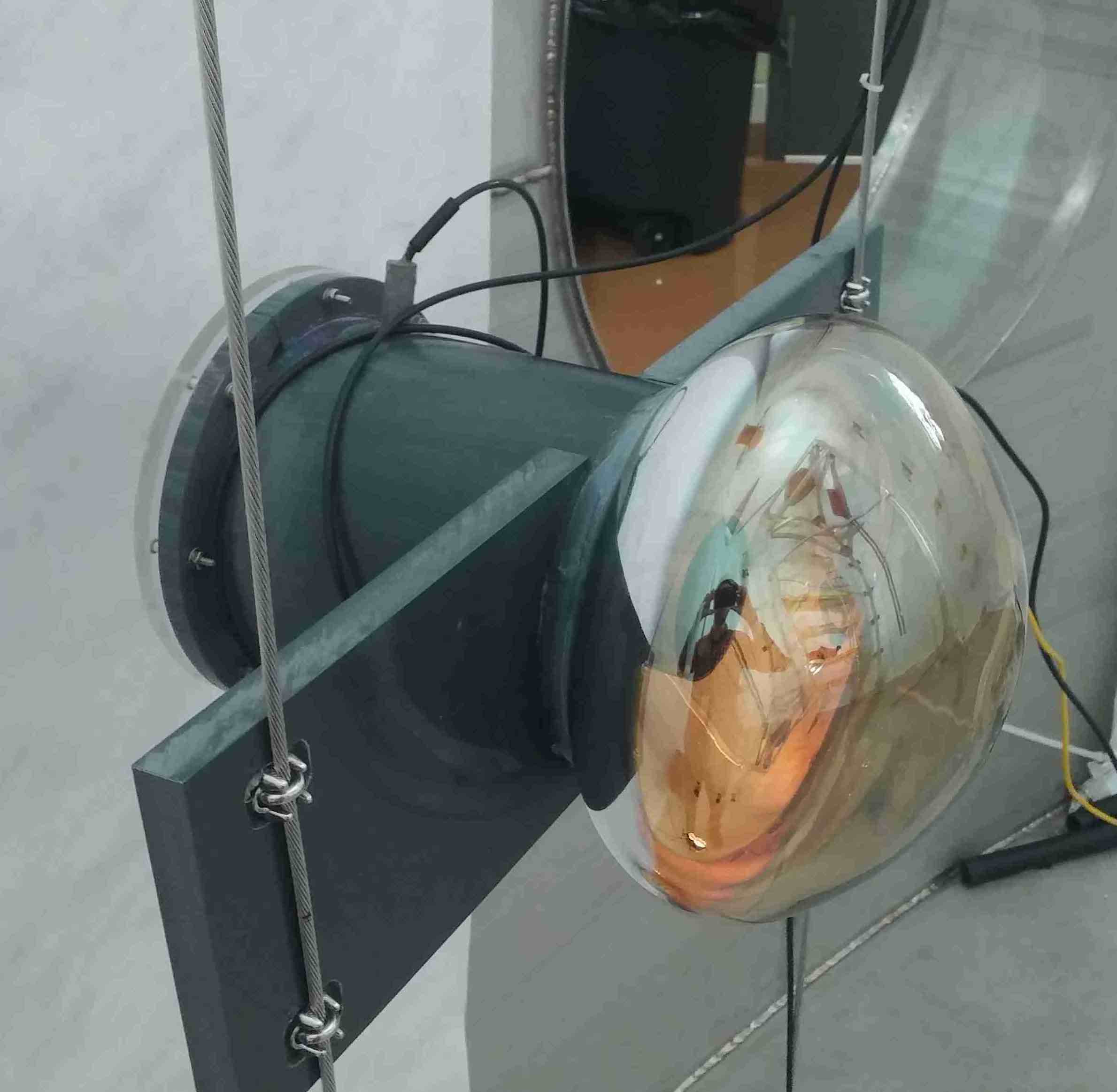}&
                \includegraphics[width=0.3\linewidth]{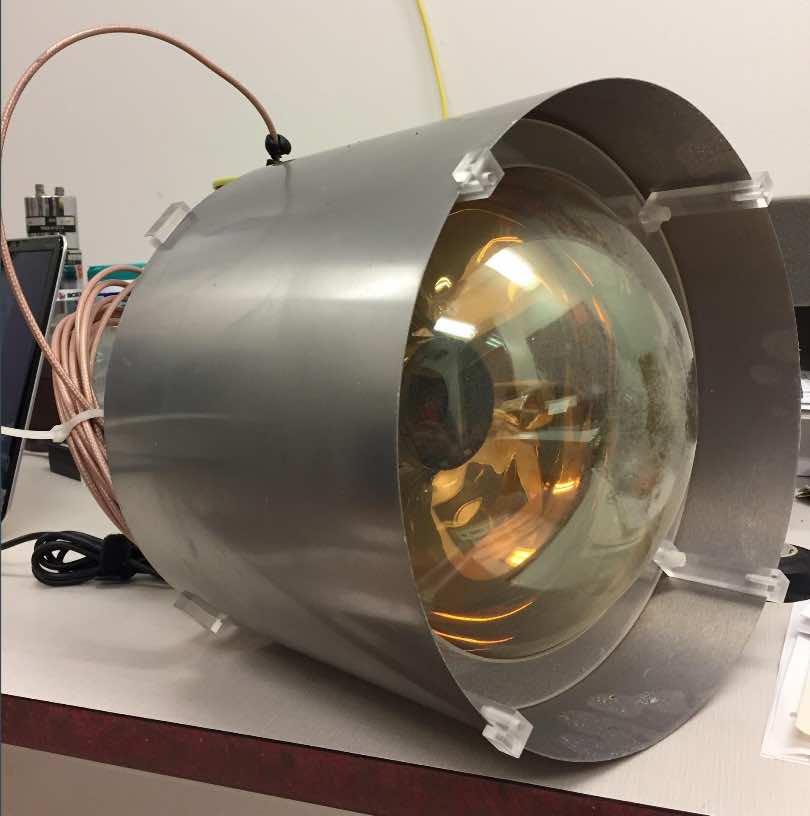}\\
         \end{tabular}         
	\caption{From left to right: ETEL 11-inch D784KFLB tube, Hamamatsu 10-inch R7081 tube from LUX, Hamamatsu 10-inch R7081 tube from Watchboy.}
	\label{fig:pMTs_pictures}
\end{figure}

In Phase~II, the NCV will be removed, and neutrons will be detected via the Cherenkov emission resulting from capture on Gd. A larger number of photodetectors will thus be necessary in order to efficiently tag and reconstruct those neutron captures. 
The ANNIE collaboration has already acquired several sets of PMTs from different sources. These sets of PMTs (see Fig.~\ref{fig:pMTs_pictures}) include but will not be limited to:
\begin{itemize}
\item 65 Hamamatsu 10-inch R7081 photomultipliers: This set consists of 20 photomultiplier tubes previously used in the water veto of the LUX experiment and 45 tubes used in the WATCHBOY detector~\cite{Dazeley:2015uyd}.
\item 22 ETEL 11-inch D784KFLB photomultipliers: Initially designed by Electron Tubes Enterprises, Ltd (ETEL) for the LBNE project, these photomultiplier tubes have been thoroughly tested~\cite{Barros:2015pjt} at UC Davis and Penn and are deemed fit to be used in the detector.
\end{itemize}
ANNIE is requesting DOE funding to purchase to purchase 40 new Hamamatsu high quantum efficiency (HQE) R5912 8-inch PMTs (for a total of 127) as replacements for the stock of 60 8-inch phototubes borrowed from UC Irvine for use in Phase I. The collaboration is exploring several options to borrow additional PMTs, which would further reduce the Run II budget request and could expand photocathode coverage beyond the baseline. 
\\

\subsection{LAPPD Housing and Deployment}
\label{sec:lappds}


UC Davis has designed a housing  for underwater deployment of LAPPDs and is in the process of making a prototype for testing before the end of ANNIE Phase~I. Several factors come into play. The housing should:
(1) be able to dissipate the significant heat load (up to 20 watts) generated by the PSEC electronics; 
(2) be easy to deploy and retrieve in ANNIE without requiring removal of the tank inner structure;
(3) protect the LAPPD from any stress to the front photocathode or vacuum seal;
(4) be neutrally buoyant in water, yet light enough to be handled by hand when in air; and, 
(5)  constantly measure the temperature in humidity inside the housing, and alarm if these parameters go outside limits.

\begin{figure}
\begin{center}
\includegraphics[width=0.9\linewidth]{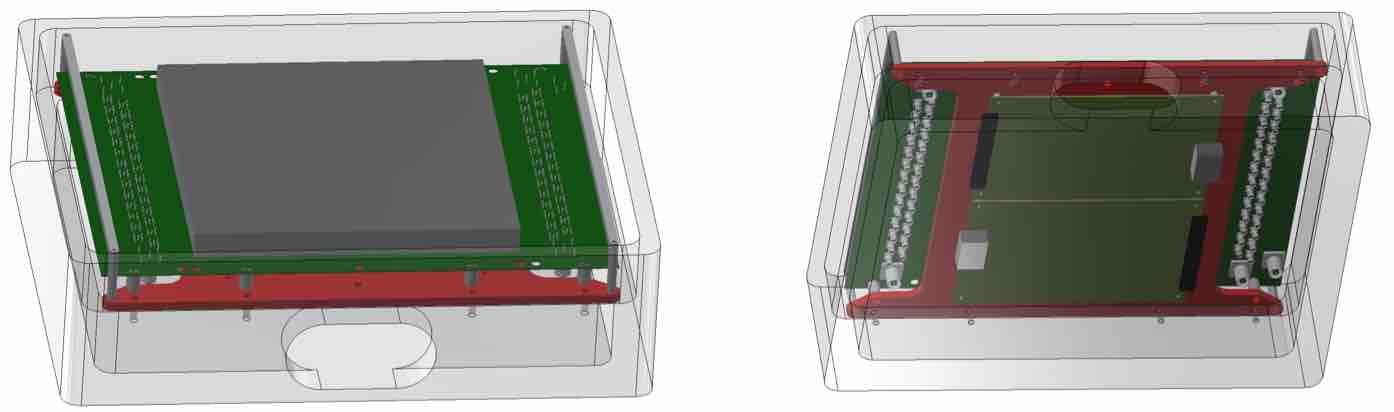}
\caption{The aluminum housing design for the LAPPD. Top view (right) and bottom view (left). The PSEC electronics sit below the MCP near a fan and temperature/humidity sensor.}
\label{LAPPD_housing}
\end{center}
\end{figure}

The housing design adopted is shown in Fig.~\ref{LAPPD_housing}. It consists of an aluminum housing (to promote thermal conductivity) with an internal brushless circulation fan and humidity and temperature sensors, plus an o-ring sealed  removable interface plate to allow changes to the  power and readout  configuration without requiring the design of a new housing. 

\begin{figure}
\begin{center}
\includegraphics[width=0.55\linewidth]{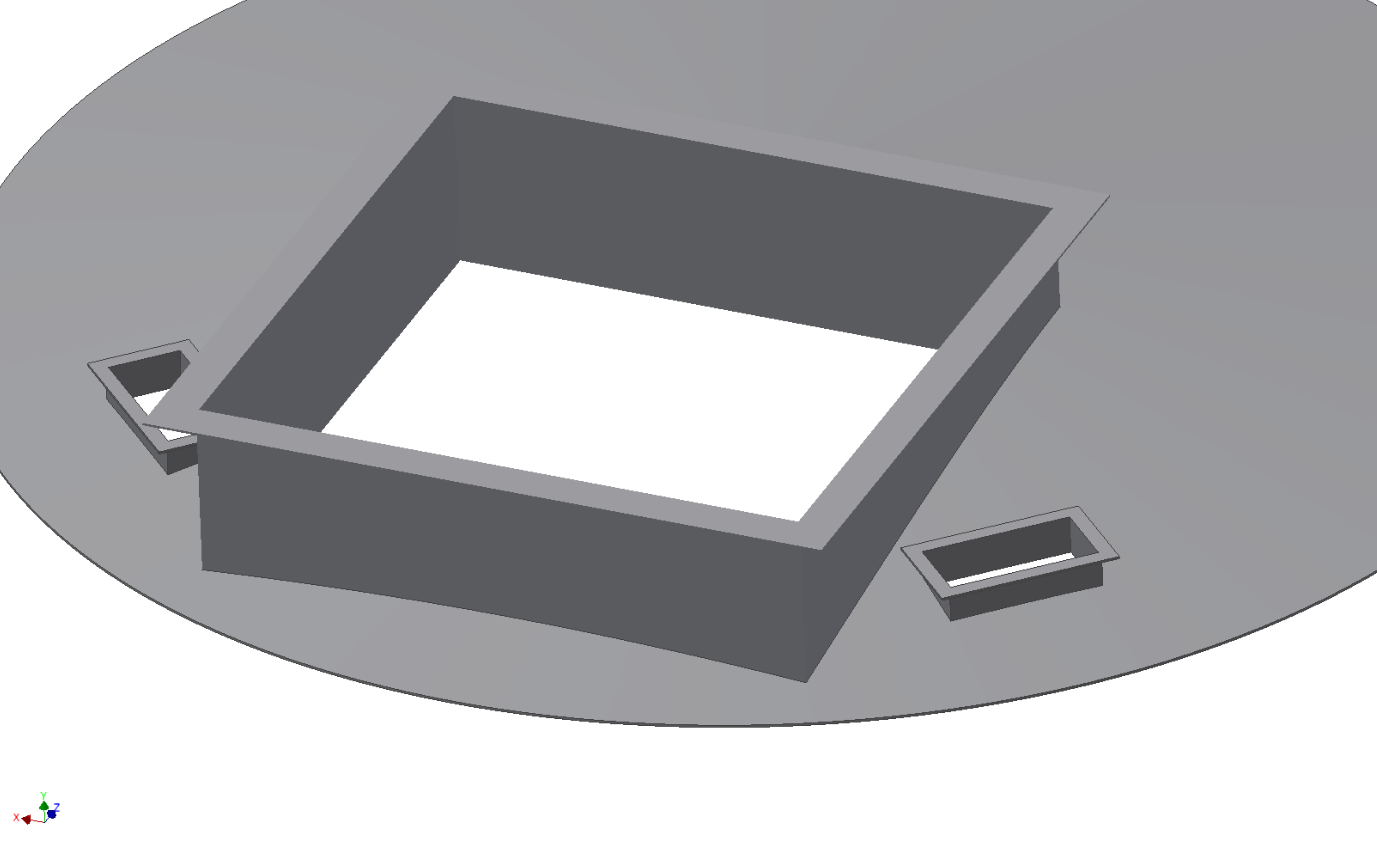}
\includegraphics[width=0.15\linewidth]{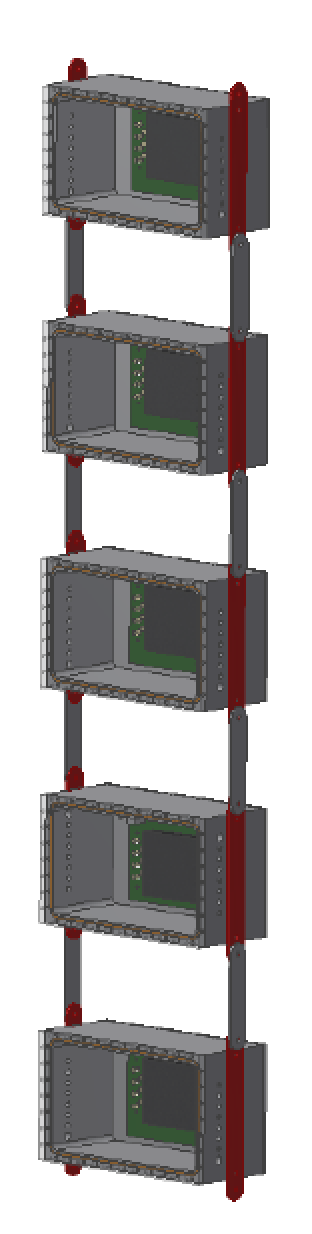}   \caption{An ANNIE LAPPD string (RIGHT) that can be lowered into the detector through the light-tight slots in the tank top (LEFT). Power and signal cables (not shown) will be attached to the side deployment lines.}
\label{LAPPD_tank}
\end{center}
\end{figure}

An important innovation in the mechanical design of the ANNIE LAPPD system is the ability to deploy LAPPDs in the already installed detector. The housings will be mounted in series on pair of cables that will fit through slots cut in the ANNIE tank top, as shown in Fig.~\ref{LAPPD_tank}. The ability to add and remove LAPPDs in-situ decouples the experiment timeline from the LAPPD delivery schedule. It also makes the most critical system for the experiment more easily accessible. 


\section{Phase~II Readout}
\label{sec:phase2_readoutupgrade}

The aim of the readout within the ANNIE water volume is to record the fast Cherenkov light flashes from a muon track sampled by LAPPDs with a timing resolution better than 100 psec, while also detecting the entire light signature from the event with conventional PMTs, including the slower and delayed ($\rm \sim 30 \mu s$) light initiated by the neutron capture. The system must also read out detector elements outside the water volume (the MRD and front veto) which provide additional constraints on the final state muon while rejecting muons that enter the tank from outside the detector.

The Phase-II ANNIE readout, developed by ISU, addresses these needs with three parallel sets of readout electronics---a custom readout based on the PSEC-4 ASIC~\cite{oberla201415gsa} developed by U Chicago for the LAPPD system, a VME-based 500MHz analog-to-digital converter~\cite{bogdan2010custom} (ADC) system developed by U Chicago for the conventional PMTs, 
and a CAMAC/NIM based system for the front veto and the MRD---
tied together by common trigger, timing and data acquisition systems. 
The readout systems are designed to run asynchronously, with GPS-derived synchronization signals and a common trigger system allowing events to be built easily offline.
All but the PSEC-4-based custom readout system are already present in ANNIE Phase-I.

\begin{figure}
	\centering
	 \begin{tabular}{c c}
                \includegraphics[width=0.37\linewidth]{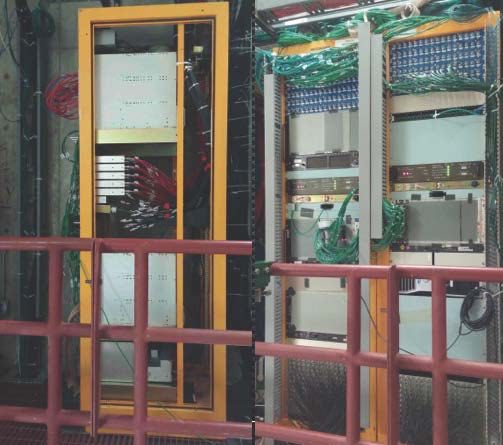}&
                \includegraphics[width=0.48\linewidth]{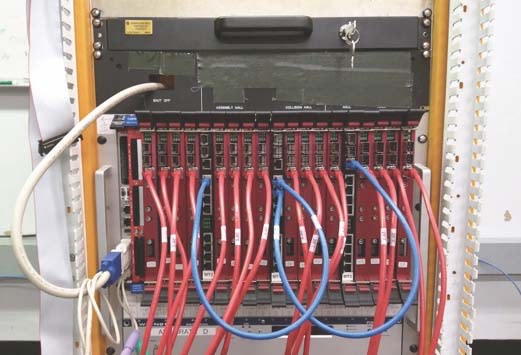}\\ 
         \end{tabular}         
	\caption{LEFT: Some of the ANNIE electronics racks in SciBooNE hall.  The leftmost rack houses two VME crates (one active and one spare), the high voltage pickoff boxes for the water-PMTs, the rack protection system and a NIM crate with level translator cards for the trigger system and discriminators for the cosmic-ray trigger.  RIGHT: The VME crate with a VME CPU card (left-most), the trigger card (2nd slot), 3 MT cards (black) and 16 ADC cards (red).}
	\label{fig:electronics_pics}
\end{figure}

The ANNIE readout electronics and data acquisition (DAQ) rely on four already installed racks on the second floor of the SciBooNE hall (pictured in Fig.~\ref{fig:electronics_pics}) and DAQ computers located on the ground level. The individual systems pictured are described in more detail below.

\label{sec:electronics}
\subsection{The ADC and PSEC Systems}
\label{sec:adcandpsec}

The requirement of picosecond time resolution and a deep buffer motivated the design of a dual readout for the water volume: PSEC4 fast digitizers for the LAPPDs and the ADC system with a deep buffer and a sampling frequency of 500MHz for the conventional photomultiplier tubes.

\subsubsection{The ADC System}
The signals from the water PMTs are digitized by 4-channel 500~MHz 6U-VME cards~\cite{bogdan2010custom} developed at the University of Chicago.  
They operate off a common clock, sync and trigger signals provided by one of the co-developed Master Trigger (MT) cards.  These cards have one upstream port and 8 downstream ports and can be cascaded to control an arbitrary number of ADC cards.  The boards are powered, configured and read-out over VME.  They also have facilities for an optional 3.1 GHz optical link.  The Phase~I configuration had 16 ADC cards and 3 MT cards in a 21 slot VME64x crate, controlled by a single VME CPU.  In Phase~II, additional ADC and MT cards will be installed into two additional VME crates with accompanying crate controllers.  The DAQ is already capable of using multiple crate controllers to control cards across multiple VME crates.

\subsubsection{The PSEC System}
The basic unit of the PSEC data acquisition system consists of a front end digitizing board, also known as the \textit{ACDC} card, and a back-end central card~\cite{bogdan2016modular} that interfaces the ACDC card(s) to the data acquisition system and provides rudimentary trigger logic. 
Each central card is capable of controlling and receiving data from up to 8 downstream devices, either additional central cards or ACDC cards.

Each 30-channel front end \textit{ACDC} card has 5 PSEC4 sampling ASICs, a clock jitter cleaner, and a control FPGA. 
Every channel has a configurable-threshold discriminator connected to the FPGA to generate board-level trigger decisions.  
Each PSEC4 ASIC has a 256 sample buffer, giving an effective buffer depth of 25~ns at 10 GHz.
A sampling-hold function (where the chip pauses between sampling and digitization) permits the PSEC4 chips to effectively exceed this buffer depth and accommodate larger trigger latencies. 
In normal Phase~II operation, each ACDC card will have a low board-trigger threshold, which will temporarily put the card into sampling-hold while forwarding the trigger information to the central card. The central card will use inputs from all available ACDC cards to determine if a triggerable physics event has occurred, and issue a digitize command to each participating ACDC card. If not each card will automatically return to sampling mode after a set amount of time has passed.

The PSEC4 chip and the ACDC card have been extensively tested. Some firmware and software work is needed to fully enable the trigger scheme described above and some software development is required to integrated it with the existing DAQ. Work is already underway at ISU to complete the development and integration of this system. 

Sufficient stock of PSEC-4 chips and spares is owned by collaborator M. Wetstein. Phase~II funding will be requested from the DOE to purchase and populate the 40 ACDC cards and 7 central cards necessary to instrument the system for up to 20 LAPPDs.


\subsection{The MRD and Front Veto Readout}
 
The MRD and front veto require less precision than the readout for the water volume. Reconstruction of stopped muons and vetoing of rock muons require only crudely time-stamped signals indicating which scintillator paddles pulsed. These requirements can be met with the CAMAC-based Time to Digital Converter (TDC) system developed and successfully operated by the SciBooNE collaboration, using equipment borrowed from FNAL PREP. All of the needed crates are already installed, along with sufficient TDC cards to read the limited number of 82 Phase~I channels: (1) 26 front veto channels (2 layers of 13 paddles), (2) 26 MRD layer 2 channels (2 sets of 13 paddles), and (3) 30 MRD layer 3 channels (2 sets of 15 paddles). The cards needed to accommodate the remaining 306 MRD channels are available from PREP.

\subsection{Triggering and Timing}

Triggering of ADC, PSEC, and MRD systems of the detector is centrally managed by the Triggering and Timing system.  This system produces trigger signals for the entire readout and distributes synchronization signals to the ADC and PSEC systems. 

Trigger logic and distribution in ANNIE is handled by a CAEN V1495 general purpose FPGA VME card, running custom purpose-built firmware.  
The trigger logic is implemented as a logical-OR of maskable triggers, with priority given to BNB (beam) triggers. The non-BNB triggers include but are not limited to cosmic ray and LED flasher triggers for calibration and multiplicity triggers from the ADC and PSEC systems during a configurable time window after the BNB. The ADC multiplicity trigger is used for detection of neutron captures within the beam window in the NCV in Phase~I and will be used for the same purpose in the entire fiducial volume in Phase~II. The PSEC multiplicity trigger, which has yet to be implemented, will be used to identify muons (PSEC; Phase~II) within the beam window.

Each trigger is timestamped by the trigger card and tagged with its source, so that it can be used offline to associate the recorded data with the source of the trigger.  
The timestamps are coarsely referenced to special time-synchronization events produced by the driver, which also records the CPUs Real-Time Clock for each event.  They are more precisely referenced to a GPS-derived one pulse-per-second signal at the start of each UTC second.  The ADC uses these same synchronization signals to timestamp recorded events, as will the PSEC system.

\subsection{DAQ}
\label{sec:daq}

\begin{figure}
        \begin{center}
                \includegraphics[width=0.55 \linewidth]{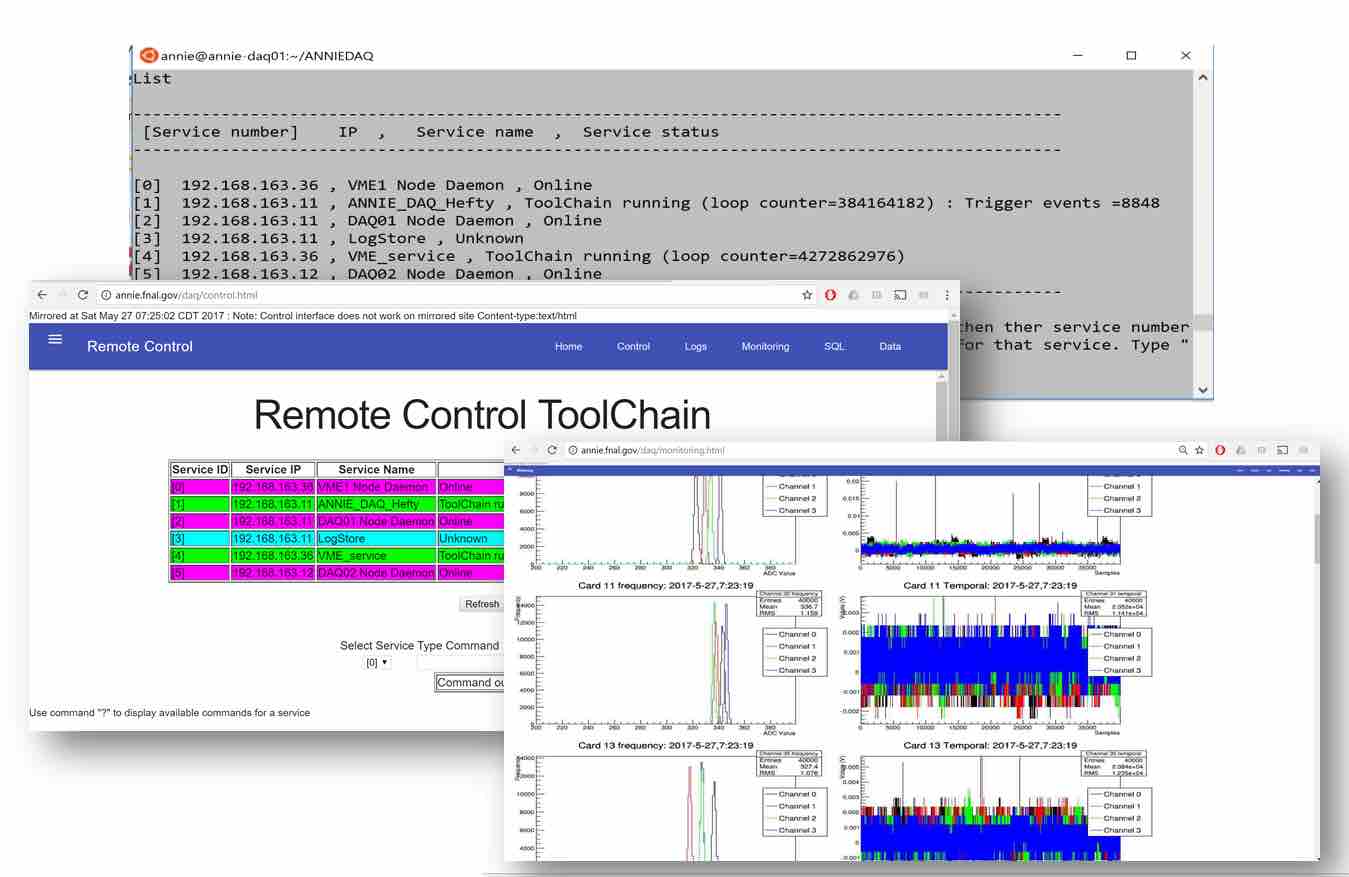}
        \end{center}
        \caption{ A sample of DAQ Web and command line interfaces and monitoring}
                \label{fig:DAQWebpage}
\end{figure}

The ANNIE DAQ is built on the ToolDAQ framework by Ben Richards from QMUL. This C++ DAQ framework has inbuilt dynamic service discovery and fault tolerant communication infrastructure.
The modular nature of its run and node control tools facilitates development and allows for dynamic runtime changes to the DAQ framework. Taken together these features allow the DAQ to be easily scaled to any size across multiple network nodes and data streams.
The low number of dependencies also makes it compatible with a wide range of modern and legacy hardware. 
ToolDAQ also provides a native web monitoring and control interface that has been expanded upon and customized for use by ANNIE (see Fig.~\ref{fig:DAQWebpage}).  This allows all aspects of the DAQ 
to be managed from remote web browsers.

\begin{figure}
        \begin{center}
                \includegraphics[width=0.55 \linewidth]{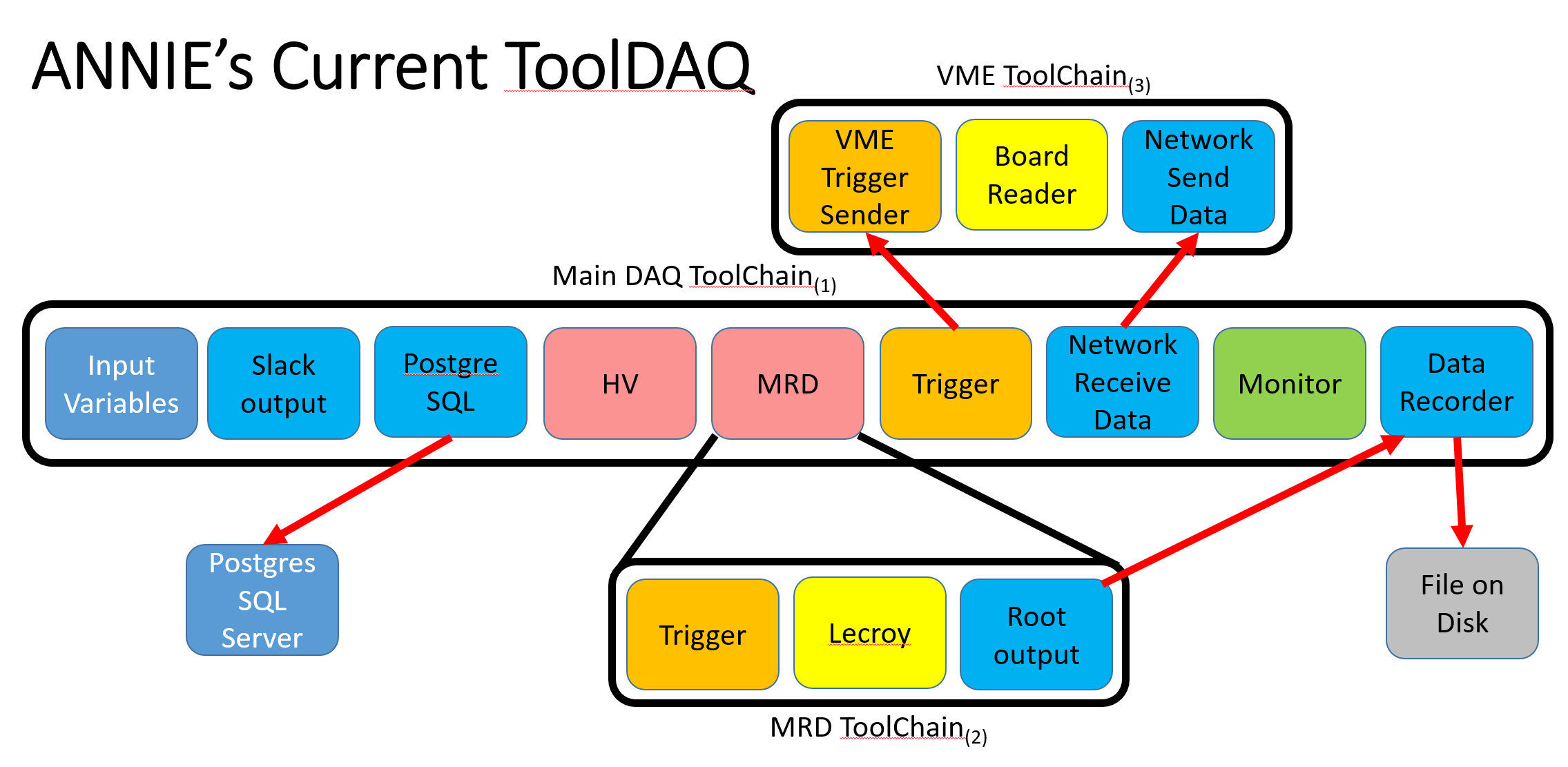}
        \end{center}
        \caption{ A Schematic of ANNIES DAQ softare showing modular Tools and ToolChains}
                \label{fig:ANNIEsToolDAQ}
\end{figure}

ANNIE's phase I ToolDAQ implementation consists of three ToolChains (objects that manage the execution of a set of modular tools) (Fig.~\ref{fig:ANNIEsToolDAQ}).
The main ToolChain$_{(1)}$ runs on the rack mounted server with a redundant backup, and handles a number of key functions including run configuration and entry into the PostgreSQL run database, 
trigger configuration and logging, 
automated control and log recording of the HV system, 
and monitoring plot production. It also communicates with the other two ToolChains and is responsible for concatenating the four asynchronous data streams (Water PMT data, MRD data, Trigger card data, configuration and log record data) into a single ROOT output file.
%
The remaining two ToolChains manage communication with the water-volume readout and MRD electronics and run respectively on the VME CPU module and in parallel threads on the rack mounted server.

The data from the online DAQ system then passes through an offline post processor (ANNIEPOST) which is also built with ToolDAQ. This application adds information about the system geometry (e.g. PMT locations), applies calibrations (e.g. energy scale, pedestal subtraction), and separates the ADC readout buffer into unique triggers.

The asynchronous and modular nature of the Phase~I ANNIE DAQ means that with additional interface hardware (more CCUSBs and VME CPUs) it can automatically accommodate the additional readout electronics and data streams required for Phase~II.
The lone exception is for the LAPPD readout electronics, which do require a small amount of alteration to the existing DAQ in terms of drivers to read the hardware and code to handle a new type of asynchronous data stream. 
Hooks for these alternatives already exist in the design and structure of the existing DAQ. The ISU test stand will be used to develop and test these additions before deployment.  


\section{High Voltage and Slow Controls}\label{sec:phase2_hvandsc}
The high voltage (HV) supply system for the Phase 1 detector was constructed using equipment obtained from Fermilab's Physics Resource Equipment Pool (PREP), which loans equipment donated by past experiments free of charge. The system currently consists of a single CAEN SY527 crate, with five A938 24-channel negative HV cards for the front veto and instrumented MRD layers, and five A734 16-channel positive HV cards for the tank PMTs. In Phase~II the number of channels required increases significantly. 
In order to minimize cost while accommodating the space constraints of the experimental hall, ANNIE plans to supply 160 positive HV channels with the CAEN SY527 crate and 480 negative HV channels with a more modern, higher-channel-density SY4527. These numbers exclude the LAPPDs, for which HV is currently planned to be generated using separate custom supply cards. The Sheffield group has responsibility for this system. 

Monitoring and control of the high voltage system is performed via a LabVIEW based program on a dedicated computer. The SY527 system is read over a RS232 serial communication link; the SY4527 will use Ethernet. 

\begin{table}[]
  \centering
  \caption[]{VME device count for the Phase~I and two possible Phase~II scenarios.}
  \label{tab:vme_card_table}
  \begin{tabular}{lccc}
    \toprule
    ~      &  ~              & \# for 128 PMTs & \# for 150 PMTs \\
    Device & \# for Phase~I  & and 5 LAPPDs    & and 20 LAPPDs \\
    \midrule
    ADC Cards & 16 & 32 & 38 \\
    MT Cards (Level 2 + Master)  & 2+1 & 4+1 & 5+1 \\
    ANNIE Central Cards (double-width) & 0 & 3 (6 slots) & 7 (14 slots) \\
    Trigger Cards & 1 & 1 & 1 \\
    Crate Controllers & 1 & 3 & 3 \\
    VME Crates (21 slots each) & 1 & 3 & 3 \\
	\midrule
	{\bf Total Slots Used:} & 21 & 47 (16 free) & 62 (1 free) \\
	\bottomrule
  \end{tabular}
\end{table}

\section{Conclusion}

Following the recommendation of the Physics Advisory Committee, the Fermilab directorate granted approval to begin work on ANNIE Phase~I. The Phase~I detector was installed and commissioned by April 2016 and the collaboration has since been collecting neutron background data. The collaboration has made rapid progress in analyzing these data and has demonstrated that the neutron backgrounds in the hall are low enough to allow the planned physics measurement. 

In 2016 the DOE approved funding for ANNIE through the Intermediate Neutrino program (INP). This funding enabled ANNIE to purchase required electronics and to perform work needed to ready LAPPDs for use in ANNIE, specifically: (1) development of the LAPPD readout electronics, (2) design and fabrication of the water proof housing, and (3) testing LAPPD prototypes from the manufacturer, Incom Inc.  The collaboration has made significant progress on each of these tasks that demonstrate LAPPD readiness by the ANNIE experiment.  

With working LAPPD modules now being tested by the ANNIE collaboration and with the completion of ANNIE’s Phase~I data taking, the collaboration will submit a proposal for the ANNIE Phase~II physics measurement to the Department of Energy in early Fall 2017. 

The design and construction of the Phase~I detector established all of the major infrastructural components of the ANNIE experiment. The Phase~II modifications will consist of increasing the number and modifying the arrangement of PMTs, and introducing the LAPPD modules. Phase~I also established the layout and overall architecture of the ANNIE readout and DAQ. Work is underway to finish integrating the PSEC4-system which will read out the LAPPDs, whereas the rest of the system upgrade will consist of adding more channels. The requested Fermilab support for Phase II is an incremental evolution to the work already invested in ANNIE. 

ANNIE brings to FNAL a new technology direction and its success supports new physics that will aid the understanding of neutrino interactions and therefore have a direct impact on the flagship long-baseline neutrino oscillation program as well as on other interesting physics. 




\newpage
\bibliographystyle{unsrt}
\bibliography{bibliography}


\end{document}